\documentclass[prd,aps,preprint,tightenlines,superscriptaddress,nofootinbib,showpacs,showkeys]{revtex4}
\usepackage{mathrsfs}
\usepackage{amsfonts}
\usepackage{array}
\usepackage{epsfig}

\usepackage{amsmath}    
\usepackage{graphicx}   
\usepackage{verbatim}   
\usepackage{color}      
\usepackage{subfigure}  
\usepackage{hyperref}   

\begin{document}
\newcommand{\subd}{./}
\newcommand{\Dzero}{D0}

\newcommand{\DATE}{\today}

\begin{flushright}
SMU-HEP-12-23
\end{flushright}

\title{The CT10 NNLO Global Analysis of QCD}

\author{Jun Gao}
\affiliation{
Department of Physics, Southern Methodist University,
Dallas, TX 75275-0181, USA}
\author{Marco Guzzi}
\affiliation{
Deutsches Elektronensynchrotron DESY, Notkestrasse 85 D-22607 Hamburg, Germany}
\author{Joey Huston}
\affiliation{
Department of Physics and Astronomy, Michigan State University,
East Lansing, MI 48824-1116, USA}
\author{Hung-Liang Lai}
\affiliation{
Taipei Municipal University of Education, Taipei, Taiwan}
\author{Zhao Li}
\affiliation{
Institute of High Energy Physics, Chinese Academy of Sciences, Beijing 100049, China}
\author{Pavel Nadolsky}
\affiliation{
Department of Physics, Southern Methodist University,
Dallas, TX 75275-0181, USA}
\author{Jon Pumplin}
\affiliation{
Department of Physics and Astronomy, Michigan State University,
East Lansing, MI 48824-1116, USA}
\author{Daniel Stump}
\affiliation{
Department of Physics and Astronomy, Michigan State University,
East Lansing, MI 48824-1116, USA}
\author{C.--P. Yuan}
\affiliation{
Department of Physics and Astronomy, Michigan State University,
East Lansing, MI 48824-1116, USA}
\affiliation{
Center for High Energy Physics, Peking University,
Beijing 100871, China}

\begin{abstract}
{
We present next-to-next-to-leading order (NNLO) parton distribution 
functions (PDFs) from the CTEQ-TEA group. The CT10NNLO PDF fit is 
based on essentially the same global data sets used in the CT10 and
CT10W NLO PDF  
analyses. After exploring the goodness of the fits to the HERA
combined data and  
the Tevatron jet data, we present various predictions at NNLO accuracy 
for both existing and forthcoming precision measurements from the CERN
Large Hadron Collider.  
The range of variations in the gluon distribution introduced by
correlated systematic effects in inclusive jet production is also
examined. 
}

\end{abstract}




\maketitle

\tableofcontents
\newcommand{\figFtctotA}
{
\begin{figure}[tbh]
\begin{center}
\includegraphics[height=0.75\textheight]{\subd/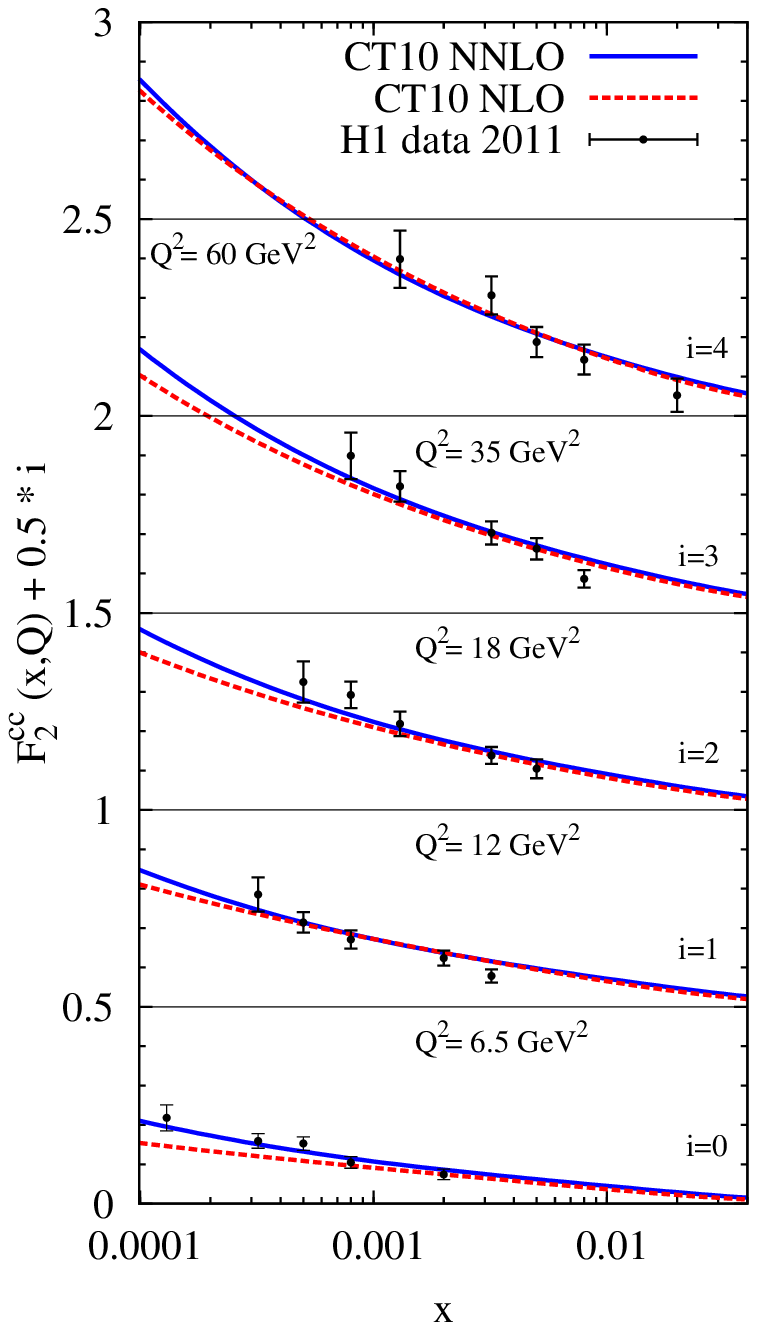}
\caption{The semi-inclusive heavy-quark function $F_{2}^{c\bar c}$ 
as a function of $x$ in different bins of $Q$. The CT10NLO, CT10NNLO
predictions and H1 data \cite{Aaron:2011gp} are compared. 
\label{fig:F2c_Qbin}}
\end{center}
\end{figure}

}

\newcommand{\figALLBANDS}
{
\begin{figure}[tbh]
\begin{center}
$
\begin{array}{c}
\includegraphics[width=6in]{\subd/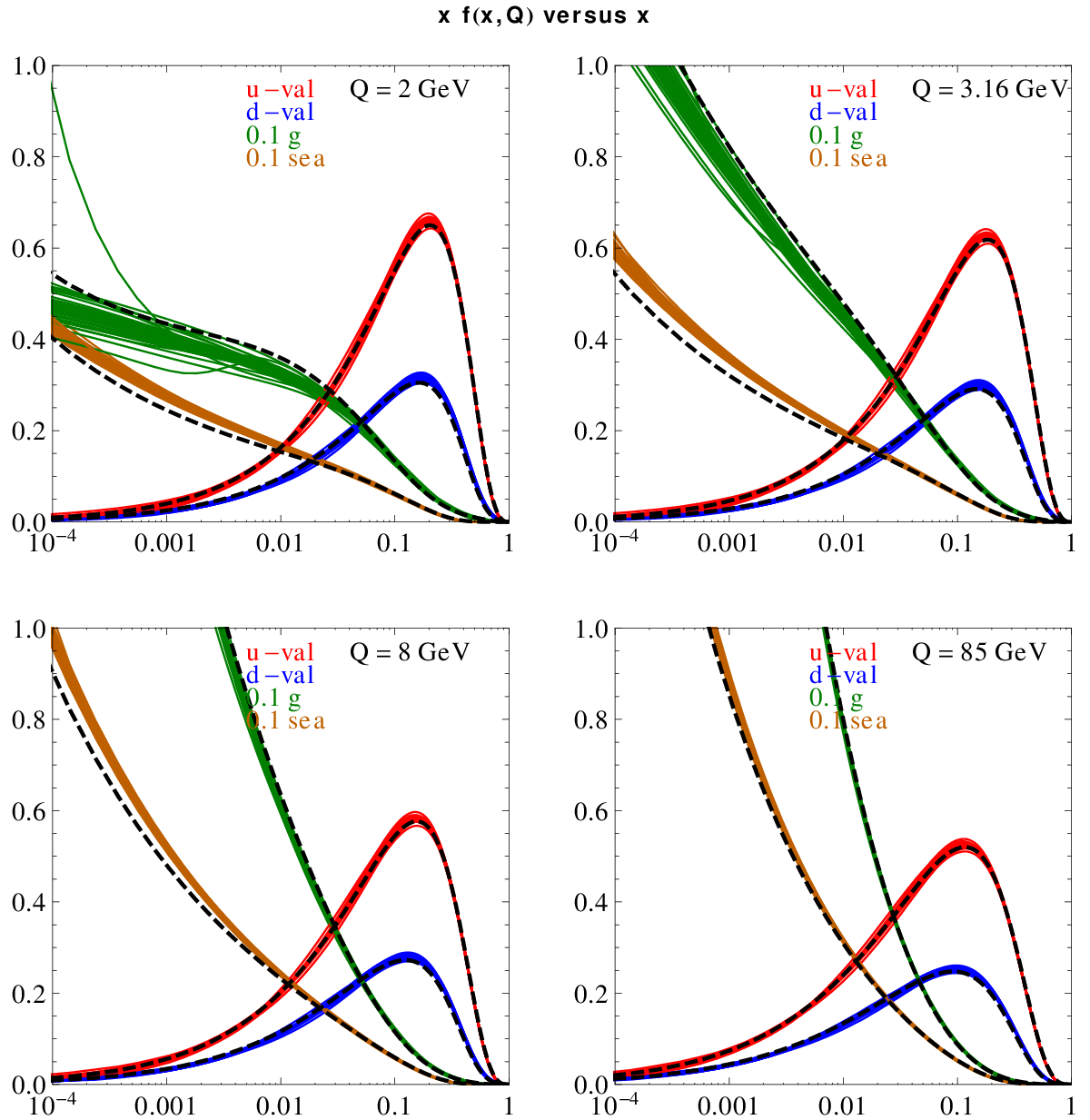}
\end{array}
$ \caption{CT10NNLO parton distribution functions. These figures
show the Hessian error PDFs from the CT10NNLO analysis. Each
graph shows $x\,u_{\rm valence} = x(u-\overline{u}), x\,d_{\rm
valence} = x(d-\overline{d}), 0.10\,x\,g$ and
$0.10\,x\,q_{\textrm{sea}}$ as functions of $x$ for a fixed
value of $Q$. The values of $Q$ are $2,~ 3.16,~ 8,~ 85$\ GeV. 
The quark sea contribution is 
$q_{\rm sea}= 2(\overline{d}+\overline{u}+\overline{s})$. The dashed
curves are the central CT10 NLO fit. \label{fig:ALLBANDS}}
\end{center}
\end{figure}
}
\newcommand{\figbestnnlovsnloA}
{
\begin{figure}[tbh]
\begin{center}
\includegraphics[width=4in]{\subd/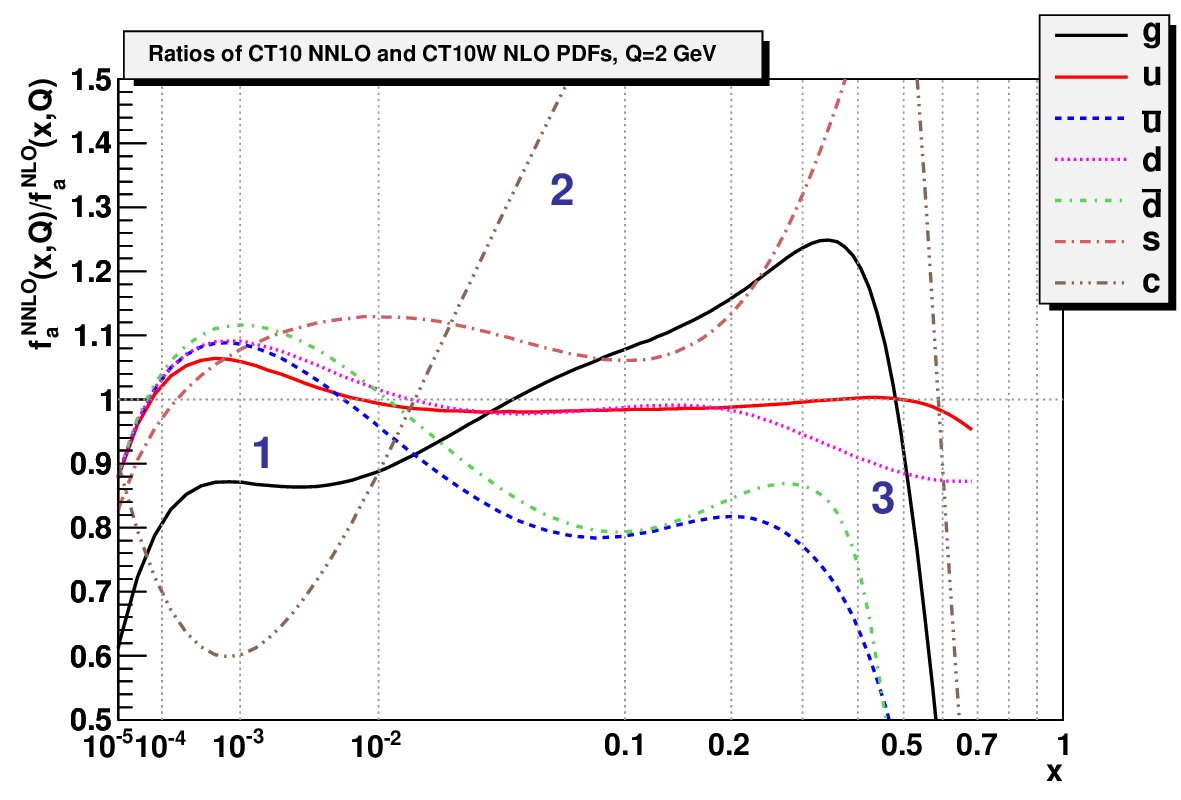}
\caption{ Ratios of various CT10NNLO central fit parton
distributions to those of the CT10W central fit, at $Q=2$\,GeV.
\label{fig:bestnnlovsnloA}}
\end{center}
\end{figure}
}
\newcommand{\figbestnnlovsnloB}
{
\begin{figure}[tbh]
\begin{center}
\includegraphics[width=4in]{\subd/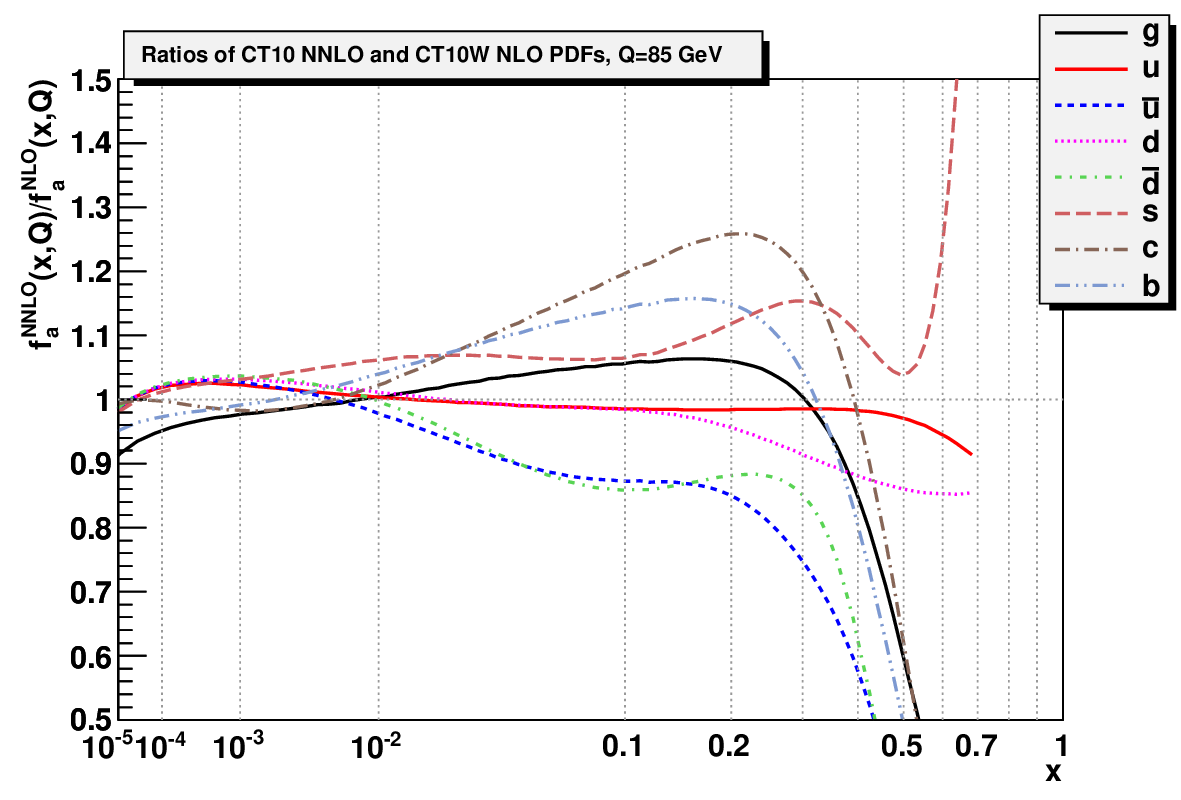}
\caption{Same as Fig.~\ref{fig:bestnnlovsnloA}, at $Q=85$\,GeV.
\label{fig:bestnnlovsnloB}}
\end{center}
\end{figure}
}
%
\newcommand{\figNNLOvsNLO}
{
\begin{figure}[h]
\begin{center}

\includegraphics[width=0.49\textwidth]{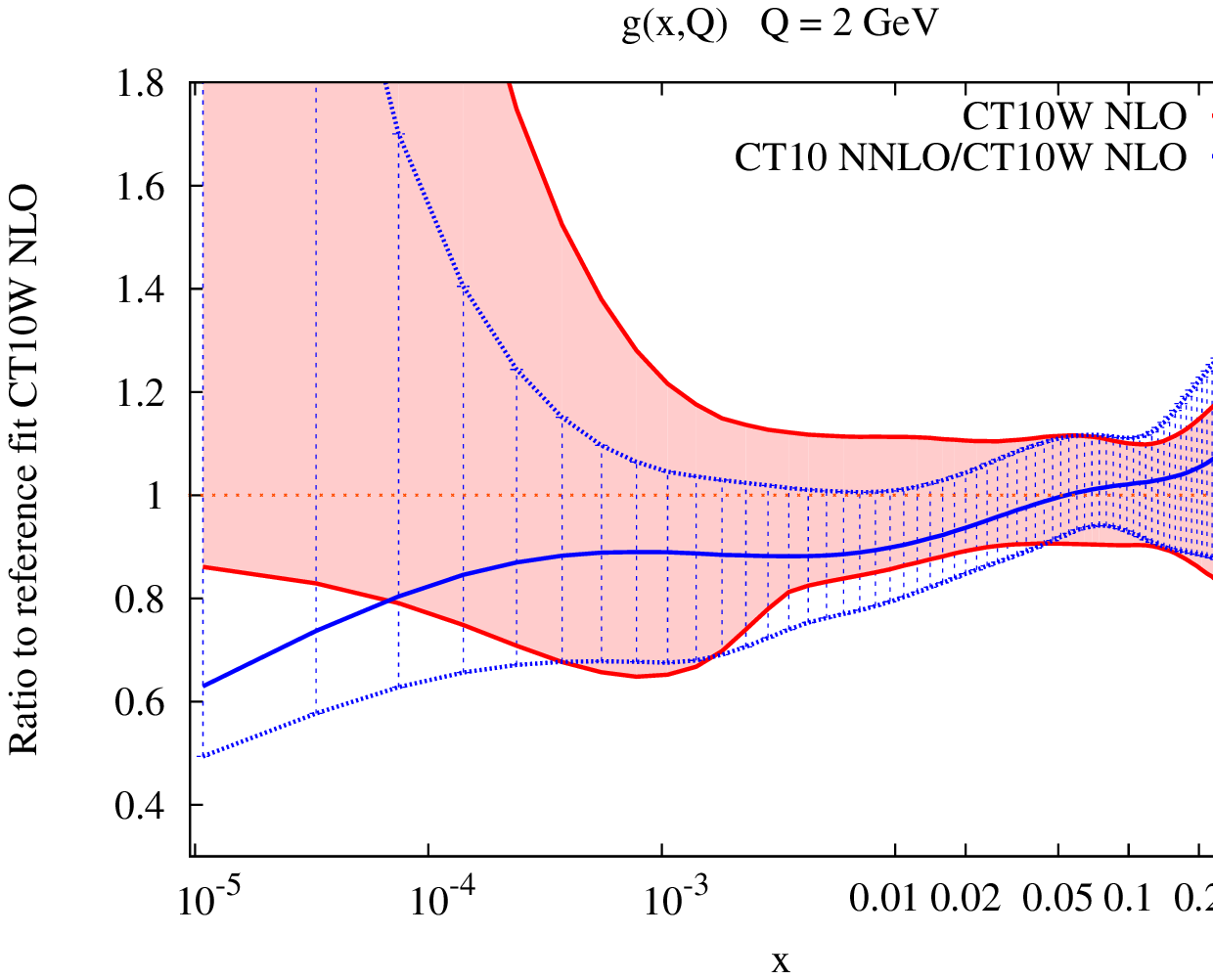}
\includegraphics[width=0.49\textwidth]{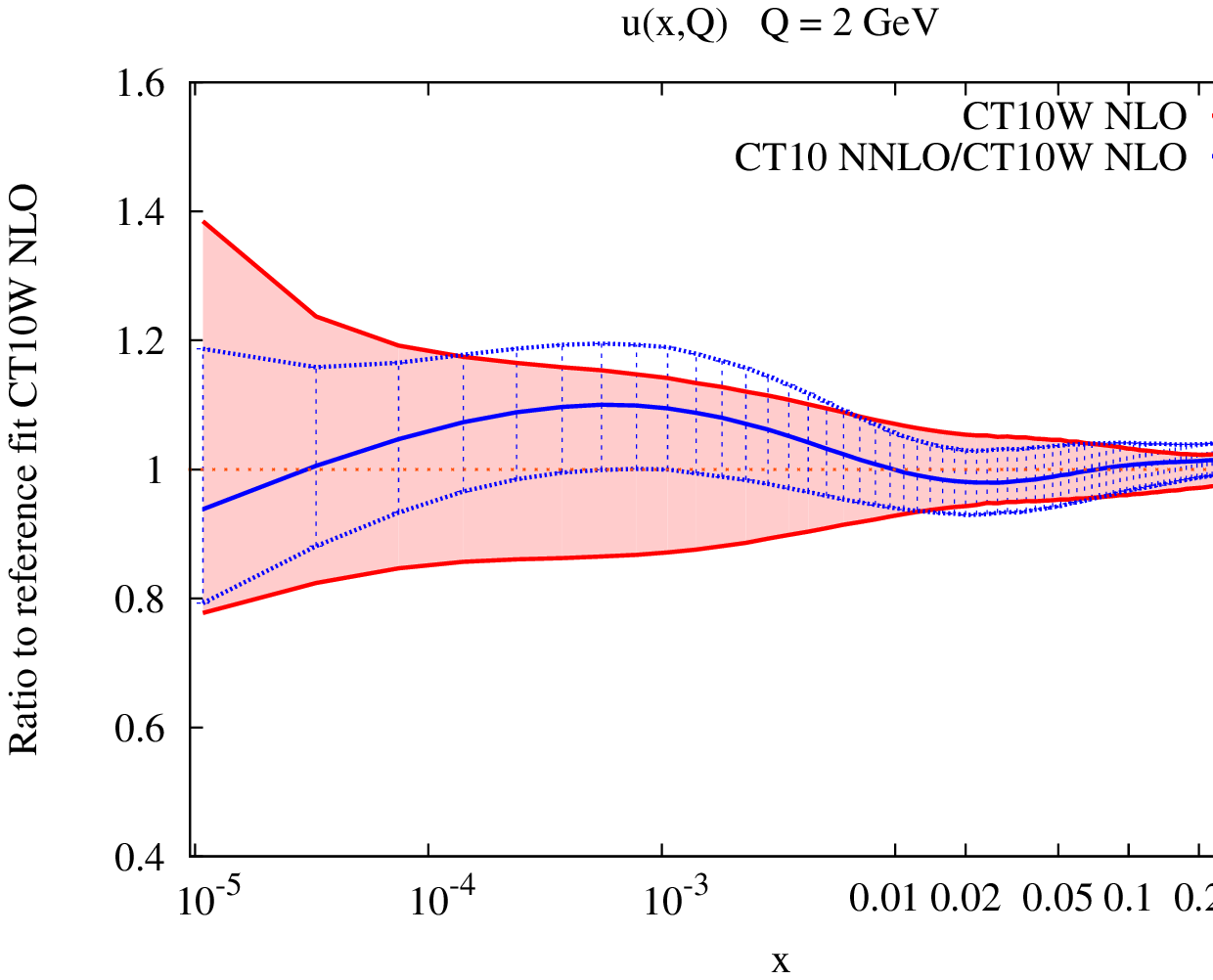}\\
\includegraphics[width=0.49\textwidth]{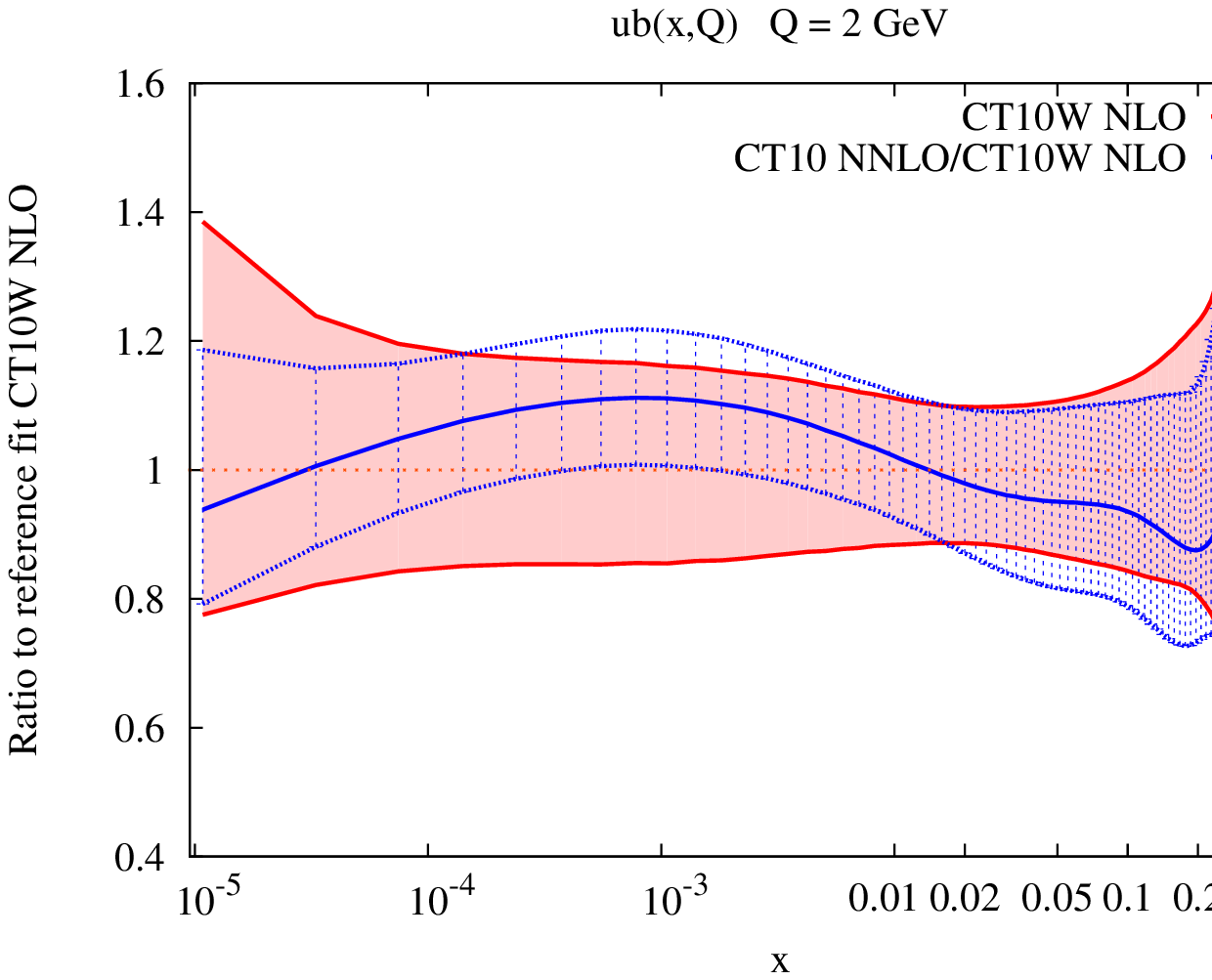}

\caption{Comparison of CT10NNLO error PDFs to CT10W (NLO) error PDFs.
All are expressed as ratios to the central fit of the CT10W 
set. \label{fig:NNLOvsNLO}}
\end{center}
\end{figure}
}
\newcommand{\fignnlovsmstw}
{
\begin{figure}[tbh]
\begin{center}
\includegraphics[width=4in]{\subd/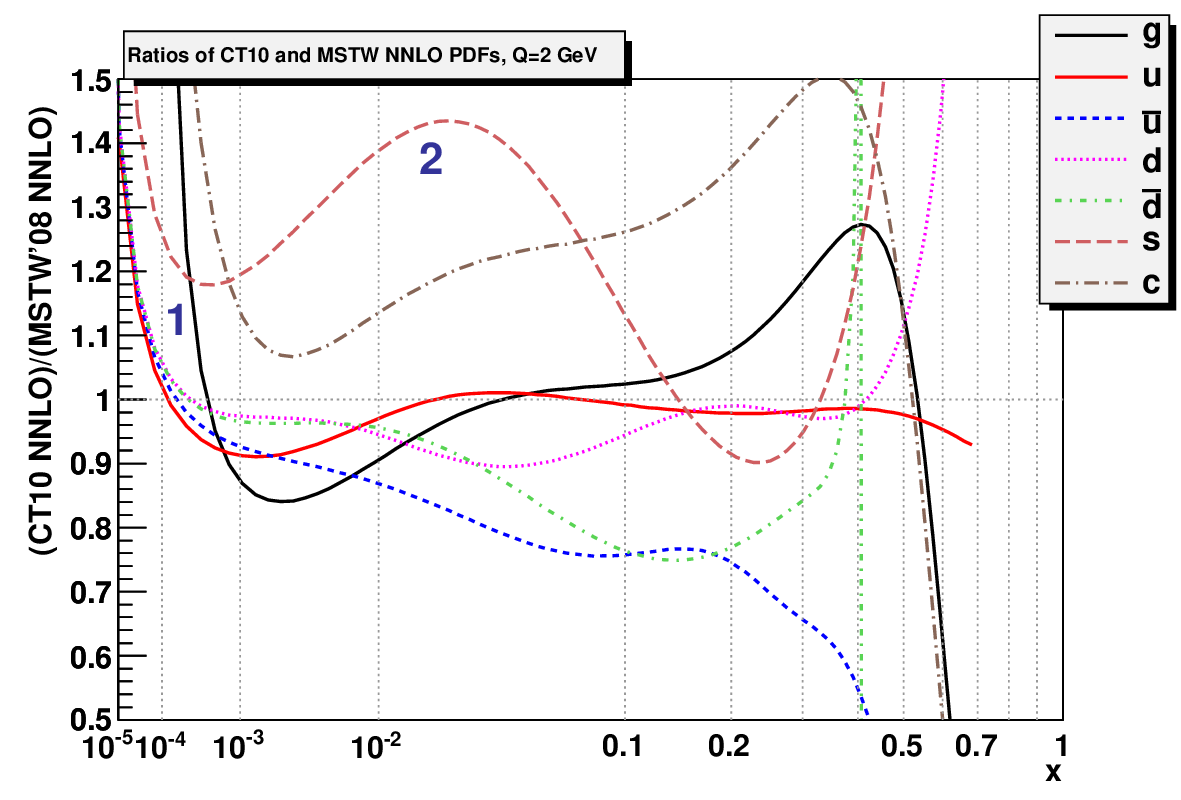}
\caption{ Ratios of various CT10NNLO central fit parton
distributions to those of the MSTW2008NNLO central fit, at
$Q=2$\,GeV. \label{fig:nnlovsmstw}}
\end{center}
\end{figure}
}
%
\newcommand{\figHERAdata}
{
\begin{figure}[tbh]
\begin{center}
$
\includegraphics[width=3.0in]{\subd/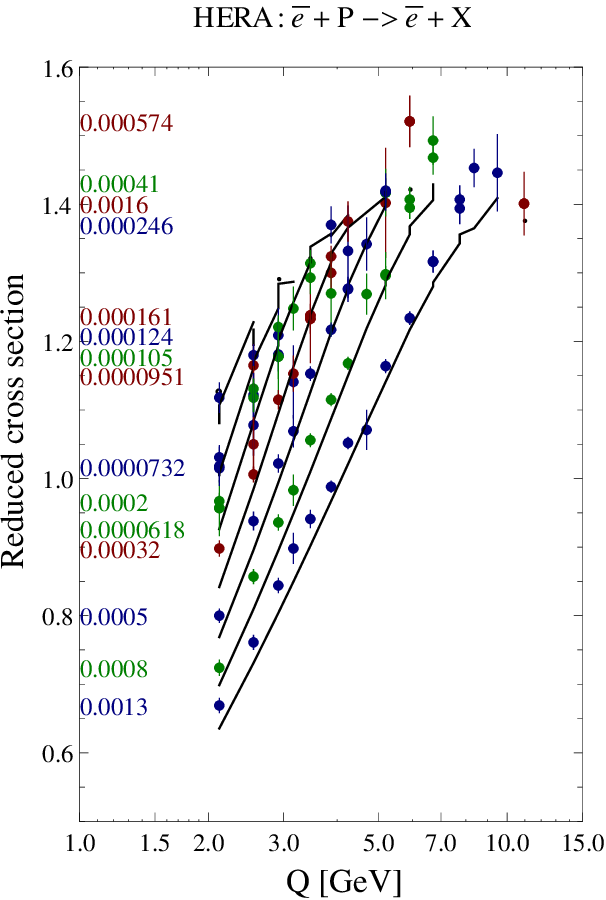}
\hfill
\includegraphics[width=3.0in]{\subd/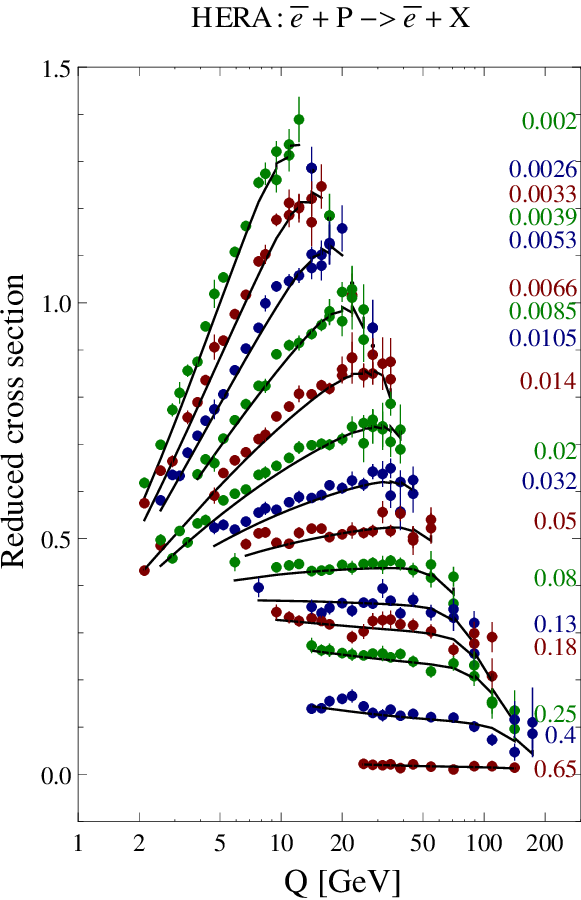}
$ \caption{The HERA combined data, with comparison to theory. The
reduced cross section $\sigma_{r}(x,Q)$ for NC DIS of positrons
is plotted as a function of $Q$ for 33 values of $x$. The graph
on the left has $x$ values in the range $0 < x <0.002$; the one on the
right has $0.002 \leq x \leq 0.65$. The points are the published
central values, while error bars correspond to the uncorrelated errors only.
The red curves show the theoretical value of the reduced cross
section $\sigma_{r}(x,Q)$ computed with CT10NNLO PDFs.
\label{fig:HERAdata}}
\end{center}
\end{figure}
}
\newcommand{\figHERADoverT}
{
\begin{figure}[tbh]
\begin{center}
$
\includegraphics[width=5.0in]{\subd/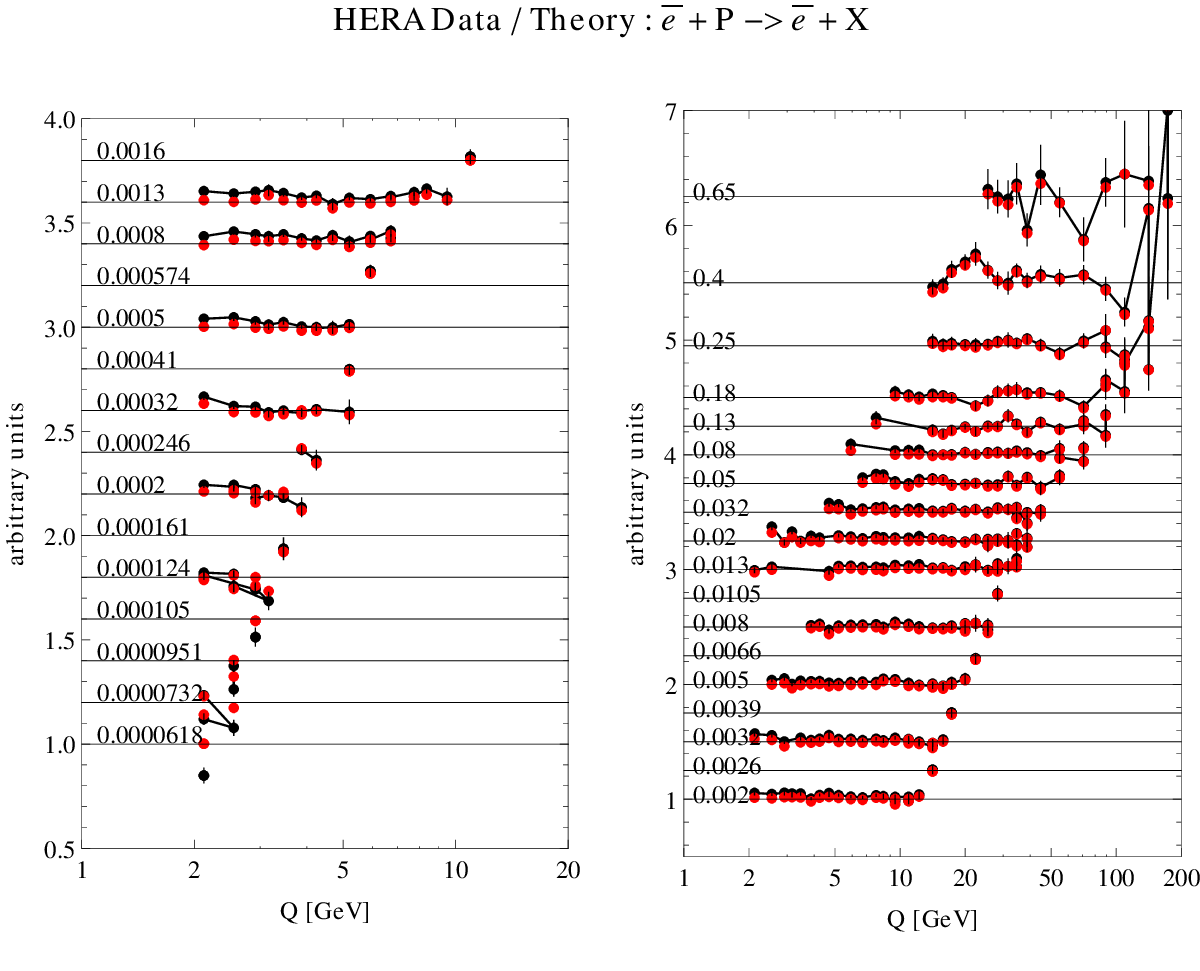}
$
\caption{
 Ratio of data divided by theory, for the reduced cross section for positron-proton
NC DIS. Theory is represented by the horizontal lines. 
Indicated in both figures are the $x$-values: $0 < x \leq 0.0016$ in the left inset; 
$0.002 \leq x \leq 0.65$ in the right inset. Black dots are experimental central
values with uncorrelated errors only. Red dots are the optimally shifted data, 
where the shifts are evaluated according to the Hessian analysis of the systematic errors.
\label{fig:HERADoverT}}
\end{center}
\end{figure}
}
\newcommand{\figHERAresiduals}
{
\begin{figure}[tbh]
\begin{center}
$
\includegraphics[width=2.8in]{\subd/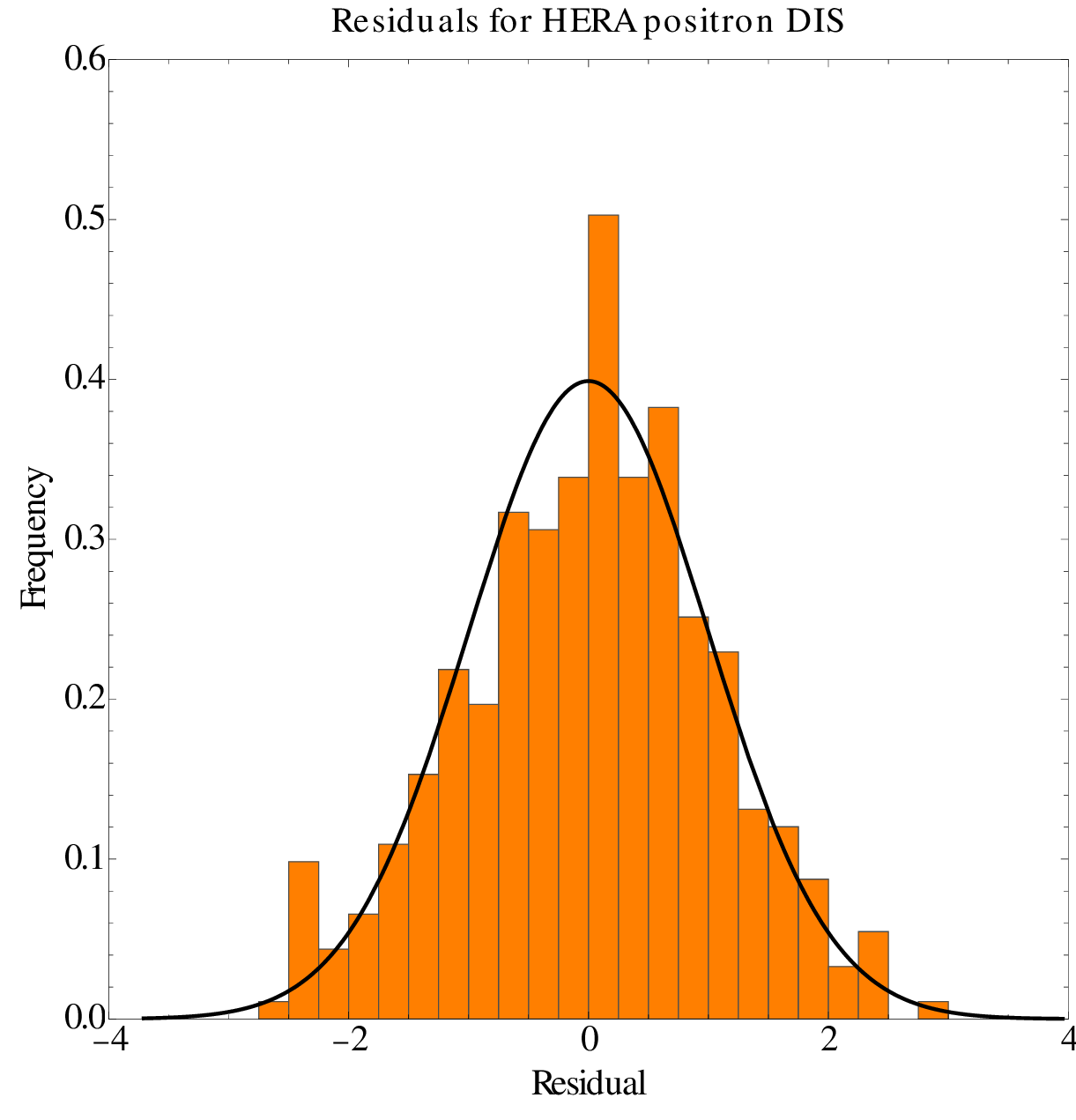}
$
\caption{
Histogram of residuals for the HERA combined data
for positron-proton NC DIS.
The residuals are defined in Eq.\ (\ref{eq:residuals}).
The solid curve is the standard normal distribution, for comparison.
\label{fig:HERAresiduals}}
\end{center}
\end{figure}
}
\newcommand{\figHERAshifts}
{
\begin{figure}[tbh]
\begin{center}
$
\includegraphics[width=2.8in]{\subd/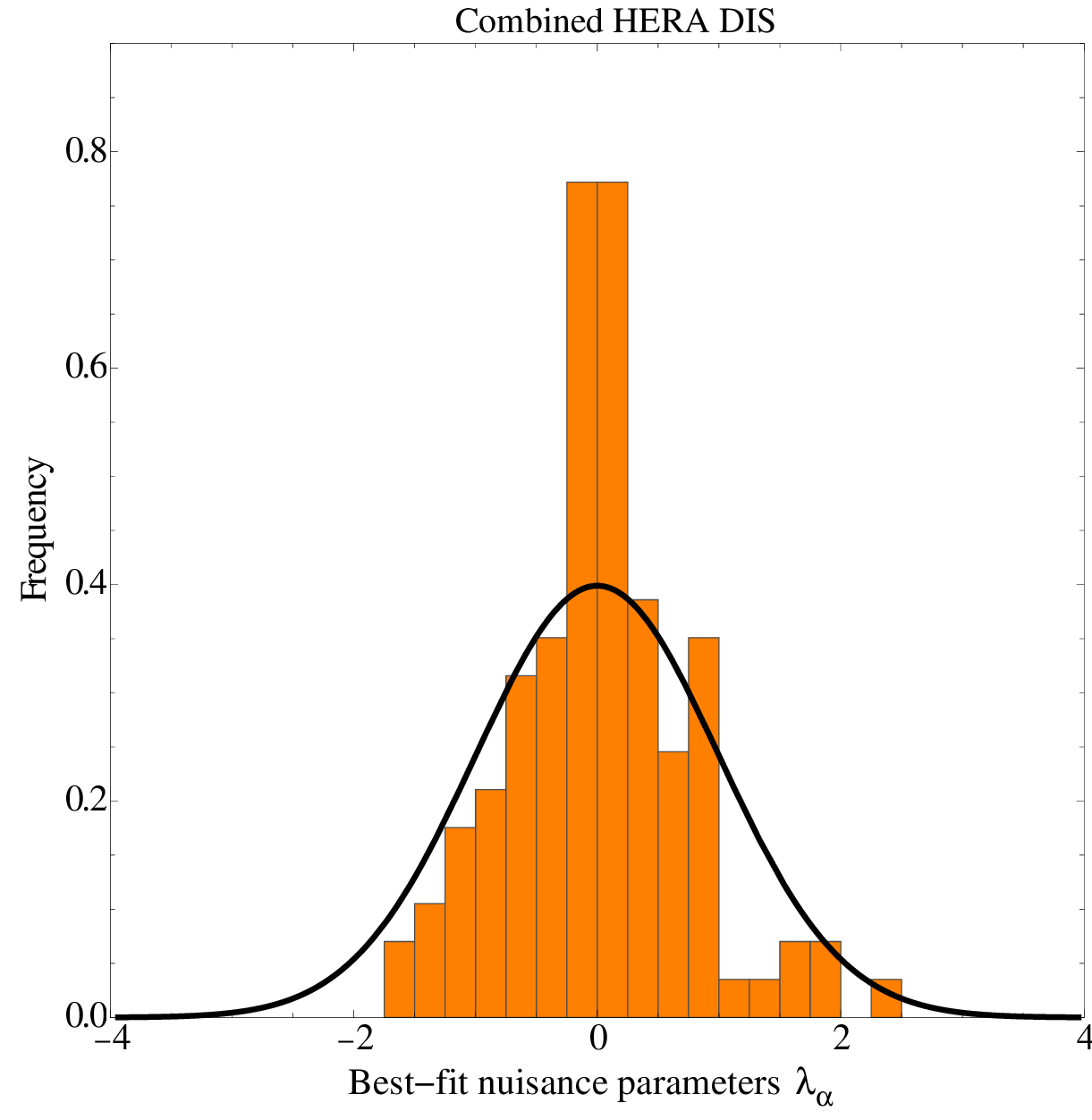}
$
\caption{
Histogram of the optimal normalized shift variables
$\left\{\overline{\lambda}_{\alpha}\right\}$
for the HERA combined data.
There are 114 systematic errors,
all of which apply to all of the 579 measurements
in the HERA combined data set.
The solid curve the standard normal distribution, for comparison.
\label{fig:HERAshifts}}
\end{center}
\end{figure}
}

\newcommand{\figTevWasyResiduals}
{
\begin{figure}[tbh]
\begin{center}
$
\includegraphics[width=4.0in]{\subd/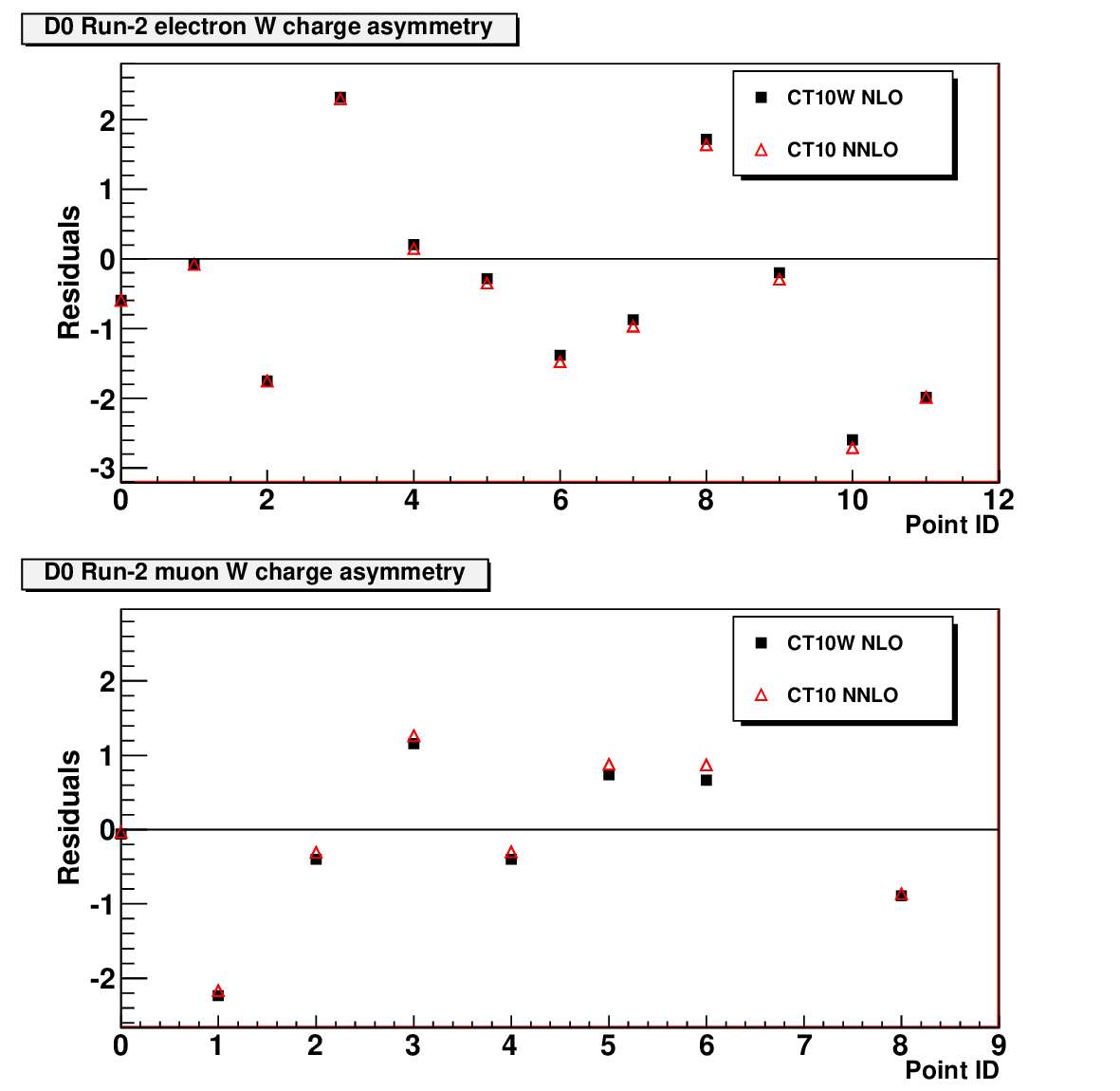}
$
\caption{
Point-by-point residuals for the D\O~ Run-2 charge asymmetry data in
the electron (upper inset) and muon (lower inset) channels. 
The residuals are defined in Eq.\ (\ref{eq:residuals}).
\label{fig:TevWasyResiduals}}
\end{center}
\end{figure}
}
\newcommand{\figDzeroJthy}
{
\begin{figure}[tbh]
\begin{center}
$
\includegraphics[width=3.0in]{\subd/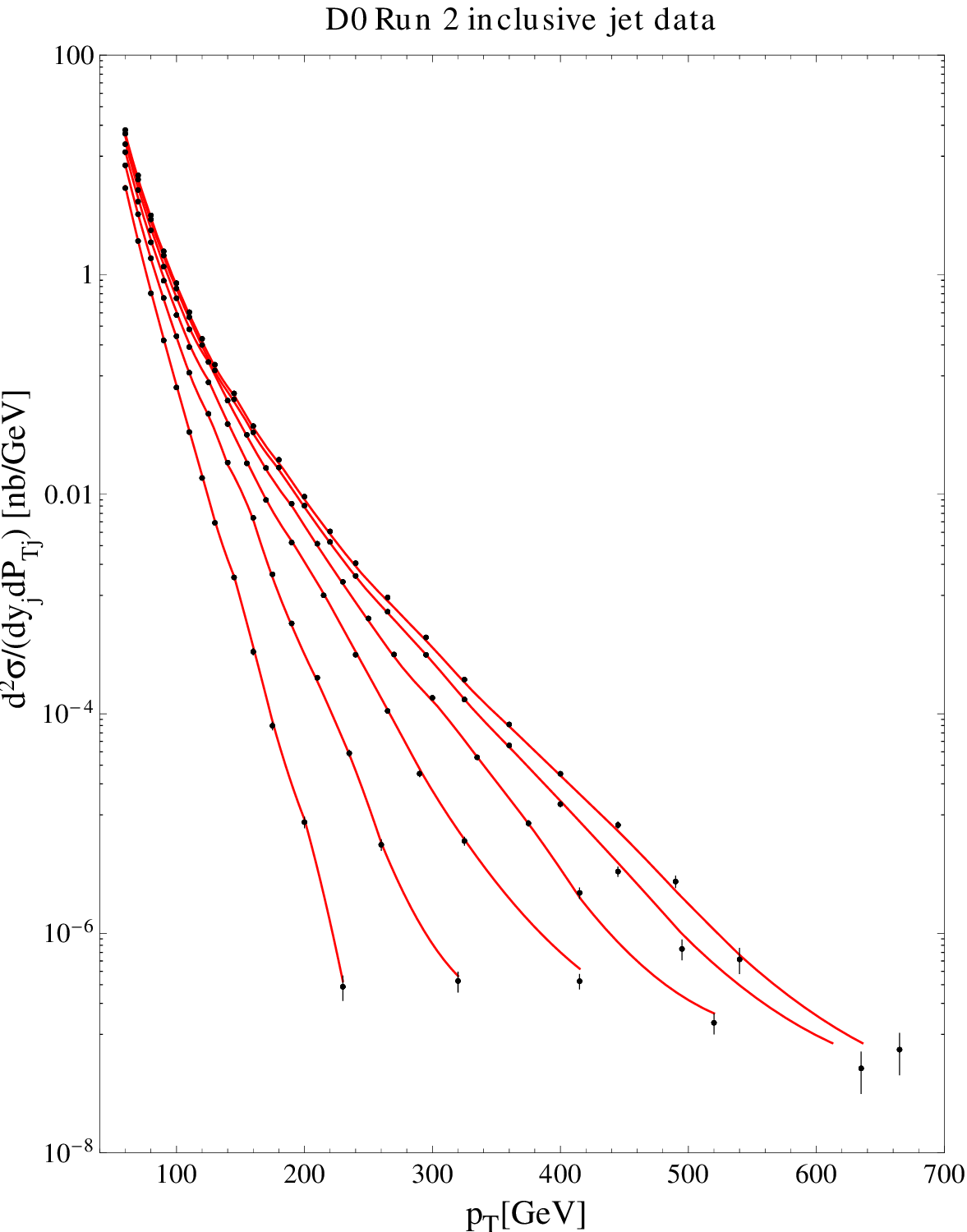}
$ \caption{Data and theory for the inclusive jet production data at D\O\ Tevatron Run-2. 
The black points represent central values of the data, error bars are
uncorrelated errors, and red curves are NLO theory obtained by using CT10NNLO PDFs.
\label{fig:DzeroJthy}}
\end{center}
\end{figure}
}
\newcommand{\figDzeroJresiduals}
{
\begin{figure}[tbh]
\begin{center}
$
\includegraphics[width=2.8in]{\subd/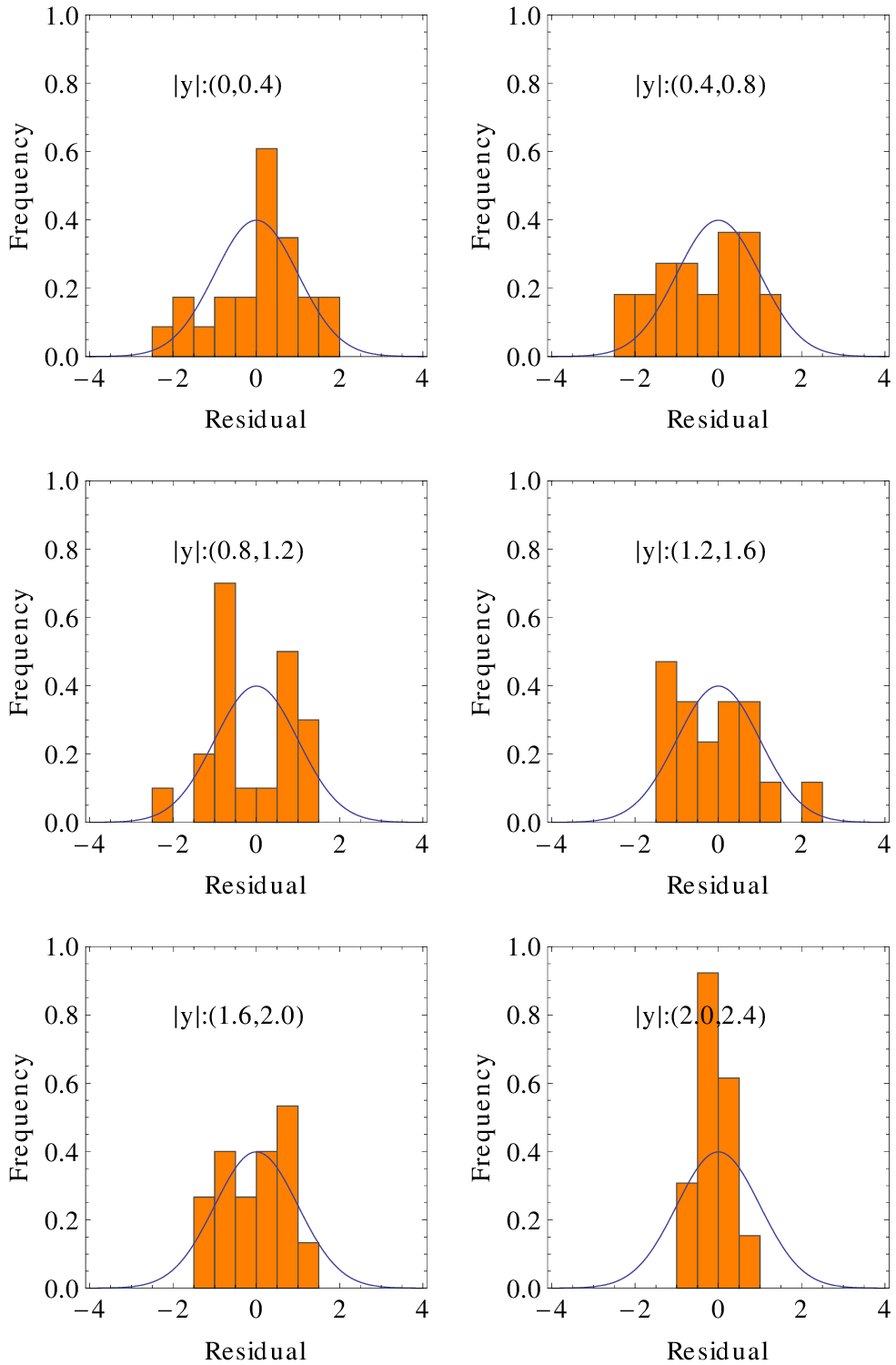}
$
\caption{
Histograms of residuals for the D\O\ Tevatron Run-2 inclusive jet production data.
Residuals are defined in Eq.\ (\ref{eq:residuals}).
The solid curve is the standard normal distribution, for comparison.
\label{fig:DzeroJresiduals}}
\end{center}
\end{figure}
}
\newcommand{\figDzeroJshifts}
{
\begin{figure}[tbh]
\begin{center}
$
\includegraphics[width=2.8in]{\subd/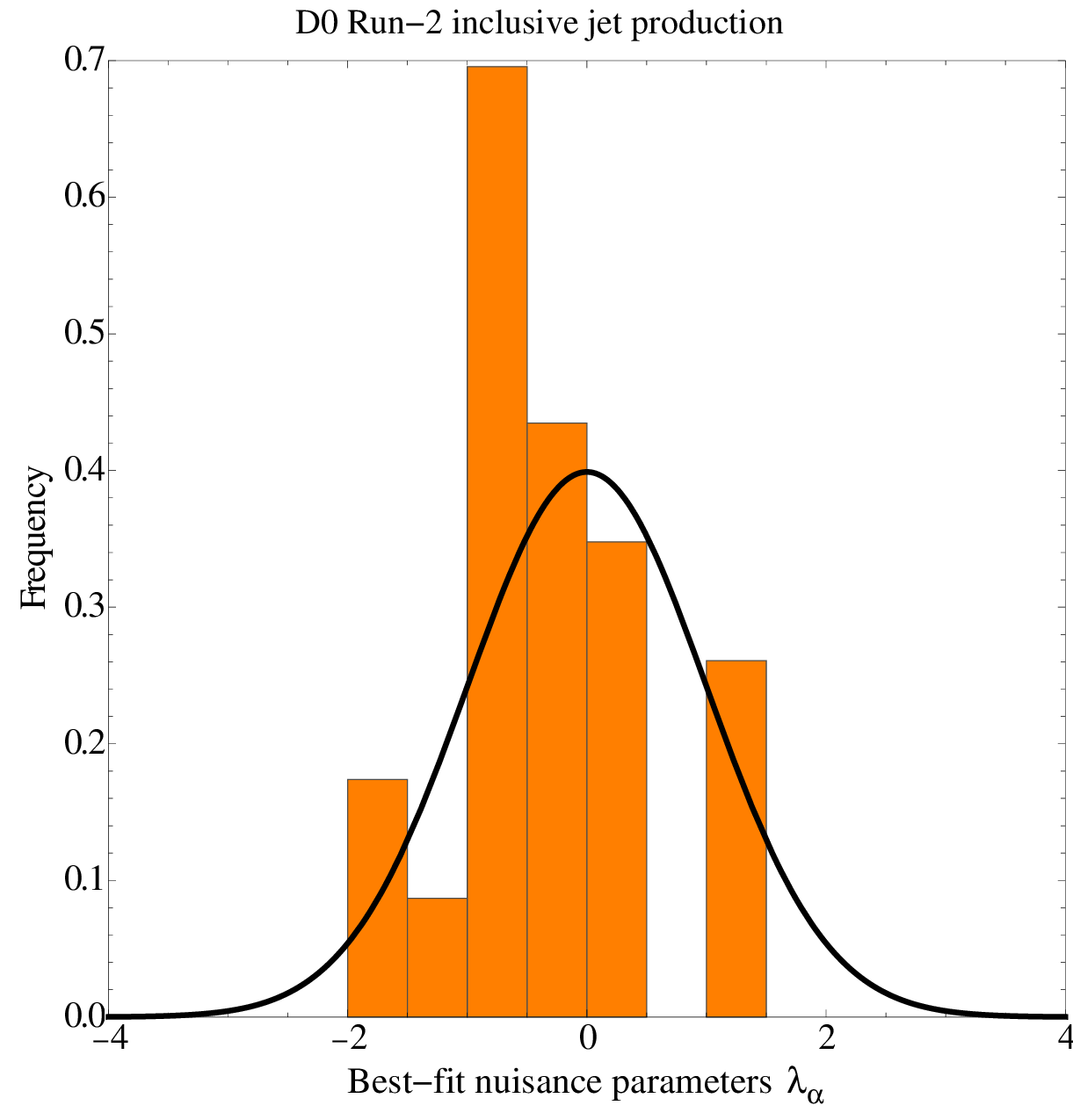}
$
\caption{
Histogram of the optimal systematic shifts $\{\overline{\lambda}_{\alpha}\}$
for the D\O\ Tevatron Run-2 inclusive jet production data.
\label{fig:DzeroJshifts}}
\end{center}
\end{figure}
}
\newcommand{\figCDFJthy}
{
\begin{figure}[tbh]
\begin{center}
$
\includegraphics[width=3.0in]{\subd/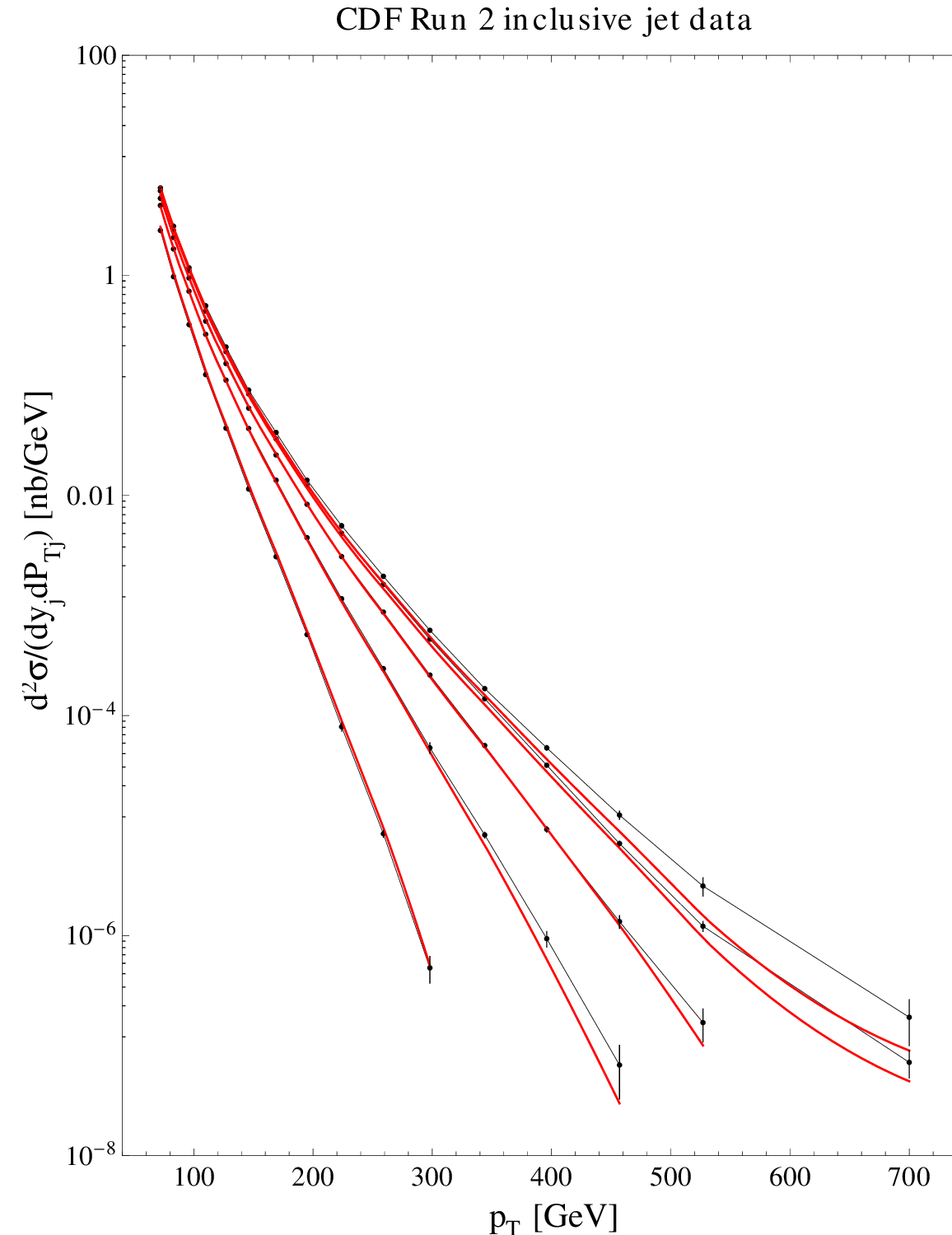}
$ \caption{Data and theory for inclusive jet production at CDF Tevatron Run-2. 
The black points are the central values of the data, the error bars are
uncorrelated errors, and the red curves represent 
the NLO predictions computed by using CT10NNLO PDFs. \label{fig:CDFJthy}}
\end{center}
\end{figure}
}
\newcommand{\figCDFJresiduals}
{
\begin{figure}[tbh]
\begin{center}
$
\includegraphics[width=2.8in]{\subd/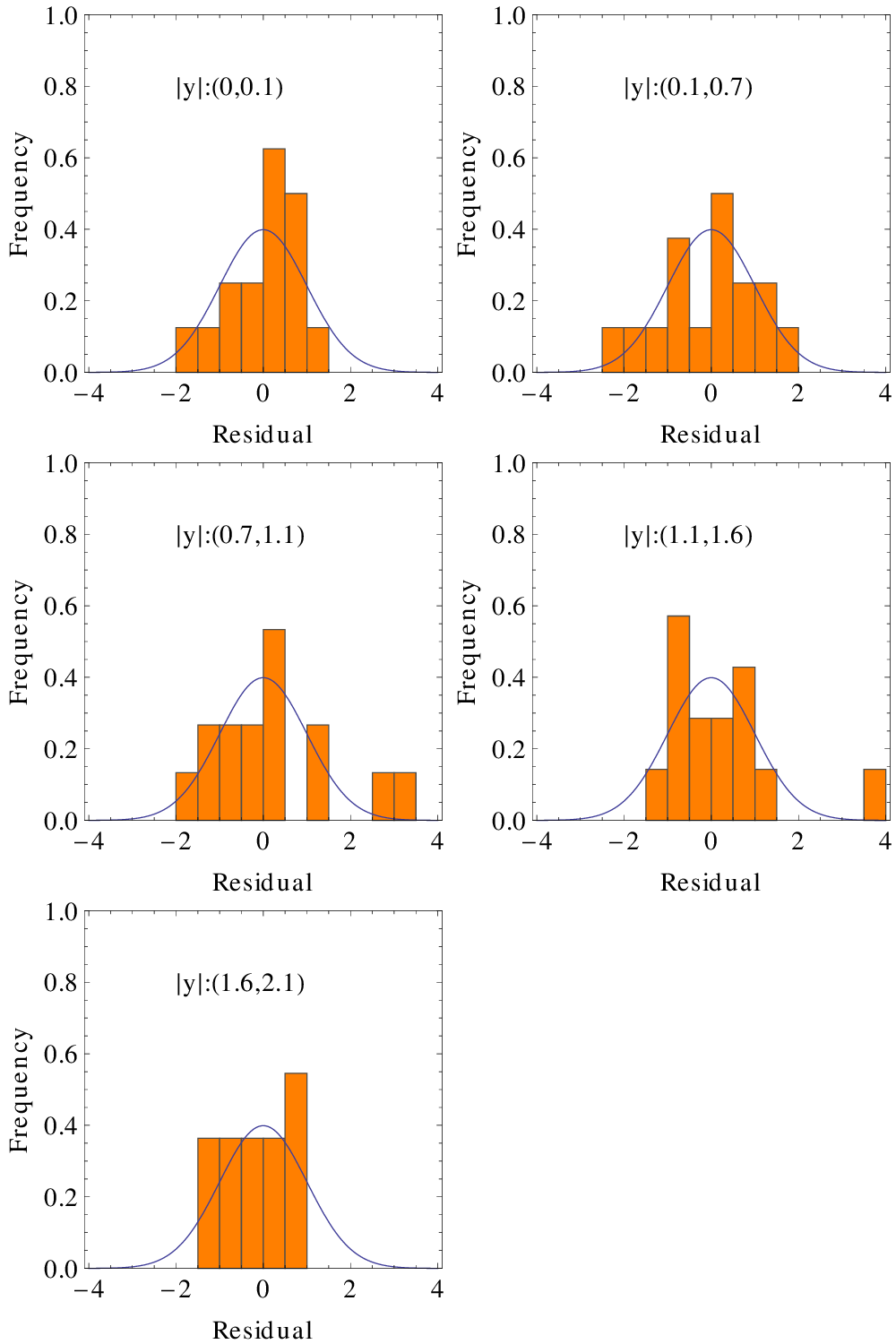}
$
\caption{
Histogram of residuals (defined in Eq.\ (\ref{eq:residuals})) 
for the CDF Run-2 inclusive jet production.
\label{fig:CDFJresiduals}}
\end{center}
\end{figure}
}
\newcommand{\figCDFJshifts}
{
\begin{figure}[tbh]
\begin{center}
$
\includegraphics[width=2.8in]{\subd/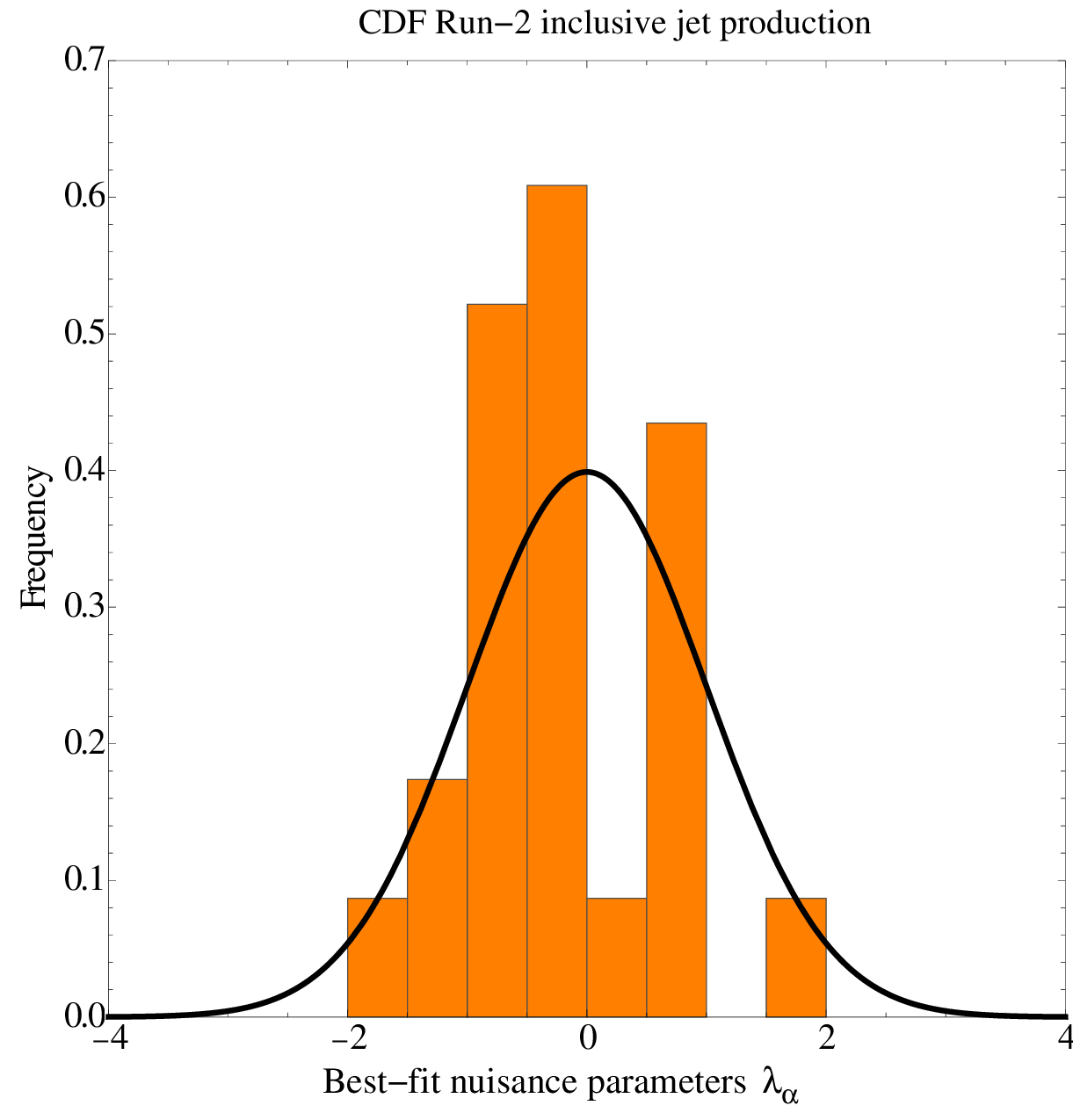}
$
\caption{
Histogram of the optimal systematic shifts $\{\overline{\lambda}_{\alpha}\}$
for the CDF Tevatron Run-2 inclusive jet production data.
\label{fig:CDFJshifts}}
\end{center}
\end{figure}
}
\newcommand{\figjetatlasA}
{
\begin{figure}[tbh]
\begin{center}
\includegraphics[width=0.7\textwidth]{\subd/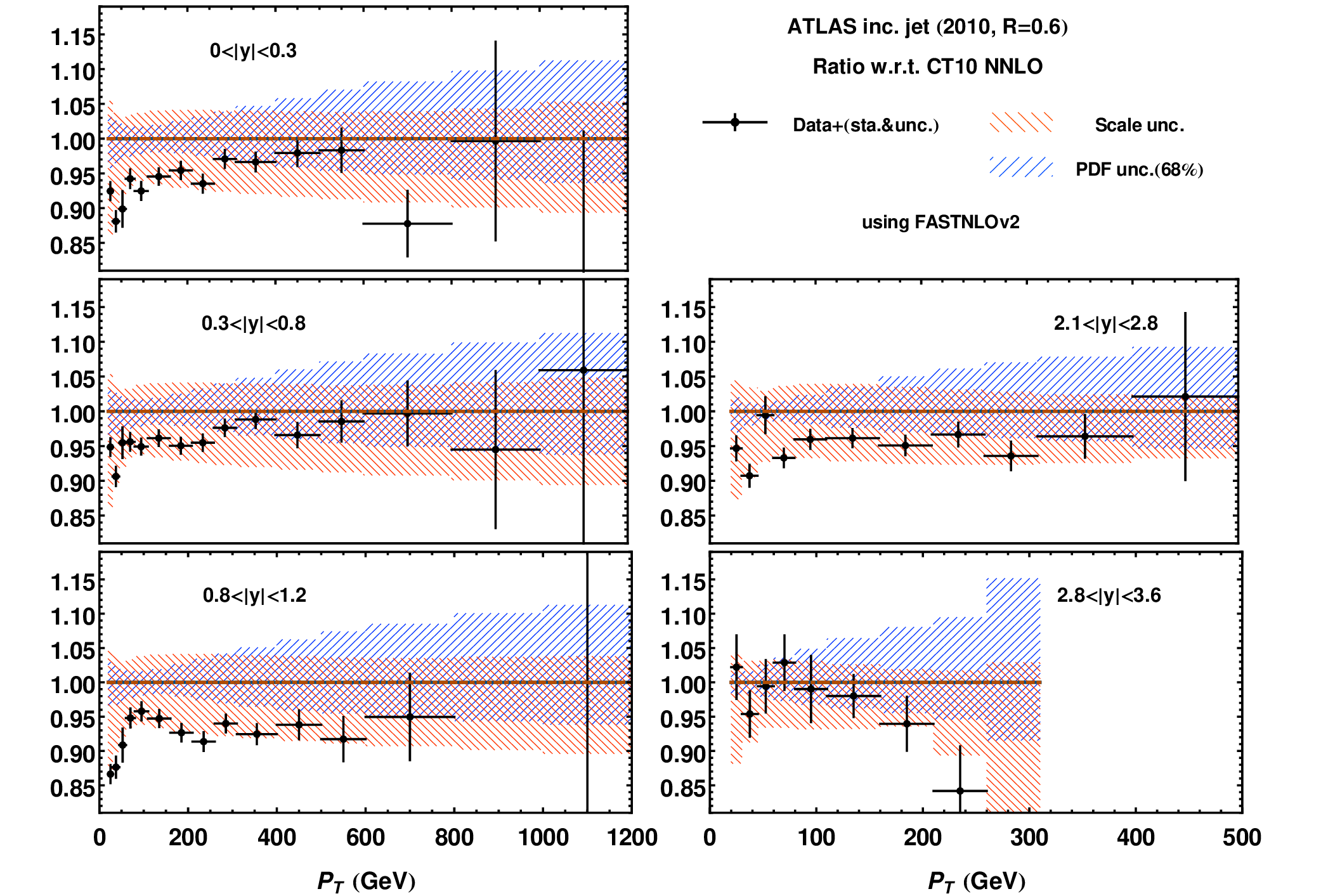}
\caption{Comparison of CT10NNLO predictions with unshifted 2010 ATLAS
inclusive jet data ($R=0.6$).
\label{fig:jetatlasA}}
\end{center}
\end{figure}
}
\newcommand{\figjetatlasB}
{
\begin{figure}[tbh]
\begin{center}
\includegraphics[width=0.7\textwidth]{\subd/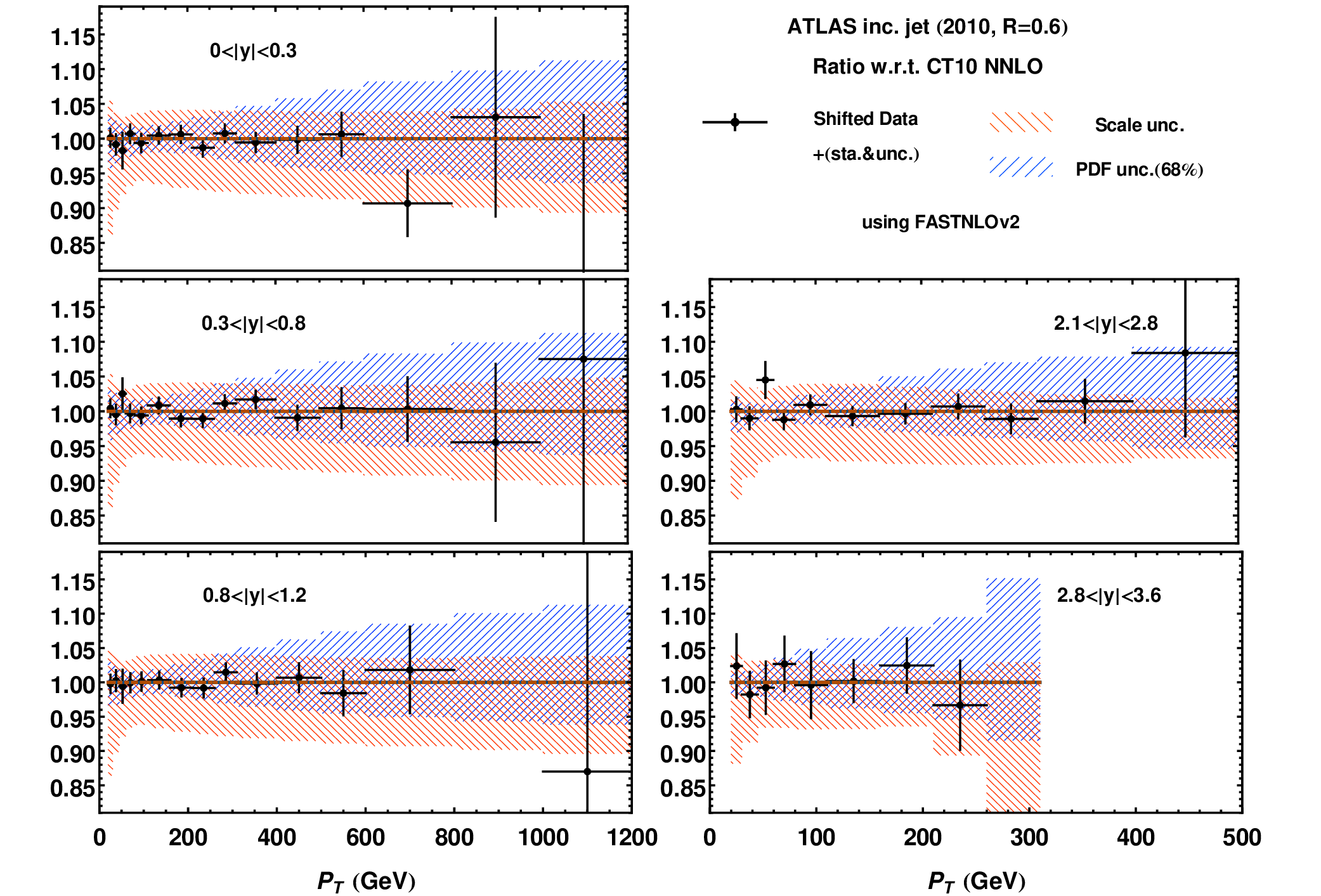}
\caption{Comparison of CT10NNLO predictions with shifted 2010 ATLAS
inclusive jet data ($R=0.6$).
\label{fig:jetatlasB}}
\end{center}
\end{figure}
}
%
\newcommand{\figHxsecA}
{
\begin{figure}[tbh]
\begin{center}
\includegraphics[width=0.4\textwidth]{\subd/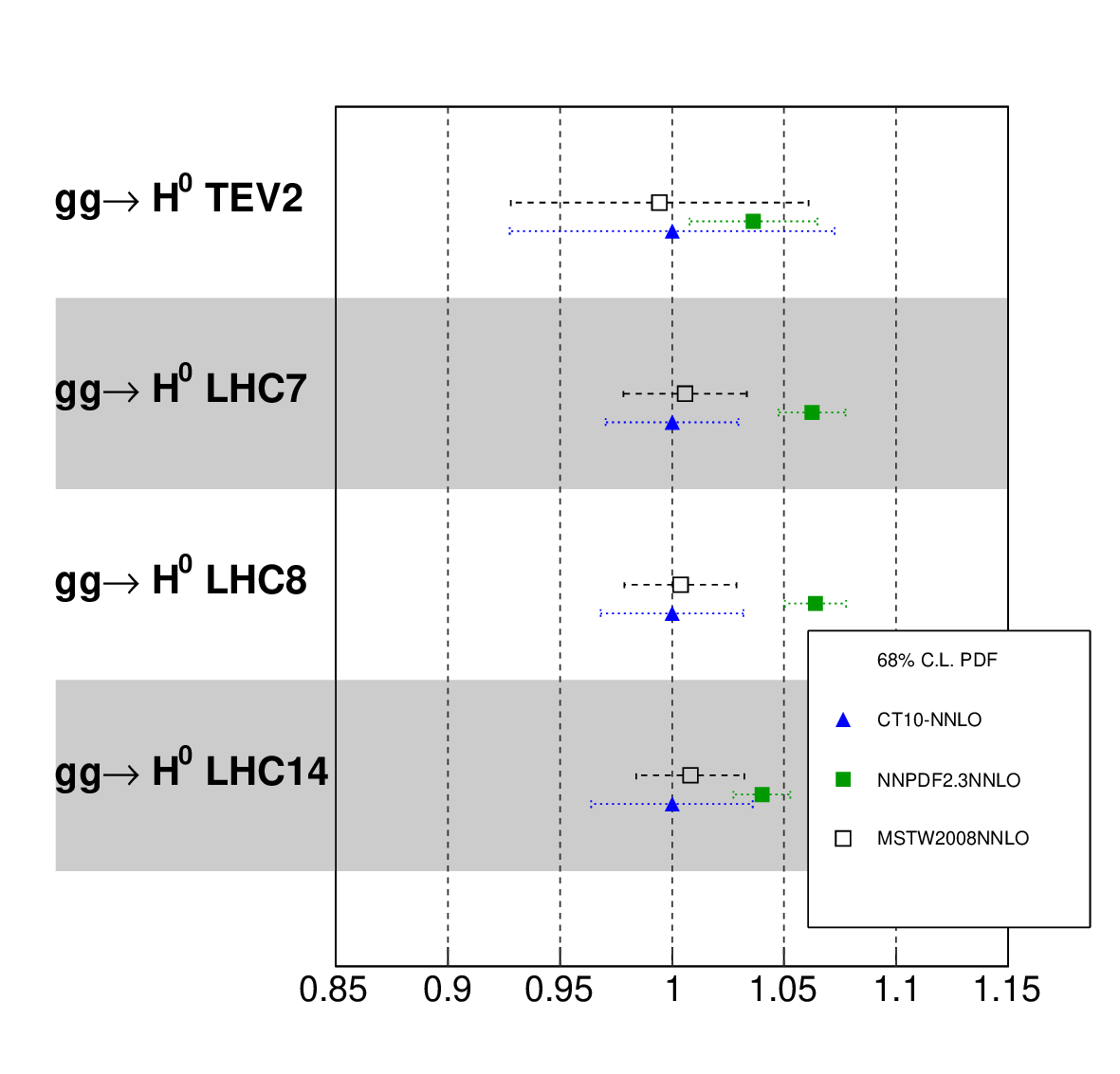}
\includegraphics[width=0.4\textwidth]{\subd/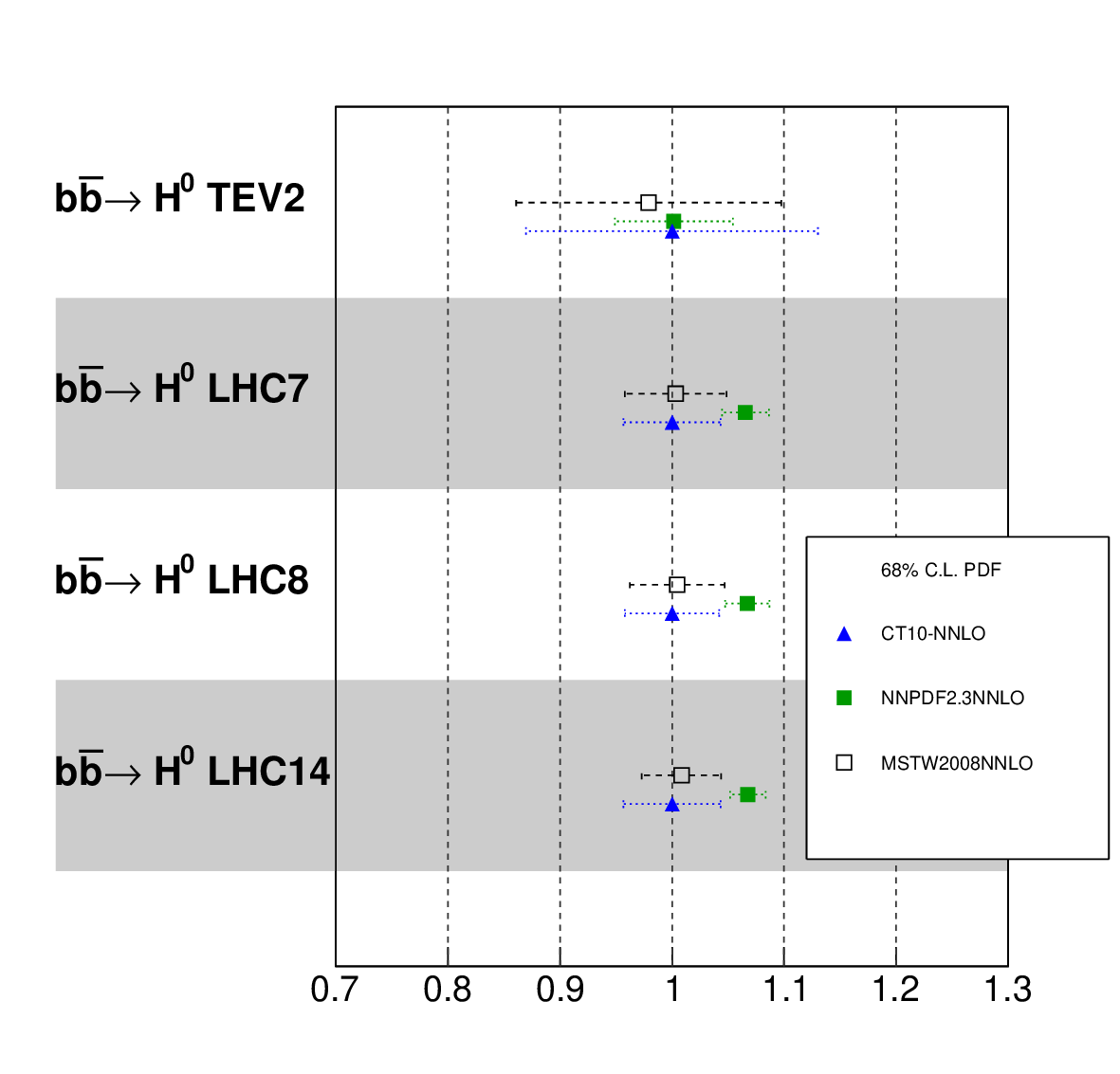}
\caption{NNLO cross sections of SM Higgs boson production via gluon
  fusion (left) and bottom quark annihilation (right), 
normalized to CT10NNLO predictions.
\label{H-xsecA}}
\end{center}
\end{figure}
}
\newcommand{\figZLxsecB}
{
\begin{figure}[h]
\begin{center}
\includegraphics[width=0.4\textwidth]{\subd/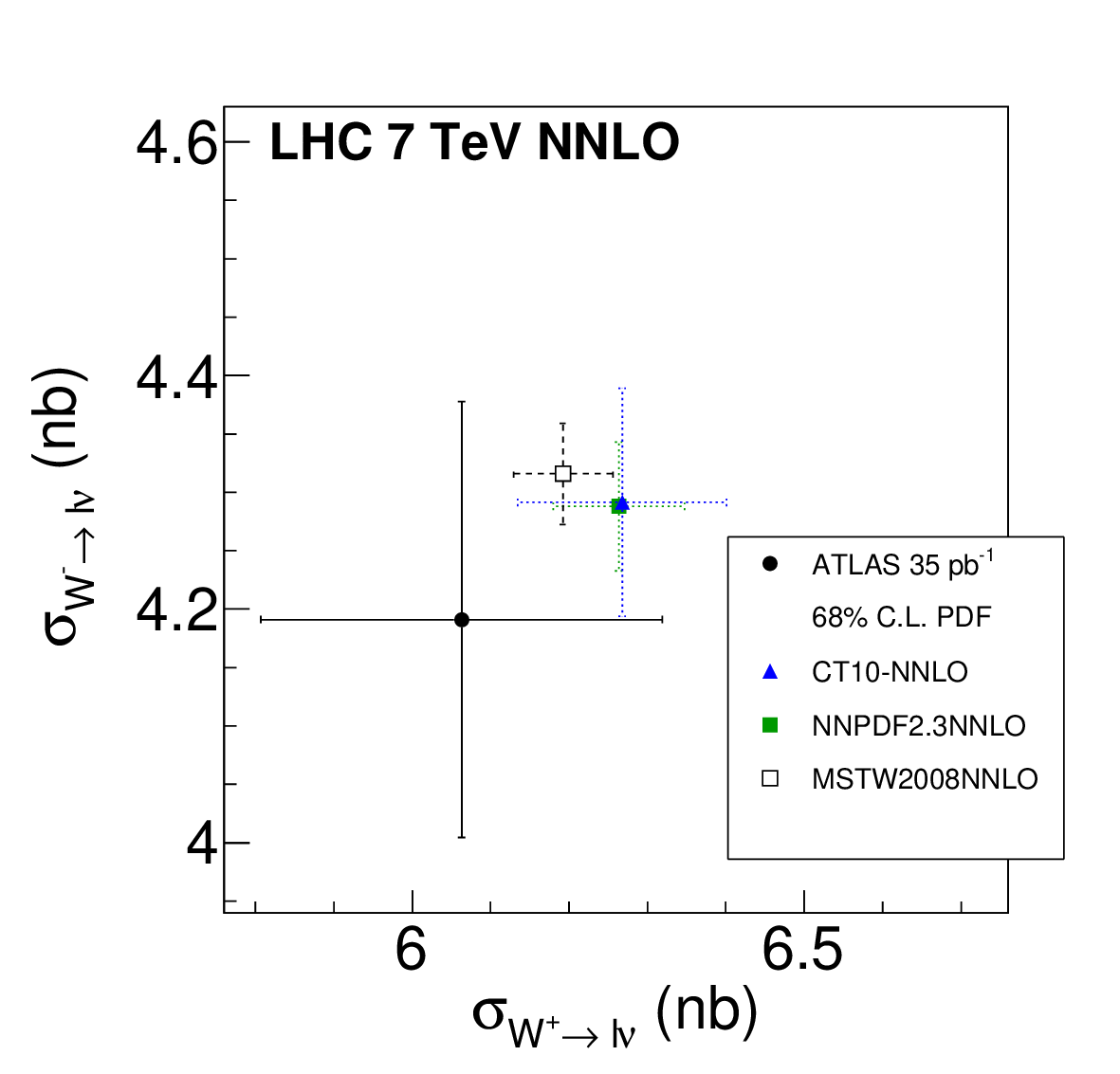}
\includegraphics[width=0.4\textwidth]{\subd/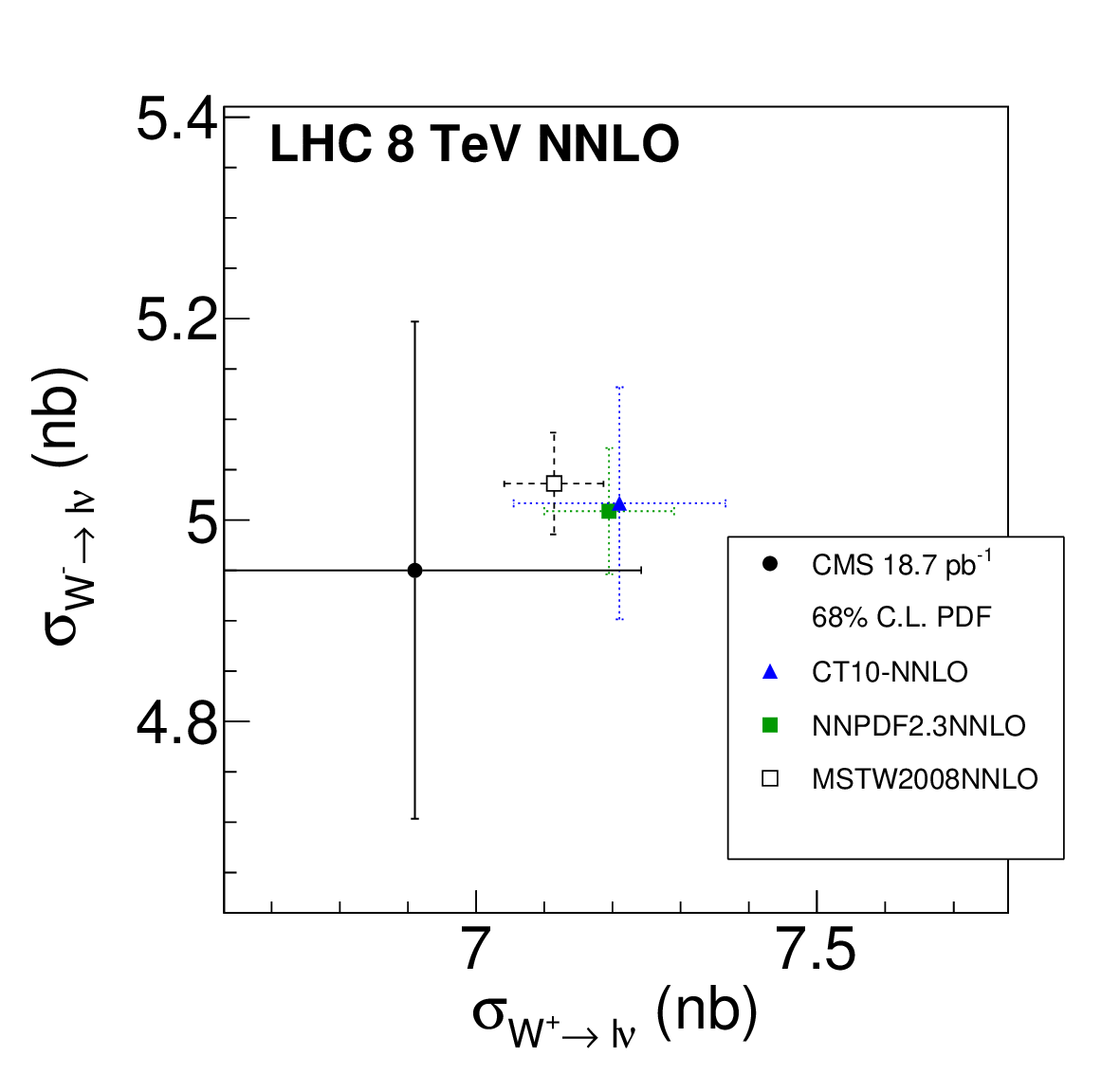}
\includegraphics[width=0.4\textwidth]{\subd/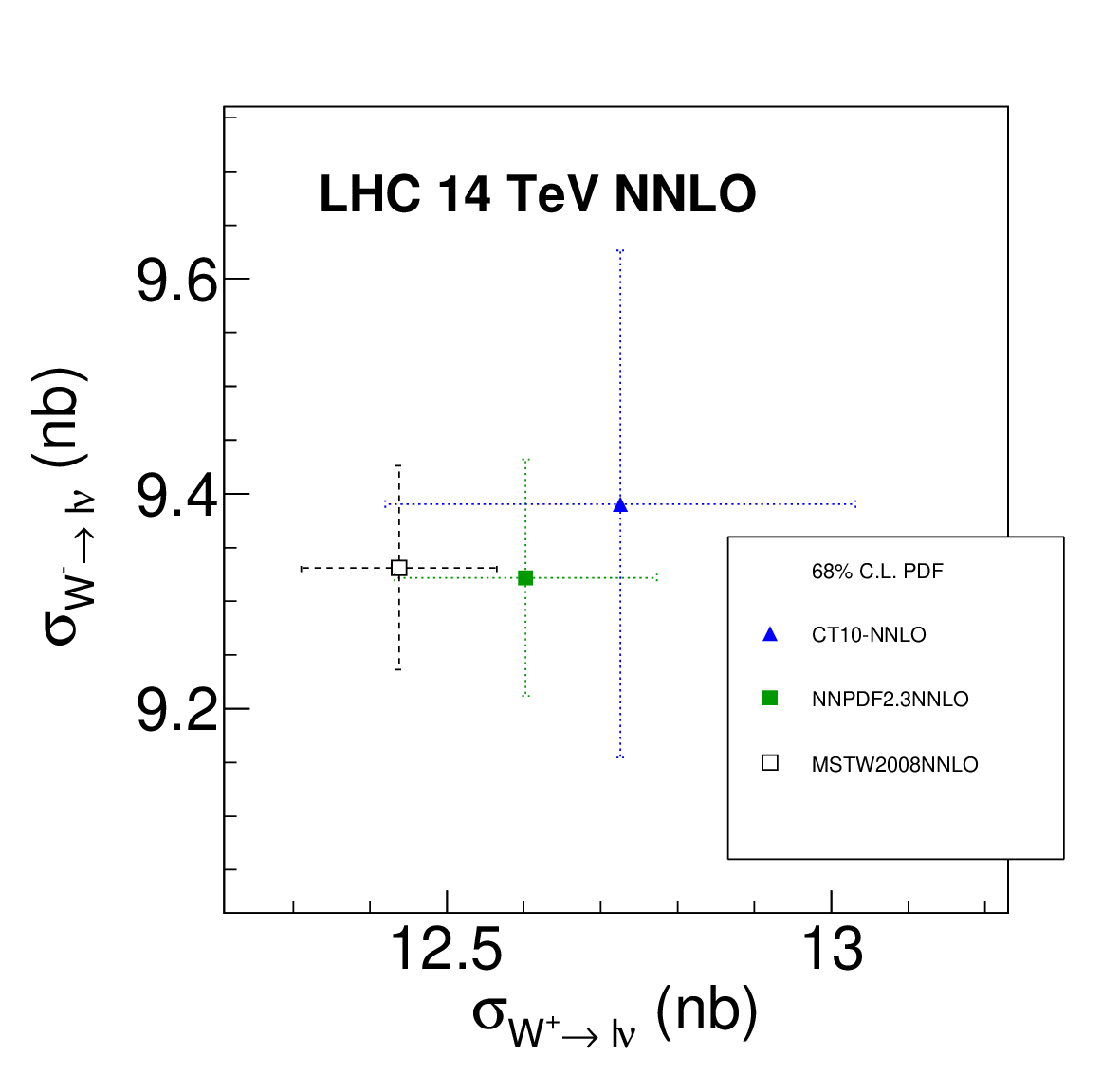}
\caption{NNLO $W^+$ and $W^-$ cross sections at the LHC.
\label{ZL-xsecB}}
\end{center}
\end{figure}
}
\newcommand{\figZLxsecC}
{
\begin{figure}[h]
\begin{center}
\includegraphics[width=0.4\textwidth]{\subd/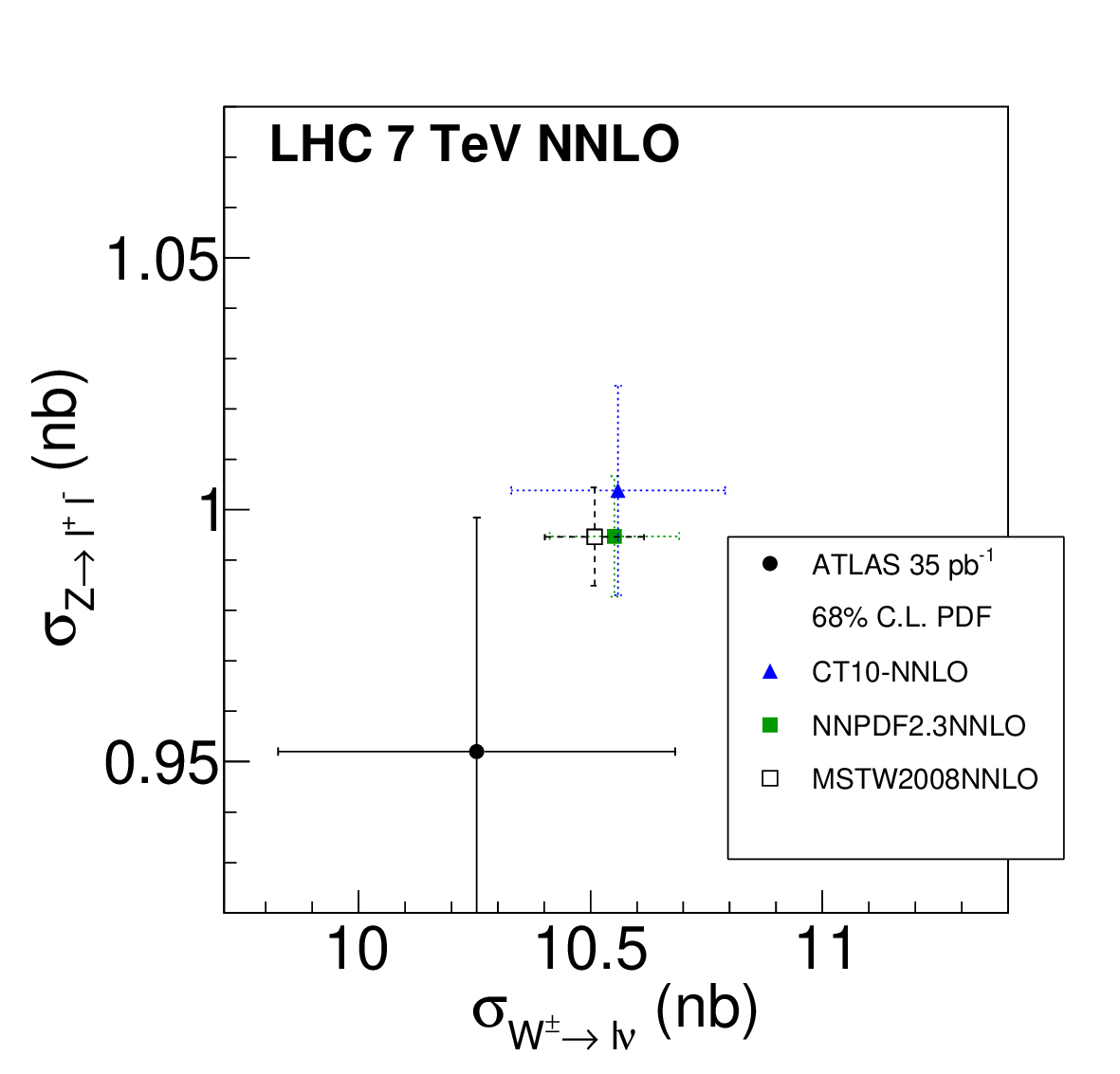}
\includegraphics[width=0.4\textwidth]{\subd/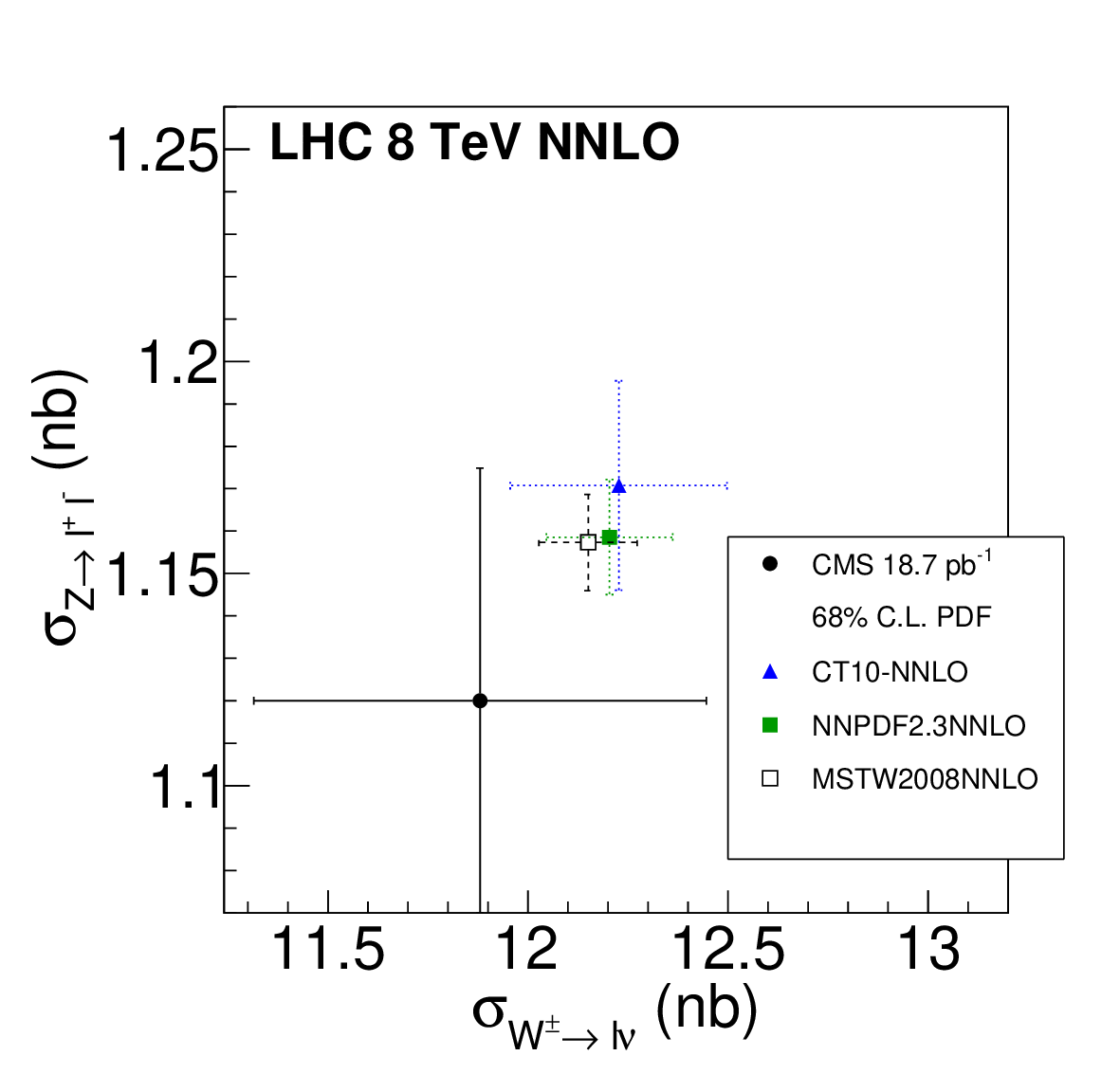}
\includegraphics[width=0.4\textwidth]{\subd/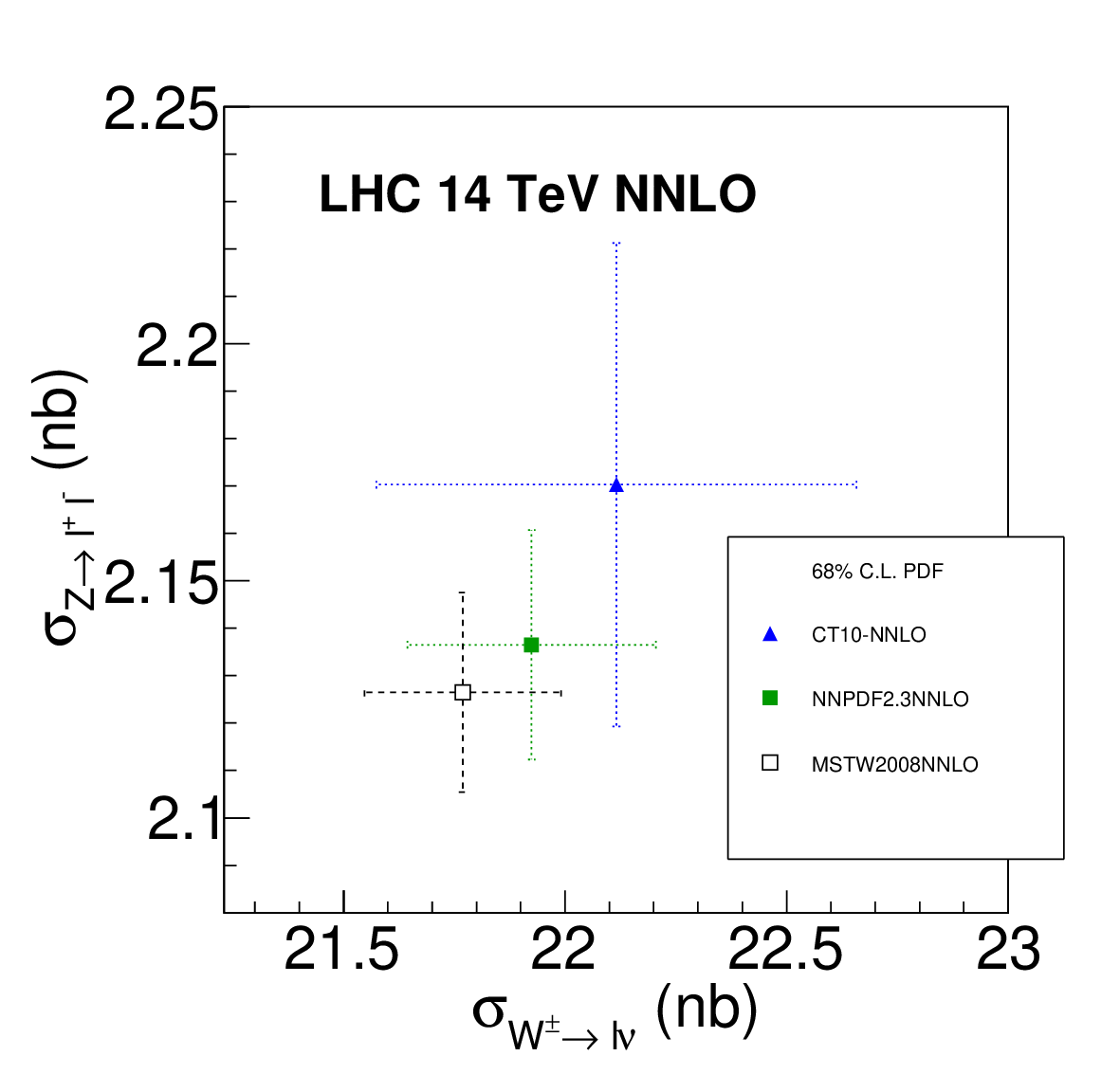}
\caption{NNLO $W^\pm$ and $Z$ cross sections at the LHC.
\label{ZL-xsecC}}
\end{center}
\end{figure}
}

\newcommand{\figZLxsecE}
{
\begin{figure}[h]
\begin{center}
\includegraphics[width=0.4\textwidth]{\subd/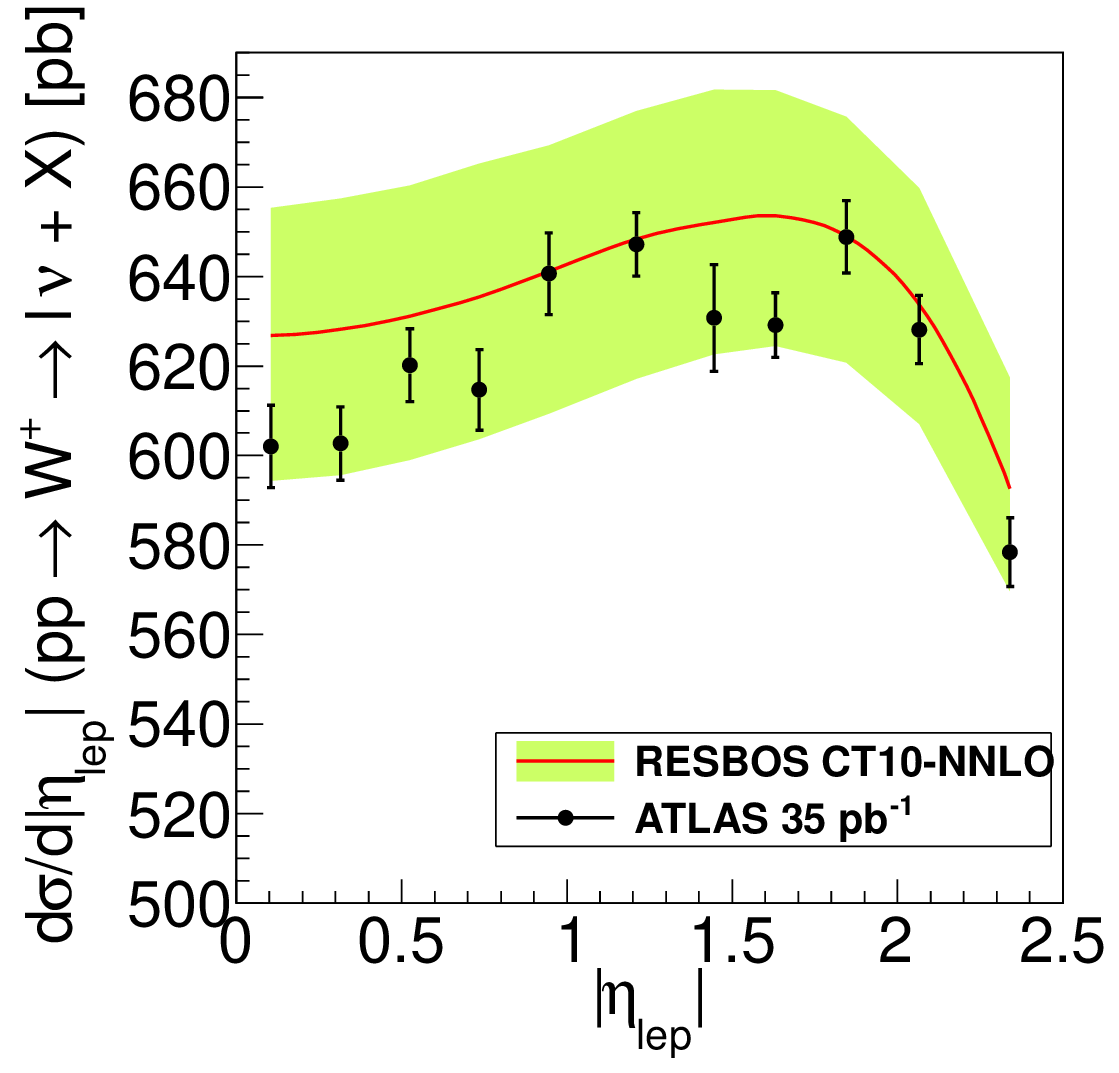}
\includegraphics[width=0.4\textwidth]{\subd/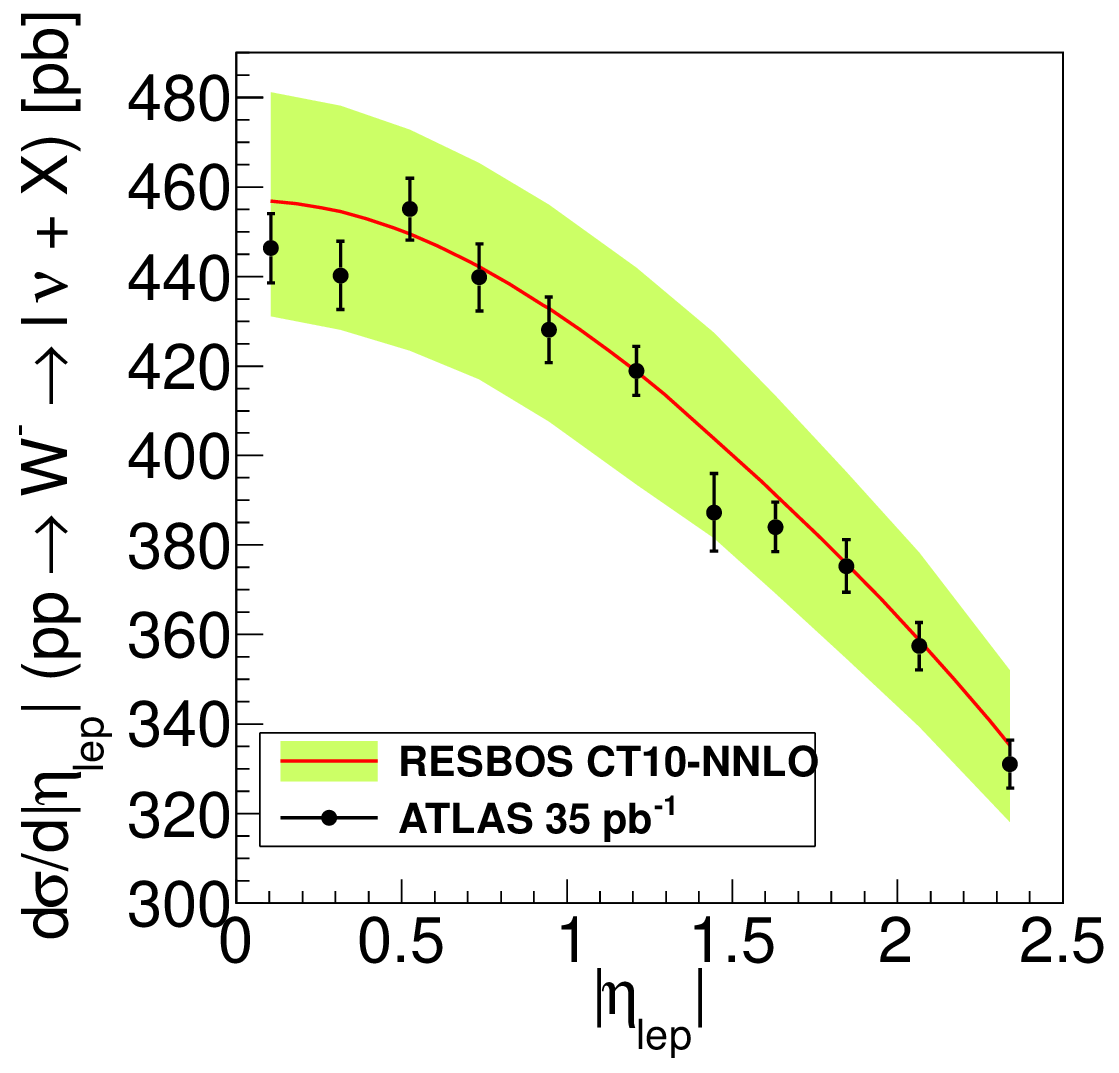}
\includegraphics[width=0.4\textwidth]{\subd/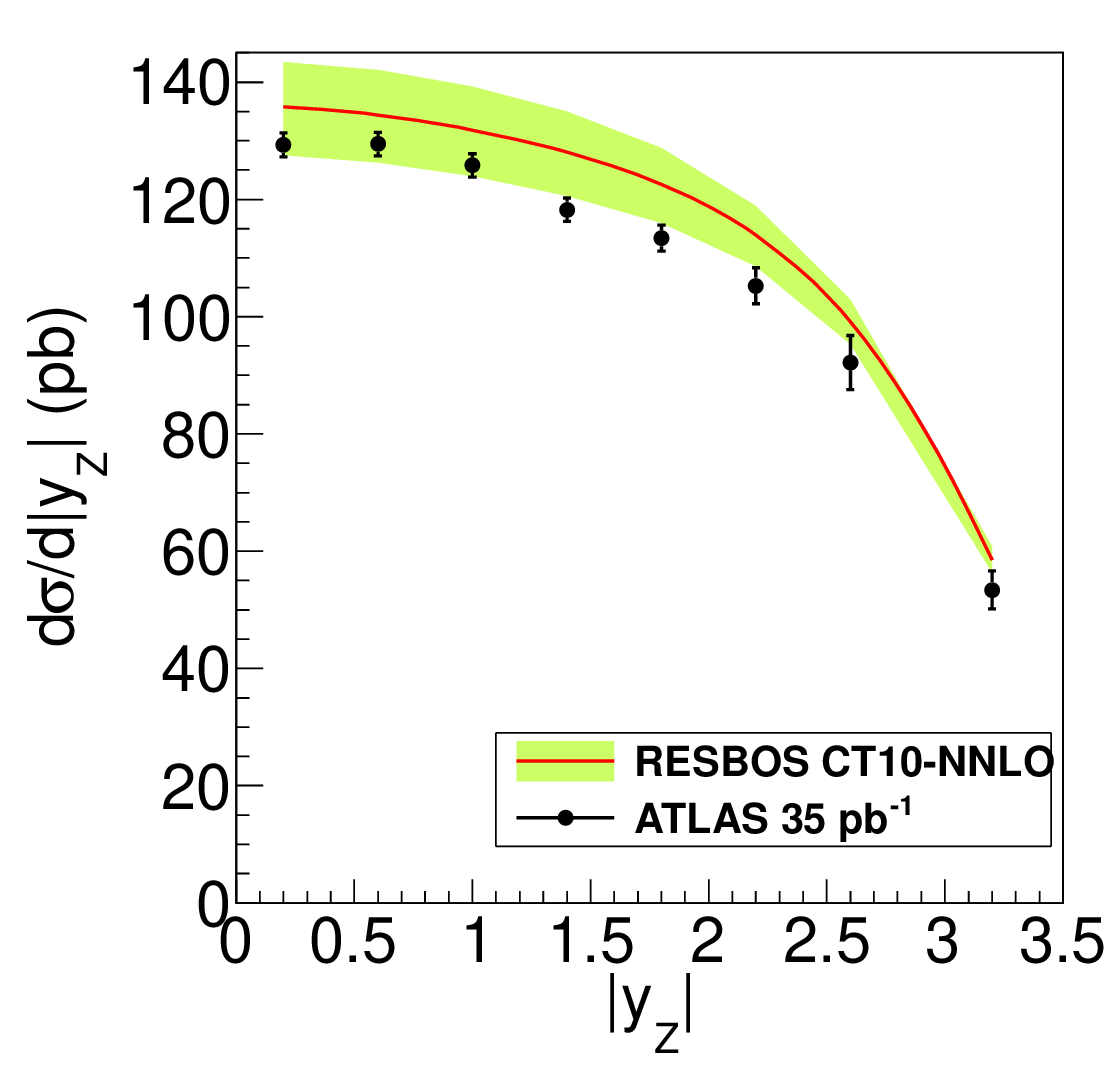}
\caption{Rapidity distributions of $Z$ and $W^\pm$ cross sections in ATLAS ($35\mbox{ pb}^{-1}$) measurements at 7 TeV.
\label{ZL-xsecE}}
\end{center}
\end{figure}
}

\newcommand{\figWasyA}
{
\begin{figure}[p]
\begin{center}
\includegraphics[width=0.48\textwidth,height=250pt]{\subd/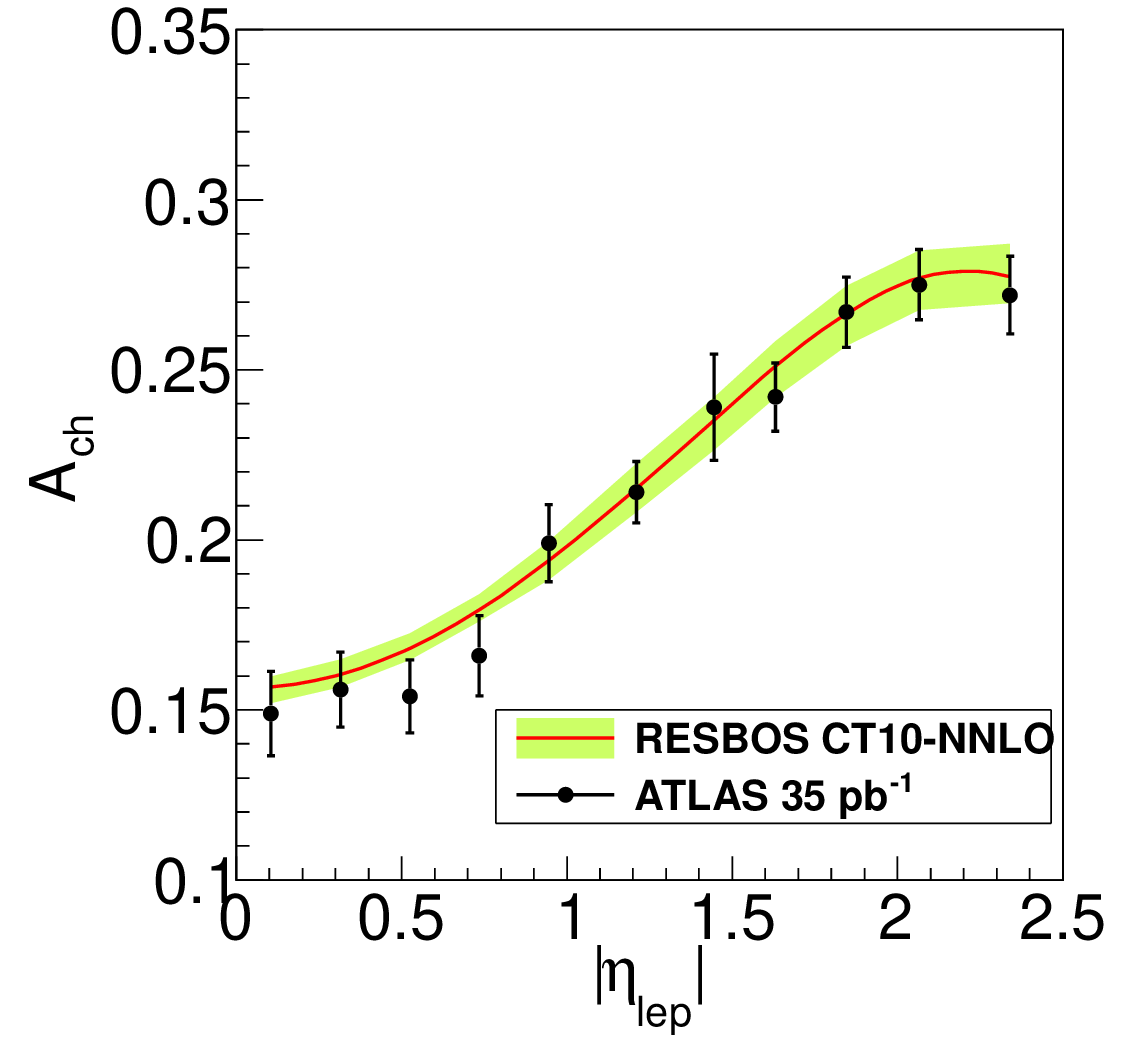}
\includegraphics[width=0.48\textwidth,height=250pt]{\subd/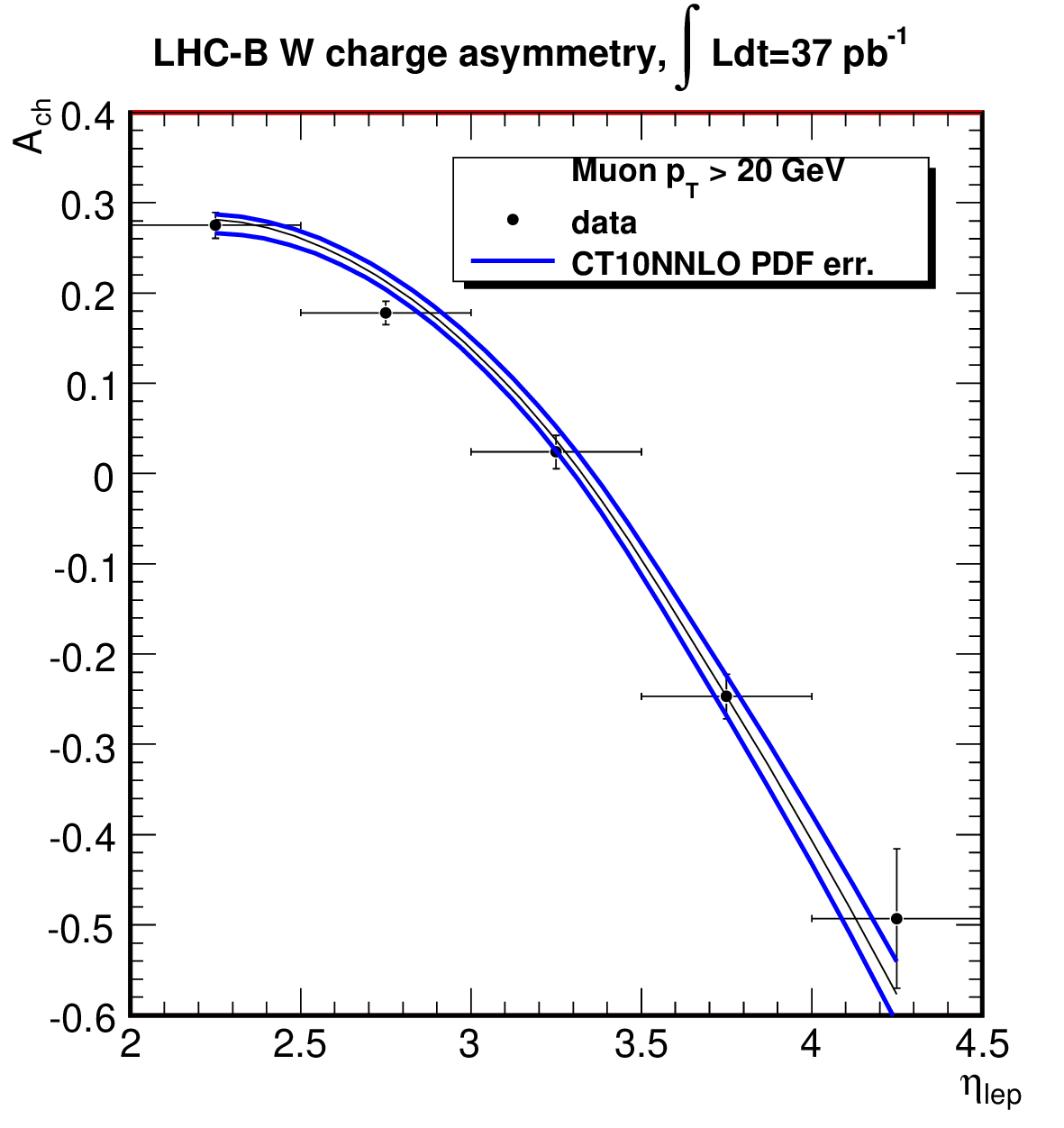}
\caption{A lepton rapidity distribution of the $W^\pm$ charge asymmetry in
the  ATLAS ($35\mbox{ pb}^{-1}$) measurement (left) 
and LHC-B ($37\mbox{ pb}^{-1}$) measurement (right) at 7 TeV.
\label{WasyA}}
\end{center}
\end{figure}
}

\newcommand{\figWasyB}
{
\begin{figure}[p]
\begin{center}
\includegraphics[width=0.48\textwidth,height=250pt]{\subd/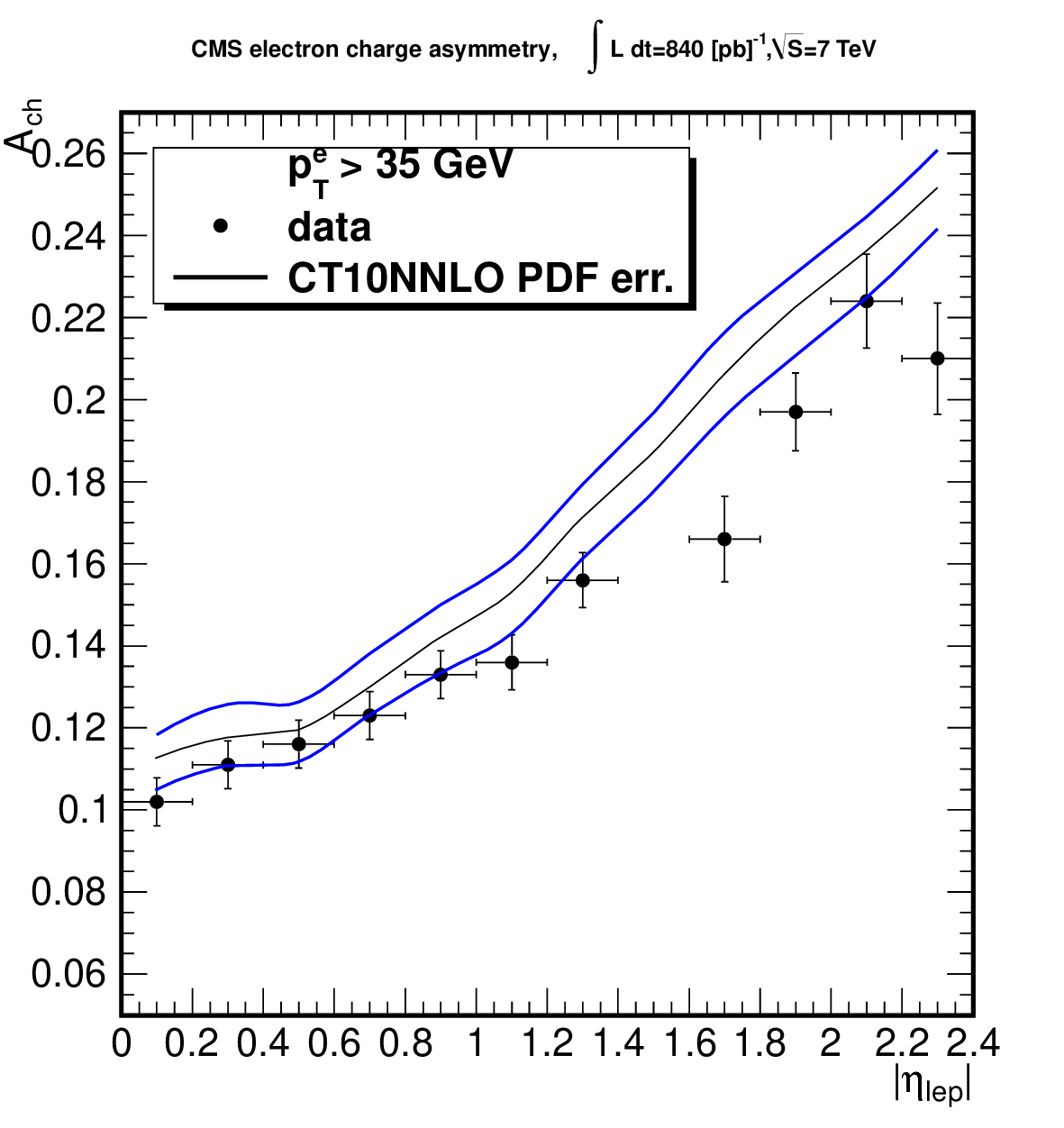}
\includegraphics[width=0.48\textwidth,height=250pt]{\subd/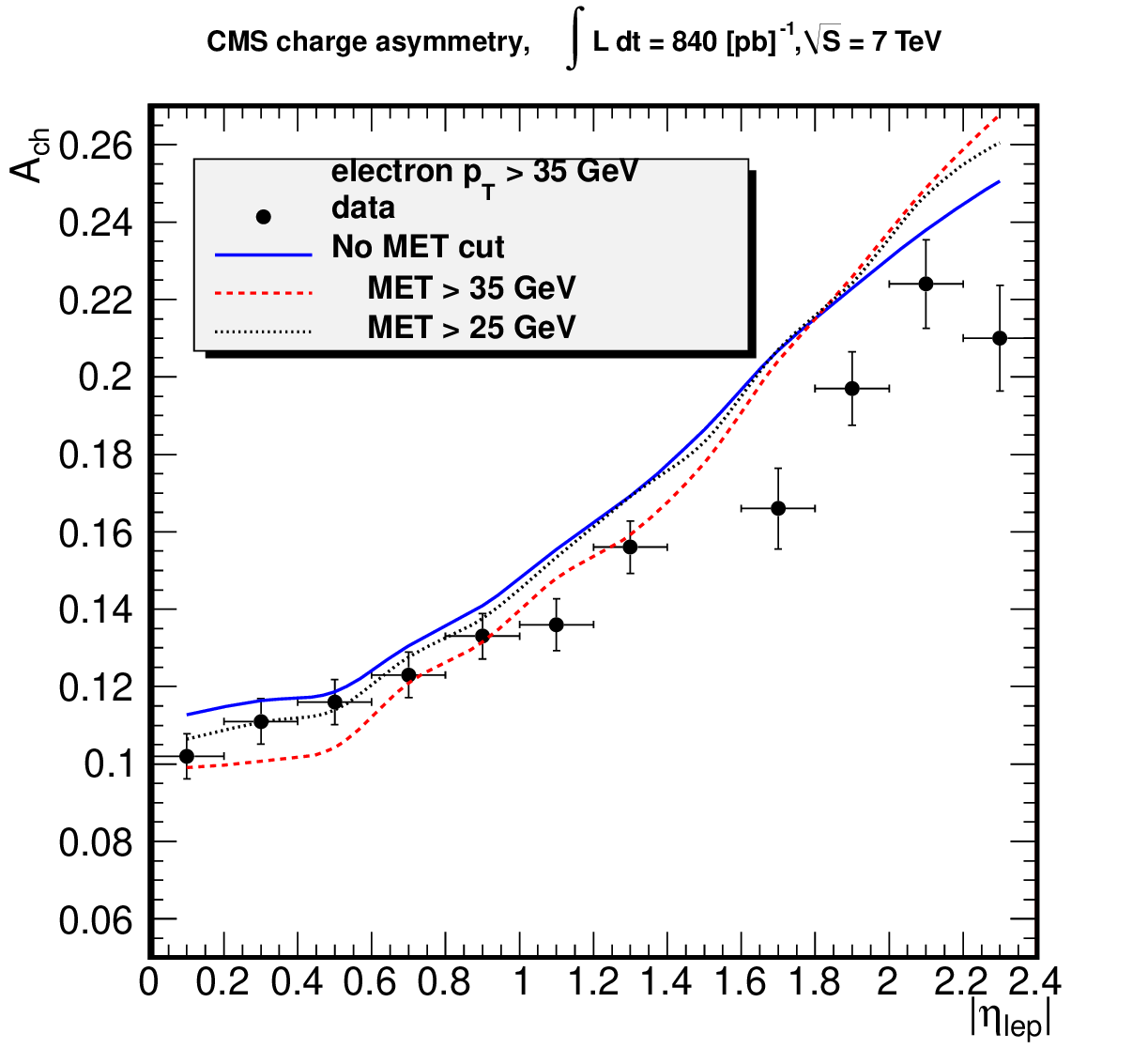}
\caption{A lepton rapidity distribution of the $W^\pm$ charge asymmetry in
the CMS
($840\mbox{ pb}^{-1}$) measurement at 7 TeV.
\label{WasyB}}
\end{center}
\end{figure}
}

\newcommand{\figWasyCorr}
{
\begin{figure}[tb]
\begin{center}
\includegraphics[width=0.48\textwidth]{\subd/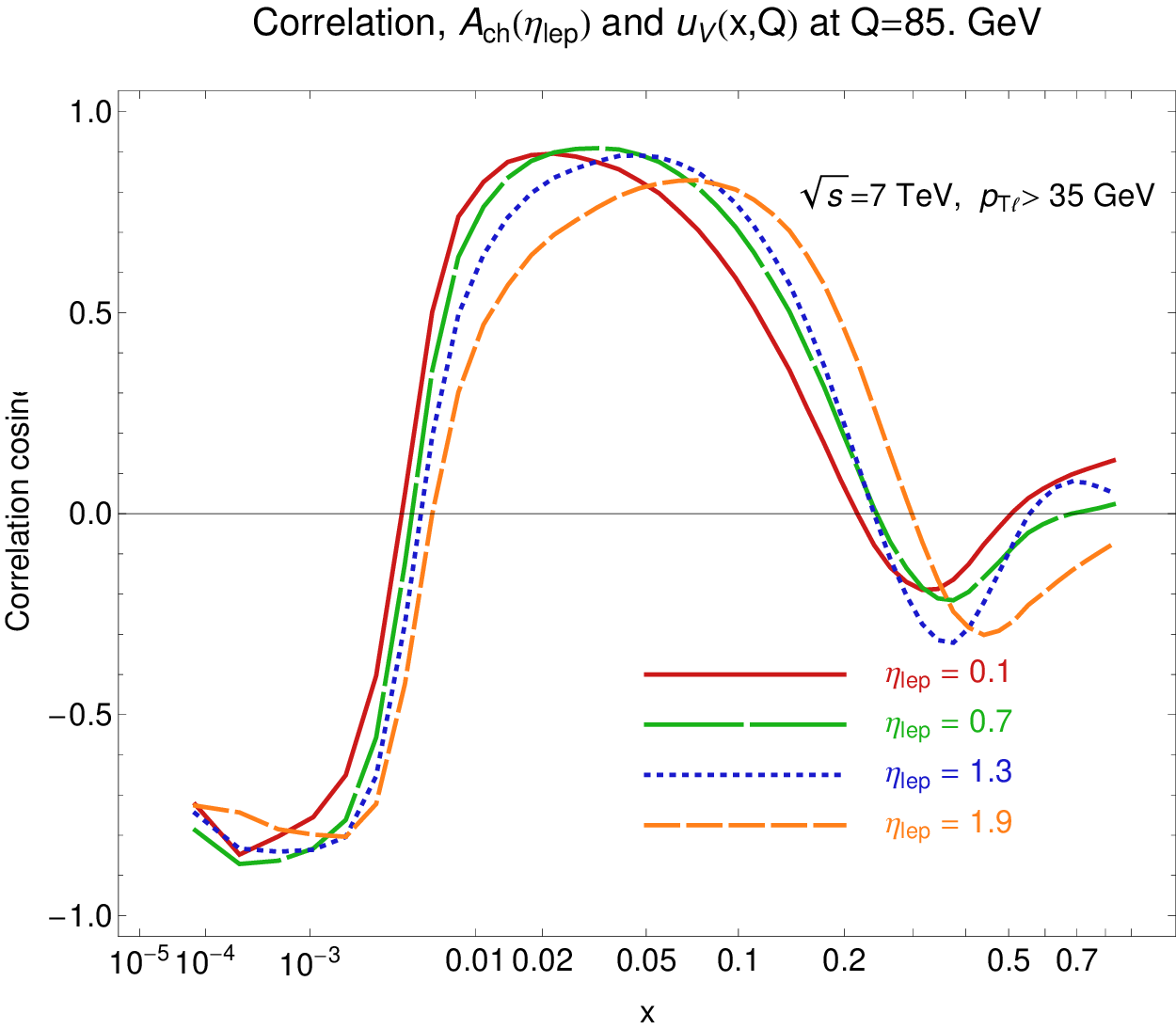}\quad\includegraphics[width=0.48\textwidth]{\subd/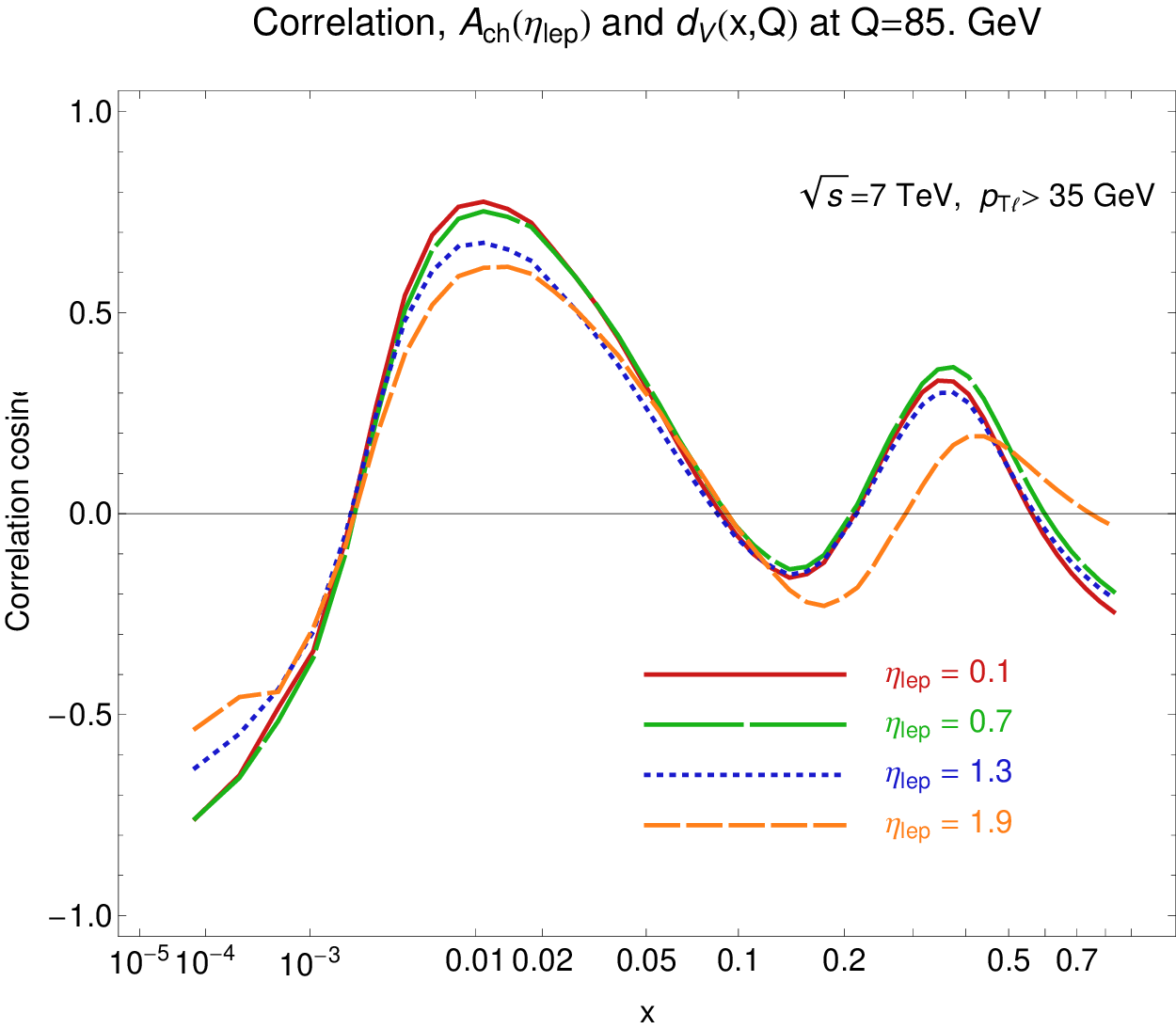}\\
\includegraphics[width=0.48\textwidth]{\subd/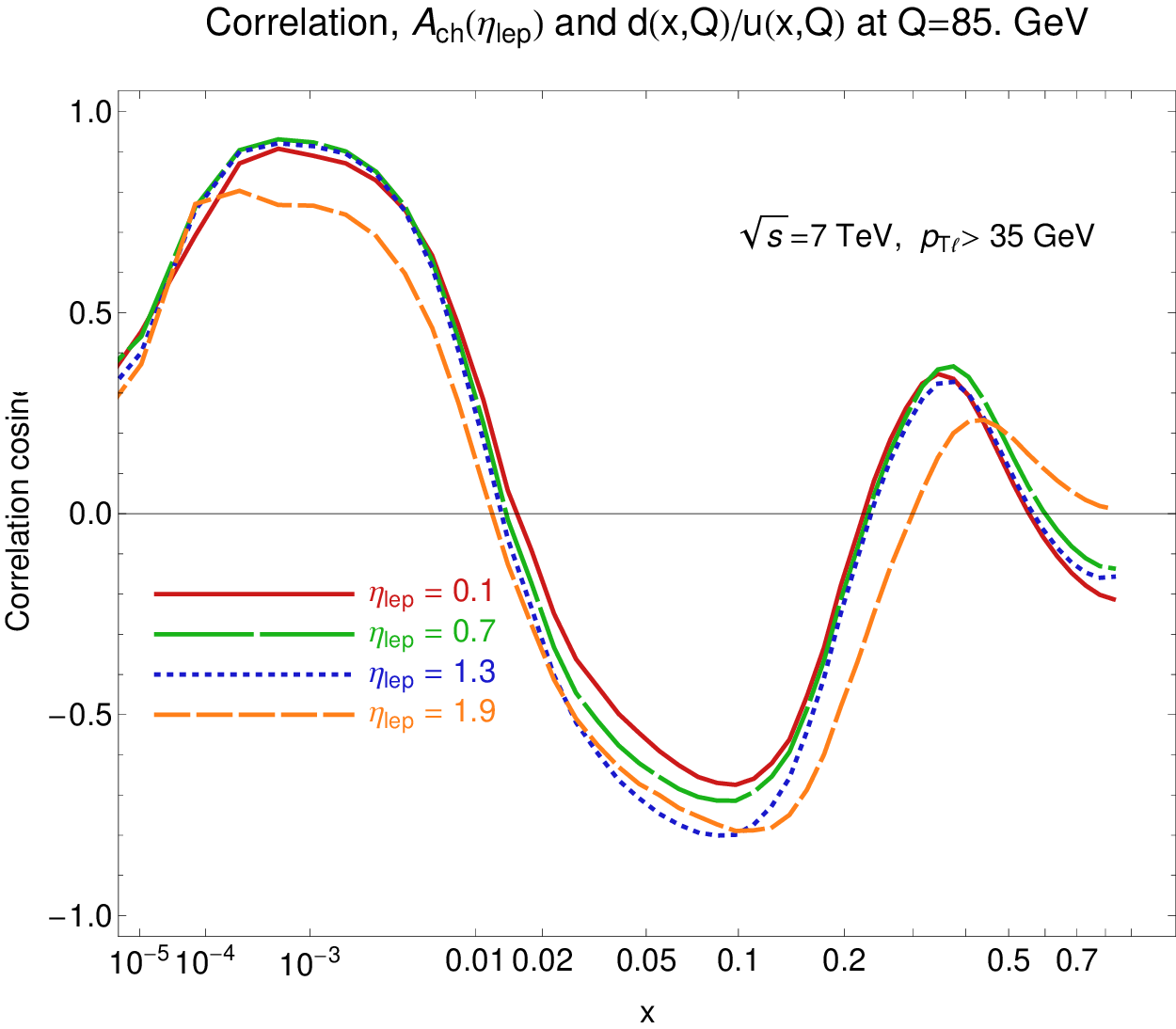}
\caption{PDF-induced correlations between $W$ charge asymmetry $A_{ch}(\eta_{lep})$ at the LHC 7 TeV and $u_v(x,Q)=u(x,Q)-\bar u(x,Q)$, $d_v(x,Q)=d(x,Q)-\bar d(x,Q$, and $d(x,Q)/u(x,Q)$ at $Q=85\mbox{ GeV}$. The charge asymmetry is evaluated at lepton rapidity values indicated in the figure.  
\label{WasyCorr}}
\end{center}
\end{figure}
}

\newcommand{\figTxsecA}
{
\begin{figure}[tbh]
\begin{center}
\includegraphics[height=0.5\textheight]{\subd/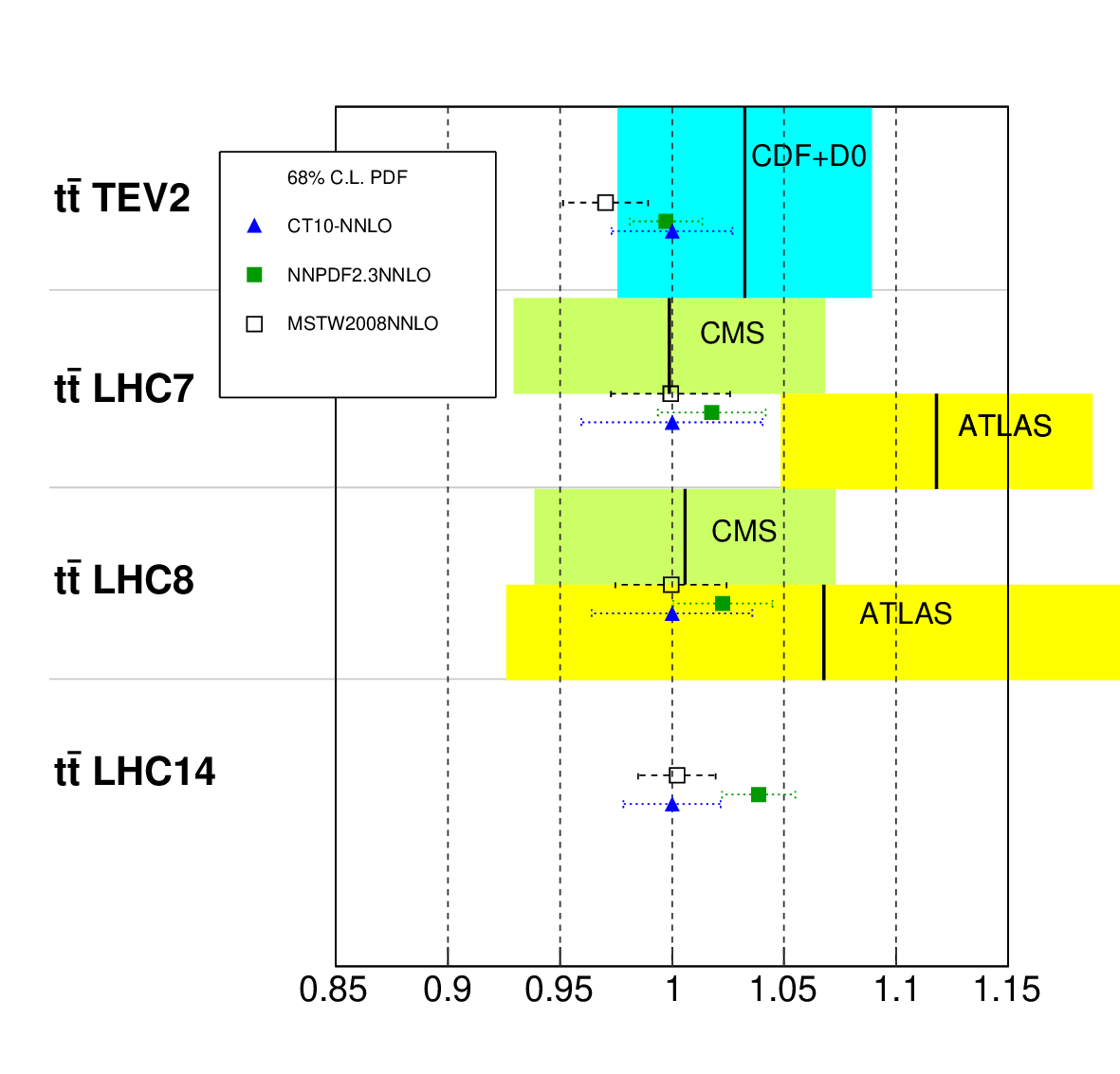}
\caption{Approximate NNLO predictions and experimental measurements
of $t\bar t$ total cross sections at the Tevatron and LHC, 
normalized to CT10NNLO predictions.
\label{T-xsecA}}
\end{center}
\end{figure}
}

\clearpage
\section{Introduction
\label{sec:intro}}

A global analysis of perturbative QCD makes use of experimental data from
many short-distance scattering processes to construct,
within some approximation,
universal parton distribution functions (PDFs) for the proton.
These PDFs can be used to calculate hadronic cross sections.
The CTEQ global analysis at the next-to-leading order (NLO) 
in the strong coupling constant $\alpha_s$ 
has been developed over decades.
Examples of general-purpose PDFs in this series include CTEQ6, 
published in 2002~\cite{Pumplin:2002vw},
followed by CTEQ6.1 in 2003~\cite{Stump:2003yu}.
Starting from CTEQ6.5 published in 2006~\cite{Tung:2006tb},
and in ensuing PDF sets such as CTEQ6.6~\cite{Nadolsky:2008zw} and
CT09~\cite{Pumplin:2009nk},  
the effects of finite quark masses on the CTEQ global analysis
have been implemented in the S-ACOT-$\chi$ factorization scheme 
\cite{Aivazis:1993pi,Collins:1998rz,Kramer:2000hn,Tung:2001mv}
at NLO accuracy. The most recent CTEQ NLO PDFs, named CT10 and CT10W,
were published in 2010~\cite{Lai:2010vv} and
are currently in wide use in phenomenological predictions  
for the Tevatron,  LHC, and other experiments.

The data from the CERN Large Hadron
Collider (LHC) cover a wide kinematic range with high expected
precision; a similar level of precision is needed for the theoretical
predictions. Thus, there is a need for using cross sections calculated
up to  next-to-next-to-leading order (NNLO) 
in the strong coupling constant $\alpha_s$ with
 parton distribution function sets that are also determined at NNLO.
Complete calculations for this order in $\alpha_s$ are available
for  the running coupling $\alpha_{\rm s}(Q)$, PDF evolution in $Q$
\cite{Moch:2004pa,Vogt:2004mw}, 
matrix elements in deep-inelastic scattering (DIS)  
\cite{SanchezGuillen:1990iq,vanNeerven:1991nn,Zijlstra:1991qc,Zijlstra:1992qd,Laenen:1992zk,Riemersma:1994hv,Buza:1995ie},
and vector boson production\,\cite{Anastasiou:2003yy,Anastasiou:2003ds}. 
NNLO matrix elements are unknown for several other processes in the global analysis.
Most notably, the key input theoretical cross section for inclusive 
jet production in $pp\!\!\! {}^{{}^{(-)}}$ collisions is still evaluated at NLO, although 
some of the NNLO radiative contributions have been already computed \cite{Ridder:2012dg,Ridder:2013mf}.\footnote{
Some other cross sections, such as for Higgs boson production and diboson 
production, have been calculated to NNLO, but to date 
have not been implemented in any global analysis. 
}
Various NNLO PDF sets have been published in the
literature\,\cite{Martin:2009iq,Ball:2012cx,Alekhin:2012ig,JimenezDelgado:2008hf,h1zeus,HERAPDF} 
that make use of NNLO matrix elements when available.

In this paper, we present a new generation of CTEQ parton distributions, at NNLO, 
named CT10 NNLO.\footnote{Parametrizations of the CT10NNLO PDFs were publicly released in 2012 \cite{Nadolsky:2012ia}.}
They are obtained from a global analysis
of QCD data in which three changes have been  made to include the calculations
at NNLO. First, the parton distribution functions $f(x,Q)$ are evolved 
according to the 3-loop Dokshitzer-Gribov-Lipatov-Altarelli-Parisi (DGLAP) 
equations. Second, the strong coupling
$\alpha_{\rm s}(\mu_R)$ evolves in the renormalization scale $\mu_R$
according to the 3-loop beta function. Third, the hard matrix
elements for DIS and vector boson production are 
calculated up to two QCD loops.
 A detailed implementation of neutral-current DIS cross sections in the S-ACOT-$\chi$ scheme,
 which performs a consistent treatment of non-zero masses of heavy partons, up to this order
 of accuracy 
is documented in Ref.~\cite{Guzzi:2011ew} and recapped in Sec.~\ref{sec:HQ}.

The experimental data sets included in the CT10NNLO fit are
essentially the same as in the CT10 and CT10W NLO fits, 
with the exceptions of Tevatron Run-1 inclusive jet data and a subset of the
Tevatron Run-2 lepton charged asymmetry
data from $W$ boson decays. The changes in the selection of the experimental data sets are summarized in Sec.~\ref{sec:DATA}. 
No LHC data  are included
in the CT10 NNLO analysis, which can therefore be used to make predictions 
based exclusively on the pre-LHC data. [The impact of the HERA and LHC data
published after 2010 will be investigated in the post-CT10 fits.]

Since the CTEQ4 analysis released in 1996
\cite{Lai:1996mg,Huston:1995tw}, hadron-hadron collider measurements on
production of hadronic jets are included in CTEQ fits to provide
pivotal information about the high-$x$ gluon distribution.
To achieve good agreement with Tevatron inclusive jet cross sections,
it is generally necessary to assume a larger gluon PDF that would be
preferred solely on the basis of DIS experiments; 
and the uncertainty in the gluon
PDF at $x>0.1$ is reduced dramatically in the global fits that utilize the
collider jet data than in the fits without them \cite{Pumplin:2009nk}.

The inclusive jet cross sections play as a prominent role in the CT10 NNLO
analysis, however, as stated previously, they are still evaluated 
with NLO matrix elements, hence require special scrutiny.
The experimental data on the inclusive jet cross sections are
statistically very precise but have significant systematic
uncertainties. The partial NNLO contributions to the LHC jet production 
reach 10-20\% \cite{Ridder:2013mf}, their magnitude is about the same
as those of the experimental systematic
effects, and possibly indistinguishable from the latter.
We examine variations in the gluon and other PDFs
caused by the QCD scale dependence and by various treatments 
of systematic uncertainties in inclusive jet production. 
The impact of jet-related uncertainties is
compared to those from other sources. 

For consistency with the CT10 NLO PDFs, 
CT10 NNLO assumes the same value of the QCD coupling strength 
$\alpha_s(M_Z)=0.118$ 
and pole masses for heavy quarks of $m_c=1.3$ GeV and $m_b=4.75$ GeV. 
A commentary on the choice of the heavy-quark masses and description
of heavy-quark production in DIS is presented in Sec.~\ref{sec:charmpheno}.
The CT10NNLO eigenvector  PDF sets are available on the CTEQ web
site~\cite{cteqweb}, and in the LHAPDF standard
format~\cite{lhapdfweb}. Together with the PDF eigenvector sets 
for the central value of $\alpha_s(M_Z)=0.118$, an additional PDF
series in which  $\alpha_s(M_Z)$  
is varied in the range 0.112-0.127
are provided. These PDFs are sufficient for computing the correlated
PDF+$\alpha_s$ uncertainty  
by adding the PDF and $\alpha_s$ uncertainties in quadrature, as
explained in Ref.~\cite{Lai:2010nw}.

The outline of the paper is as follows.
Section \ref{sec:HQ} reviews our NNLO implementation 
of DIS cross sections, placing an emphasis 
on the treatment of heavy-quark contributions.
Section~\ref{sec:SE} summarizes the statistical procedure 
of the global analysis, most notably the definition of the log-likelihood 
function and the implementation of correlated systematic errors.   
Section \ref{sec:DATA} 
lists the experimental data used in the CT10NNLO analysis.
Section \ref{sec:pdfplots} 
describes some features of the resulting CT10NNLO PDFs, while
Section \ref{sec:comparisons} 
presents detailed comparisons of data and theory.
Section \ref{sec:predictions} presents NNLO predictions for collider measurements 
based on the CT10NNLO PDFs, and
Section \ref{sec:conclusion} contains our summary and conclusions.

\section{Heavy-flavor scheme in the CT10NNLO fit \label{sec:HQ}}

\subsection{QCD factorization for heavy quarks in the S-ACOT-$\chi$
scheme}

A consistent implementation of contributions from the massive quarks ($c$ and $b$)  
is a prerequisite and challenge for a viable NNLO PDF analysis.
The mass dependence of the  heavy-quark DIS contributions to PDF fits affects
QCD precision observables in a wide range of energies
\cite{Tung:2006tb}. Heavy-quark mass effects were studied in 
PDF fits since mid-1990's in the context of several theoretical approaches, or
{}``heavy-quark schemes''.
 The S-ACOT-$\chi$ factorization scheme
\cite{Aivazis:1993pi,Collins:1998rz,Kramer:2000hn,Tung:2001mv} has been adopted
in the recent NLO fits CTEQ6HQ \cite{Kretzer:2003it}, 6.5 \cite{Tung:2006tb},
6.6 \cite{Nadolsky:2008zw}, and CT10 \cite{Lai:2010vv}. For the
present work, the S-ACOT-$\chi$ scheme has been extended to NNLO
accuracy, \emph{i.e.}, ${\cal O}(\alpha_{s}^{2})$, in the computation
of neutral-current DIS cross sections \cite{Guzzi:2011ew}.\footnote{
The charged-current DIS cross sections, 
for which some ${\cal O}(\alpha_{s}^{2})$
massive amplitudes are not available, are evaluated at NLO.} 
The alternative TR$'$ scheme \cite{Thorne:1997ga,Thorne:1997uu} is
used in the MSTW and HERAPDF fits, while the FONLL scheme \cite{Cacciari:1998it,Forte:2010ta}
has been adopted by the NNPDF collaboration. The BMSN
scheme \cite{Buza:1996wv,Chuvakin:1999nx,Bierenbaum:2009zt} 
and the fixed-flavor number
(FFN) schemes are used by the ABM and GJR groups, respectively.

The general-mass variable flavor number
(GM-VFN) schemes replace the zero-mass approximation, 
which is no longer adequate for describing the DIS data. 
They evaluate the coefficient functions using the exact dependence on heavy-quark mass $m_h$, 
while heavy-quark PDFs provide an approximation for collinear production
of $h\bar h$ pairs when $Q^2\gg m_h^2$. They are valid 
across the whole range of $Q$ values accessed 
in the global fits.\footnote{A complementary approach 
(an intermediate-mass scheme) uses approximate quark mass dependence 
in all scattering channels \cite{Nadolsky:2009ge}.
DIS cross sections in this approach utilizing the exact $O(\alpha_{s})$
massive ACOT terms and approximate $O(\alpha_{s}^{2})$ and $O(\alpha_{s}^{3})$
massive terms have been recently published \cite{Stavreva:2012bs}.}

The heavy-quark schemes are brought into better
consistency among themselves 
when going from NLO to NNLO calculations \cite{Binoth:2010ra,Guzzi:2011ew}.
For example,
the spread of theoretical predictions for the standard-candle $W$
and $Z$ boson production cross sections at the LHC has shrunk from
6-8\% at NLO \cite{Nadolsky:2008zw} (with the largest deviations
observed with the zero-mass PDFs) to less than 3\% at 
NNLO \cite{Aad:2010yt,CMS:2011aa,Watt:2011kp,Thorne:2012az,Ball:2013gsa,Gao:2013wwa}.
Quark-mass effects continue to be important. Their non-equivalent
treatment by various PDF analysis groups gives rise to residual uncertainties
in the standard candle predictions and in the observables sensitive
to the gluon or heavy-quark scattering, such as Higgs boson production.

The NNLO calculation in the S-ACOT-$\chi$ scheme in 
Ref.~\cite{Guzzi:2011ew} focused on two issues
that had not been earlier addressed. First, we clarified the connection
of the GM-VFN scheme at NNLO to the proof of QCD factorization
for DIS with massive quarks presented
by Collins \cite{Collins:1998rz}. It was demonstrated 
that the S-ACOT-$\chi$ scheme can be derived to all orders 
from Collins' approach and is validated by a QCD factorization theorem.

Second, we documented an algorithm that organizes the NNLO S-ACOT-$\chi$
calculation in close analogy to the zero-mass VFN computation. As
a result, the NNLO DIS cross sections can be constructed step-by-step
from the components that can be found in literature. 
In the S-ACOT-$\chi$ scheme,
all elements arise from the all-order
factorization formalism, which is not always the case in other frameworks.
These features
distinguish the S-ACOT-$\chi$ scheme from the TR$'$ and FONLL schemes 
that follow different implementation approaches. 

As a result, the structure of the S-ACOT-$\chi$ cross sections is 
readily reproducible; the universality of the PDFs follows from the Collins 
proof of QCD factorization \cite{Guzzi:2011ew}. 
The $Q$ dependence of the QCD coupling and PDFs is found
by numerical QCD evolution that assumes one shared $N_{f}$ value
in each $Q$ range; QCD quantities for $N_{f}$ and $N_{f}+1$ active
flavors are related at the switching energy scales through the matching
conditions. The NNLO radiative contributions are assembled straightforwardly
from the matrix elements in the massive FFN \cite{Laenen:1992zk,Riemersma:1994hv,Buza:1995ie}
and zero-mass VFN schemes \cite{SanchezGuillen:1990iq,vanNeerven:1991nn,Zijlstra:1991qc,Zijlstra:1992qd},
as well as from the mass-dependent operator matrix elements for the
heavy-quark PDFs \cite{Buza:1996wv}.

The S-ACOT-$\chi$ scheme reduces to the FFN scheme at the heavy-quark threshold
$Q^{2}\approx m_{h}^{2}$ and to the zero-mass $\overline{\rm MS}$ scheme
at $Q^{2}\gg m_{h}^{2}$, without additional renormalization. The
matching of the GM-VFN cross sections to the FFN cross sections near
the mass threshold is generally not automatic. To realize it, the TR$'$ and FONLL
scheme introduce additional elements (constraints on the $Q$ dependence
of the heavy-quark DIS contributions in the TR$'$ scheme and the {}``damping
factor'' in the FONLL framework) that are not stipulated by the 
QCD factorization theorem. In the S-ACOT-$\chi$ scheme, threshold
matching is rather a consequence of the energy conservation condition
that suppresses the difference between the GM-VFN and FFN results when $Q^{2}$
approaches $m_{h}^{2}$. This is achieved by restricting the allowed form for the
approximate coefficient functions that describe QCD scattering off an initial-state heavy
quark, so that they comply with energy conservation.
It leads to the effective rescaling
($\chi$ rescaling) of the light-cone momentum fraction in the approximate
heavy-quark scattering terms \cite{Tung:2001mv,Barnett:1976ak}.

The derivation of mass-dependent contributions  from the QCD
factorization theorem, which is shown to be compatible with the
rescaled terms, leads to more confident constraints on the PDFs. Our fitting 
code implements these NNLO Wilson coefficient functions in the 
S-ACOT-$\chi$ scheme together with the HOPPET program for the evolution 
of $\alpha_{s}$ and PDFs \cite{Salam:2008qg}, in which the
switching points between the active flavors can be expressed either
in terms of the $\overline{\rm MS}$ masses or pole masses.\footnote{The authors
thank Gavin Salam for his help with the setup of this code.} The fitting
program can read either the pole masses or $\overline{\rm MS}$ masses as
an input.  

In the latter case, the $\overline{\rm MS}$ masses 
are converted into the pole masses when needed, {\it e.g.}, to evaluate
those operator matrix elements $A_{ab}^{(2)}$ that are published in terms
of the pole masses. Although the  $\overline{\rm MS}$ masses are known more
precisely than the pole masses, using 
the $\overline{\rm MS}$ mass at the NNLO level 
does not lead to a more accurate fit, 
as the the $\overline{\rm MS}\rightarrow\mbox{
  pole}$ conversion in the DIS cross sections introduces an additional
perturbative uncertainty that overrides the precision of the
$\overline{\rm MS}$ input. In the CT10 NNLO fit, the pole masses $m_c=1.3$
GeV and $m_b = 4.75$ GeV have been assumed. Complementary fits in
which the $\overline{\rm MS}$ charm mass was chosen as the input and
constrained by the global data were also carried out
\cite{Gao:2013wwa}.

\subsection{The S-ACOT-$\chi$ scheme at NNLO in a nutshell}

In the S-ACOT-$\chi$ scheme,  a generic \emph{inclusive} structure function $F_{2,L}\equiv F$ takes form of a convolution product ($\otimes$) of the Wilson coefficient 
functions $C_{ia}$ and the parton distribution functions $f_{a/p}(\xi,\mu)$:\begin{align}
F(x,Q) & =\sum_{i=1}^{N_{f}^{fs}}e_{i}^{2}\sum_{a=0}^{N_{f}}\int_{x}^{1}\frac{d\xi}{\xi}\, C_{ia}\left(\frac{x}{\xi},\frac{Q}{\mu},\frac{m_{h}}{\mu},\alpha_{s}(\mu)\right)\,f_{a/p}(\xi,\mu)\nonumber \\
 &
\equiv\sum_{i=1}^{N_{f}^{fs}}e_{i}^{2}\sum_{a=0}^{N_{f}}\left[C_{ia}\otimes
  f_{a/p}\right](x,Q),\label{Ffactorization}\end{align}
where $\xi$ is the light-cone momentum fraction,
$\mu$ is the factorization scale, $N_{f}$ indicates the number
of active flavors, and $N_{f}^{fs}$ is the number of the produced
final-state flavors (most generally, $N_{f}^{fs} \neq N_{f}$). 
The structure function can be also written as\begin{equation}
F=\sum_{l=1}^{N_{l}}F_{l}+\sum_{h=N_{l}+1}^{N_{f}^{fs}}F_{h},\label{FSeveralHeavyFlavors}\end{equation}
where $l$ and $h$ are the indices of light-quark and heavy-quark
flavors probed by the photon, respectively \cite{Forte:2010ta}. [Note that the S-ACOT-$\chi$ scheme can simultaneously
account for several flavors with non-zero masses]. 
On the right-hand side,  \begin{equation}
F_{l}=e_{l}^{2}\sum_{a}\left[C_{l,a}\otimes f_{a/p}\right](x,Q),\hspace{0.4cm}F_{h}=e_{h}^{2}\sum_{a}\left[C_{h,a}\otimes f_{a/p}\right](x,Q).\label{Fheavy}\end{equation}
The ${\cal O}(\alpha_{s}^{2})$ radiative contributions, $F_{l}^{(2)}$
and $F_{h}^{(2)}$, are \begin{eqnarray}
 &  & F_{l}^{(2)}=e_{l}^{2}\left\{
  C_{l,l}^{NS,(2)}\otimes(f_{l/p}+f_{\bar{l}/p})+c^{PS,(2)}\otimes\Sigma+c_{l,g}^{(2)}\otimes
  f_{g/p}\right\} ,\label{F2l}\\
 &  & F_{h}^{(2)}=e_{h}^{2}\left\{ c_{h,h}^{NS,(2)}\otimes(f_{h/p}+f_{\bar{h}/p})+C_{h,l}^{(2)}\otimes\Sigma+C_{h,g}^{(2)}\otimes f_{g/p}\right\} ,\label{F2h}\end{eqnarray}
where the lower-case notation $c_{a,b}^{(2)}$ indicates a zero-mass
Wilson coefficient function, the uppercase notation $C_{a,b}^{(2)}$
indicates a massive coefficient function,
and $\Sigma(x,\mu)=\sum_{i=1}^{N_{f}}\left[f_{i/p}(x,\mu)+f_{\bar{i}/p}(x,\mu)\right]$
denotes the singlet-quark PDF. These equations have the same form
as the factorized expressions for the zero-mass structure functions.
Their components are listed explicitly in Ref.~\cite{Guzzi:2011ew}.
In this derivation, we employ a rescaling convention \cite{Tung:2001mv}
to construct the LO, NLO, and NNLO coefficient functions with initial-state
heavy quarks, $c_{h,h}^{(k)}$ with $k=0,1,$ and $2,$ and the associated
subtraction terms. They are obtained by evaluating the zero-mass expressions
as a function of the rescaling variable $\chi$:
\begin{equation}
c_{h,h}^{(k)}\left(\frac{x}{\xi},\frac{Q}{\mu},\frac{m_{h}}{Q}\right)=c_{h,h}^{(k)}\left(\frac{\chi}{\xi},\frac{Q}{\mu},m_{h}=0\right)\,\theta(\chi\leq\xi\leq1),\label{ChhChi}\end{equation}
where \begin{equation}
\chi=x\,\left(1+\frac{(\sum_{fs}m_{h})^{2}}{Q^{2}}\right),\label{chi}\end{equation}
and $\sum_{fs}m_{h}$ is the sum of the heavy-quark masses in the
final state (equal to $2m_{h}$ in the lowest-order $c\bar{c}$ pair
production). These rescaled coefficient functions obey energy conservation
and vanish near the production threshold, so that the FFN result is
reproduced in this limit.

Eqs.~(\ref{F2l}) and (\ref{F2h}) apply to
the inclusive DIS functions $F_2$, $F_L$, etc. 
In the case of semi-inclusive DIS production of heavy quarks, the
definition of the semi-inclusive (SI) structure functions $F_{h,SI}$, 
such as $F_{2}^{c\bar c}$ measured at HERA, requires additional
care in order to obtain infrared-safe results at all $Q$ \cite{Chuvakin:1999nx}.
In the CT10NNLO global fit, the following approximation for $F_{h,SI}$
has been adopted \cite{Forte:2010ta,Guzzi:2011ew}. At moderate $Q$
values accessible at HERA, it is defined as \begin{eqnarray}
F_{h,SI}^{(2)}(x,Q)=F_{h}^{(2)}(x,Q)+\sum_{l=1}^{N_{l}}e_{l}^{2}L_{I,q}^{NS,(2)}\otimes(f_{l/p}+f_{\bar{l}/p}),\label{FhSI}\end{eqnarray}
where $F_{h}^{(2)}(x,Q)$ is the ${\cal O}(\alpha_{s}^{2})$ contribution
to the inclusive DIS function $F(x,Q)$ arising from photon scattering
on a heavy quark, cf. Eq.~(\ref{F2h}). The function $L_{I,q}^{NS,(2)}(\xi,Q/\mu,m_{h}/\mu)$
is the non-singlet part of the light-quark component $F_{l}(x,Q)$
that contains radiation of a $h\bar{h}$ pair in the final state,
computed in Ref.~\cite{Buza:1995ie}. In the kinematic region $Q<10$
GeV, which supports  most of the HERA charm production data, we
observe that $L_{I,q}^{NS,(2)}$ contributes about 0-3\% on the semi-inclusive
charm cross section, {\it i.e.}, it is small compared to the
typical experimental errors.

\subsection{A phenomenological illustration\label{sec:charmpheno}}

In Fig.~\ref{fig:F2c_Qbin}, the CT10 NLO/NNLO predictions for $F_{2}^{c\bar{c}}$
are compared to a recent data set from the H1 collaboration
\cite{Aaron:2011gp}, in which the charm structure function $F_{2}^{c\bar{c}}$
was extracted from the $D^{*}$ meson distribution. Predictions
for $F_{2}^{c\bar{c}}$ are computed as a function of the momentum
fraction $x$ for five different bins of $Q$. The dashed red lines
and solid blue lines represent the S-ACOT-$\chi$ predictions at NLO
and NNLO, respectively. 
Although
this specific data set \cite{Aaron:2011gp} was not included in the
CT10 fits, the overall agreement is very good. The NLO and NNLO
predictions are close to one another for the most part, at least in the intermediate
$10^{-3}\leq x\leq10^{-2}$ kinematic region, in which the bulk of
the data was collected. However, the NLO prediction is shown for a particular
factorization scale that improved agreement with the data,
$\mu=\sqrt{Q^{2}+m_{c}^{2}}$, while a more typical scale choice
$\mu=Q$ was taken at NNLO. In general, the NLO predictions for $F_{2}^{c\bar{c}}$ are characterized by a
wide band of the scale dependence uncertainty. This band is reduced
significantly when going to NNLO, and the NNLO predictions in the S-ACOT-$\chi$
scheme are closer to those in the FONLL-C and TR$'$ schemes \cite{Guzzi:2011ew}.
The NNLO predictions
are therefore more robust compared to NLO. 

\figFtctotA

Soon after the CT10 NNLO PDFs were released, several 
DIS charm production cross section measurements by H1 and ZEUS that
we use
\cite{Breitweg:1999ad,Chekanov:2003rb,Aktas:2005iw,Aktas:2006py}
were combined into one set \cite{Abramowicz:1900rp}. We compared NNLO fits 
to the separate and combined HERA charm data sets.
Their resulting PDFs and uncertainties turned out to be very similar
\cite{Gao:2013wwa}. The combined HERA charm data agrees well with the
CT10 NNLO prediction: $\chi^2/N_{pt}=55.4/47=1.18$
for the default charm pole mass $m_c^{pole}=1.3 $ GeV that we use.

The pole mass of 1.3 GeV is compatible with the preferred $\overline{\rm MS}$
charm mass $m_c(m_c)$ determined from a simultaneous fit of PDFs 
and $m_c(m_c)$ in \cite{Gao:2013wwa}. In that paper, 
the figure-of-merit function
$\chi^2$ for the global hadronic data (including the combined HERA
charm data set) was examined as a function of $m_c(m_c)$. 

The preferred value was determined to be
$m_c(m_c) = 1.19^{+0.08}_{-0.15}$ GeV at 68\% c.l.,
where the error is a quadrature sum of PDF and theoretical
uncertainties.\footnote{In contrast, the fit is hardly sensitive
to the bottom quark mass, given the smallness of
 bottom-scattering contributions to the fitted cross sections.}
This value, constrained primarily
by a combination of inclusive and charm
production measurements in HERA deep-inelastic scattering,
translates  into $m_c^{pole}=1.31^{+0.09}_{-0.16}$ GeV and
$1.58^{+0.08}_{-0.15}$ GeV if using the conversion  
formula in Eq. (17) of \cite{Chetyrkin:2000yt} at one and two loops with
$\alpha_s(M_Z,N_f=5)=0.118$. 
Either converted value is 
compatible at $2\sigma$ with $m_c^{pole} = 1.3$ GeV assumed by CT10 NNLO. 

It also interesting to note that this best-fit $m_c(m_c)$ agrees
with the outcome of an independent PDF fit in the FFN scheme at 
the same order in $\alpha_s$ (at two loops), 
$m_c(m_c)=1.15\pm0.04(\mbox{exp})^{+0.04}_{-0.00}(\mbox{scale})$ GeV \cite{Alekhin:2012vu}.\footnote{The
disparate magnitudes of the uncertainties on $m_c(m_c)$ in two fits 
are caused mostly 
by their different definitions of the PDF uncertainties 
adopted by the ABM and CT10 groups.}
In both the S-ACOT-$\chi$ and FFN fits, the central $m_c(m_c)$ values
are lower than the world-average $1.275\pm0.025$ GeV
\cite{Beringer:1900zz}, but the remaining difference disappears upon
considering variations due to the choice of the rescaling variable in
the S-ACOT-$\chi$ scheme (quantifying to some extent the missing
higher-order contributions) or including the approximate 
three-loop contribution
to massive quark DIS in the FFN scheme \cite{Alekhin:2012vu}.

The best-fit $m_c(m_c)$ and the PDFs that come with it are sensitive
to the implementation of heavy-quark terms, such as the truncation of
the perturbative conversion of the $\overline{\rm MS}$ to the pole
mass, $\lambda$ parameter in the 
rescaling correction, and implementation of correlated effects
for the combined (charm) HERA data. The  associated 
errors in the key LHC cross sections have been examined and found to
be mild and lie 
within the usual 90\% c.l. PDF uncertainty \cite{Gao:2013wwa}.
We don't separate them from the PDF uncertainty, which has a comparable
uncertainty of its own. The corresponding variations in 
the LHC NNLO $W/Z$ cross sections are below 2\%, {\it i.e.} 
reduced comparatively to the uncertainties due 
the heavy-quark scheme observed previously at NLO \cite{Nadolsky:2008zw}.

\section{Global analysis with correlated systematic
  errors \label{sec:SE}}

For completeness, and in order to establish notations for the ensuing
discussion, in this section we summarize the statistical procedure
adopted in the CT10NNLO analysis. An important aspect of 
this procedure is to determine the {\em
uncertainties} of the PDFs, which arise from several
sources, both experimental and theoretical. 

Experimental uncertainties may be uncorrelated between different 
measurements, such as the bin-by-bin {\em statistical error} on the
measured cross section. Some experimental uncertainties are highly correlated 
between different measurements within one experiment or even among several 
experiments. The {\em luminosity error}  is an example of 
a correlated experimental uncertainty in a data set. It 
affects equally the normalization of all cross section
measurements from the experiment. Other correlated uncertainties
typically exist and may have impact that is comparable to the
luminosity error. 

Inclusion of the correlated experimental errors utilizing
their standard deviations published by the experiments 
began in the CTEQ program with the construction of the CTEQ6 parton
distributions \cite{Pumplin:2002vw}. This is achieved by constructing
an appropriate figure-of-merit function $\chi^2$ that includes errors
from both uncorrelated and correlated sources, as shown explicitly 
in Sec.~\ref{sec:Log-likelihood-function}. 

By examining the $\chi^2$ function in the neighborhood of the best
fit and its dependence on $\alpha_s$, 
the CT10 global analysis determines a family of independent 
eigenvector sets that can be used to propagate the combined
PDF+$\alpha_s$ uncertainty into theoretical predictions  
(cf. Sec.~\ref{sec:PDFeigen}). The central PDF set and PDF eigenvector
sets are generally sensitive to the implementation 
of correlated systematic errors in
the fit. Possible procedures for reconstructing the correlation error
matrix from the published standard deviations, and the expected
modifications in the PDFs that they induce, 
are discussed in Sec.~\ref{sec:CorSysErrors}. A numerical
comparison of these procedures will be presented in Sec.~\ref{sec:comparisons}.

\subsection{Log-likelihood function in the CT10 NNLO analysis
\label{sec:Log-likelihood-function}}

A typical experiment $E$ in the global fit publishes a set of measurements
$\{M_{i};i=1,2,3,...,N_{pt}\}$, consisting of a central value $D_{i}$
for the observable, a standard deviation for the \emph{uncorrelated}
experimental error $s_{i}$, and standard deviations $\beta_{k\alpha}$
for each of $N_{\lambda}$ systematic errors, where $k=1,2,3,...,N_\lambda$.
We do not know the experimental errors, but experiments provide their
standard deviations. So we write 
\begin{equation}
D_{i}=X_{i}+s_{i}\delta_{i}+\sum_{\alpha=1}^{N_\lambda}\beta_{k\alpha}\lambda_{\alpha}\label{eq:Di}
\end{equation}
 where $X_{i}$ is the ``true'' value of the 
observable.\footnote{The ``true'' value is the {\em mean value} of $D_{i}$ that would
result from a large number of independent experiments.} 
Equation (\ref{eq:Di}) defines {\em nuisance parameters}, $\delta_{i}$
and $\lambda_{\alpha}$. These will be random numbers with mean value
0 and standard deviation 1, 
\begin{eqnarray}
\langle\delta_{i}\rangle=0 & {\rm and} & \langle\delta_{i}^{2}\rangle=1;\\
\langle\lambda_{\alpha}\rangle=0 & {\rm and} & \langle\lambda_{\alpha}^{2}\rangle=1.
\end{eqnarray}
 We assume that the uncorrelated errors are completely uncorrelated
among the $N$ measurements; that is, 
\begin{equation}
\langle\delta_{i}\delta_{j}\rangle=\delta_{ij}.
\end{equation}
We also assume that the systematic errors are completely correlated
among the $N_{pt}$ measurements; that is, $\lambda_{\alpha}$ does 
not depend on $i$. However, the systematic errors are among themselves
uncorrelated, 
\begin{equation}
\langle\lambda_{\alpha}\lambda_{\beta}\rangle=\delta_{\alpha\beta}.
\end{equation}

The goal of the global analysis is to find the theoretical parameters
for which the theoretical values $T_{i}$ of the observables are as
close as possible to the ``true'' values, $X_{i}$. In Eq.\,(\ref{eq:Di})
replace $X_{i}$ by $T_{i}$. Then we are led to minimize the differences,
by defining
\begin{equation}
\chi_{E}^{2}(\{a\},\{\lambda\})=\chi_{D}^{2}+\chi_{\lambda}^{2},\label{Chi2sys}
\end{equation}
 where 
\begin{equation}
\chi_{D}^{2}\equiv\sum_{k=1}^{N_{pt}}\frac{1}{s_{k}^{2}}\left(D_{k}-T_{k}-\sum_{\alpha=1}^{N_{\lambda}}\beta_{k,\alpha}\lambda_{\alpha}\right)^{2},\label{Chi2D}
\end{equation}
 and 
\begin{equation}
\chi_{\lambda}^{2}\equiv\sum_{\alpha=1}^{N_{\lambda}}\lambda_{\alpha}^{2},
\label{Chi2lambda}
\end{equation}
 and minimizing $\chi_{E}^{2}$ with respect to the theory parameters.
However, there are additional sources of systematic uncertainties 
associated with the unknown nuisance parameters $\{\lambda_{\alpha}\}$.
Therefore, to have agreement between theory and data within the standard
deviations of the experimental errors, we also vary the $\{\lambda_{\alpha}\}$
values, seeking to make $\chi_{E}^{2}$ small. 

Because $\chi_{E}^{2}$ is only a quadratic function of $\{\lambda_{\alpha}\}$,
we may obtain the minimum of $\chi_{E}^{2}$ with respect to $\{\lambda_{\alpha}\}$
analytically \cite{Pumplin:2001ct}. At the minimum, the nuisance
parameters take the values
\begin{equation}
\bar{\lambda}_{\alpha}=\sum_{i=1}^{N_{pt}}\frac{D_{i}-T_{i}}{s_{i}}\sum_{\delta=1}^{N_{\lambda}}\mathcal{A}_{\alpha\delta}^{-1}\frac{\beta_{i,\delta}}{s_{i}},\label{lambda0}
\end{equation}
 where 
\begin{equation}
\mathcal{A}_{\alpha\beta}=\delta_{\alpha\beta}+\sum_{k=1}^{N_{pt}}\frac{{\beta}_{k,\alpha}\beta_{k,\beta}}{s_{k}^{2}}.\label{A}
\end{equation}
 The corresponding best-fit representation for $\chi_{E}^{2}$ is
\begin{equation}
\min\chi_{E}^{2}=\sum_{i,j}^{N_{pt}}(D_{i}-T_{i})({\rm cov^{-1}})_{ij}(D_{j}-T_{j}).\label{eq:chi2}
\end{equation}
It includes the inverse of the covariance matrix

\begin{equation}
({\rm cov})_{ij}\equiv s_{i}^{2}\delta_{ij}+\sum_{\alpha=1}^{N_{\lambda}}\beta_{i,\alpha}\beta_{j,\alpha},\label{eq:covmat_cteq}
\end{equation}
given by
\begin{equation}
({\rm cov}^{-1})_{ij}=\left[\frac{\delta_{ij}}{s_{i}^{2}}-\sum_{\alpha,\beta=1}^{N_{\lambda}}\frac{\beta_{i,\alpha}}{s_{i}^{2}}\mathcal{A}_{\alpha\beta}^{-1}\frac{\beta_{j,\beta}}{s_{j}^{2}}\right].
\end{equation}
The best-fit value of $\chi_{D}^{2}$ can be also expressed in terms
of the data $D_{sh,i}$ that are ``shifted'' from the central values
by the best-fit correlated errors,
\begin{equation}
\min\chi_{D}^{2}=\sum_{i=1}^{N_{pt}}\left(D_{{\rm sh},i}-T_{i}\right)^{2}/s_{i}^{2};\label{eq:chisqshd}
\end{equation}
Here 
\begin{equation}
D_{{\rm sh},i}=D_{i}-\sum_{\alpha=1}^{N_{\lambda}}\beta_{i\alpha}\overline{\lambda}_{\alpha}.\label{eq:shifteddata}
\end{equation}
We quote the function $\min\chi_{D}^{2}$ 
in Table \ref{tab:EXP_bin_ID} as a useful measure of the agreement
between theory and data for each experiment. In Sec.~\ref{sec:comparisons}, 
we will also use $\chi_{E}^{2}=\chi_{D}^{2}+\chi_{\lambda}^{2}$
as another measure of the agreement between theory and data.

Thus far, we have considered only a single experiment. The \emph{global}
chi-square function that is minimized in the CT10 global analysis
sums over all experiments $E$ and includes a contribution $\chi_{th}^{2}(\{a\})$
specifying theoretical conditions for the PDF parameters: 
\begin{equation}
\chi_{{\rm global}}^{2}=\sum_{{\rm E}}\chi_{E}^{2}+\chi_{th}^{2}.\label{chi2global}
\end{equation}
The function $\chi_{th}^{2}$ is introduced to prevent some unconstrained
PDF parameters from reaching values that might lead to unphysical
predictions at small Bjorken $x$ (specifically, $x\leq10^{-4}$),
where experimental constraints are sparse \cite{Nadolsky:2008zw}. 
It rules out those PDF
parameter combinations that may result in negative cross sections
or unlikely flavor dependence. The specific condition imposed in the 
CT10 analysis is to 
constrain the ratio
$R_{s}(x,Q)=\left[s(x,Q)+\bar{s}(x,Q)\right]/\left[\bar{u}(x,Q)+\bar{d}(x,Q)\right]$  
of the strange PDFs to non-strange sea PDFs to be in the interval
$0.5\leq R_{s}(x,Q_{0})\leq1.5$ at the initial scale $Q_{0}$ and
$x$ below $10^{-5}$. In the current fit, the $R_s$ ratio is not
constrained in this region  
by the experimental data, hence a loose theory-motivated 
constraint needs to be imposed.

The global $\chi^{2}$ function in Eq.~(\ref{chi2global}) is constructed
by assuming that both uncorrelated and correlated errors are quasi-Gaussian
and symmetric. When an experiment provides asymmetric errors, we symmetrize
them. The exact procedure for symmetrization of errors has low impact
on the outcome of the fit, as the number of points with very asymmetric
errors typically is small compared to the total number of points and
parameters.

\subsection{PDF eigenvector sets and $\alpha_s$ uncertainty \label{sec:PDFeigen}}
Besides the central (best-fit) PDF set, the CT10NNLO release includes PDF
eigenvector sets to estimate the uncertainty range of our PDF fits 
using the Hessian method \cite{Stump:2001gu,Pumplin:2001ct} and 
either the symmetric \cite{Pumplin:2002vw}
or asymmetric \cite{Nadolsky:2001yg,Lai:2010vv} 
master formula to estimate the PDF uncertainties. The Hessian
method is based on an iterative procedure for finding 
linear combinations of the fitting 
parameters by diagonalization of the Hessian matrix.  In the CT10NNLO fit, 
we use 25 free parameters to describe the parton distributions at
$Q_0$, and hence have 25 eigenvector directions. For each of these directions,
we find a pair of eigenvector sets defined by moving away 
from the best-fit location (where $\chi^2$ takes its minimum value) 
by a distance that estimates the  boundary of the $90\,\%$ confidence
interval for each experiment. 

The 
method to find these eigenvectors is explained in \cite{Lai:2010vv}.
In addition to including an upper ``tolerance'' bound on the increase in total
$\chi^2$ that realizes the 90\% c.l. agreement on average, 
as in the CTEQ6 analysis, we include a 
penalty term in $\chi^2$ that quickly grows when the PDF
set fails to describe any specific experiment. 
The effective function $\chi^2_{\mathrm Eff}$ that is constructed this
way is scanned along each eigenvector direction until $\chi^2_{Eff}$ 
increases above the tolerance bound 
or quick $\chi^2_{Eff}$ growth due to the penalty is triggered.

The penalty term is constructed from 
statistical variables $S$ derived from $\chi^2_E$
values for individual experiments~\cite{Lai:2010vv}. 
In contrast to $\chi^2_E$, the variables 
$S$ obey an approximate standard normal distribution 
independently of $N_{pt}$, which simplifies the comparison
of confidence  levels between data sets containing widely different 
numbers of points. The quasi-Gaussian $S$ variables are found 
by using a simpler Fisher's approximation \cite{Fisher:1925} in
the CT10 NLO fit (reliable for experiments with many data points,
$N_{pt}\gtrsim 9$) 
and more accurate Lewis' approximation \cite{Lewis:1988} in the
CT10NNLO analysis. 

We note that even in the central fit,  
some data sets have $\chi^2_E/{\mathrm N_{pt}}$ 
that lie outside the $90 \,\%$ confidence level.
That is not surprising, of course, since there are  
28 data sets---naively the chance for all of them to lie with $90
\,\%$ confidence is only $0.9^{28} = 0.05$.  To allow for this, 
for each data set that has 
$\chi^2_E > N_{pt}$ in the central fit, we rescale its $\chi^2_E$ by a factor 
$N_{pt}/\chi^2_{\mathrm central fit}$ before computing the
penalty.  

In practical applications, the PDF uncertainty obtained with the Hessian
eigenvector sets must be combined with the $\alpha_s$ uncertainty. The
procedure  for combining these uncertainties 
is described in \cite{Lai:2010nw}. The central and error 
eigenvector sets of the CT10NNLO family 
assume $\alpha_s(M_Z)=0.118$, which is compatible with
the world-average value. In addition we provide best-fit PDF sets for other 
$\alpha_s(M_Z)$ values in the interval 0.112-0.127. For a theory
observable, the 90\% c.l. $\alpha_s$ uncertainty can be estimated 
as the half of the difference of predictions using PDF sets with
$\alpha_s=0.116$ and 0.120. Then, the  PDF+$\alpha_s$ uncertainty with
all correlations can be estimated by adding the PDF and $\alpha_s$ 
uncertainties in quadrature.
 
\subsection{Implementation of correlated systematic errors \label{sec:CorSysErrors}}

In the treatment of correlated systematic errors, there is another
subtlety concerned with the distinction between additive and multiplicative
systematic uncertainties. The correlated systematic errors fall into
two classes: additive errors, for which the experiment can determine
the absolute value $\beta_{i,\alpha}$ of the standard deviation;
or multiplicative errors, for which only the relative fraction $\sigma_{i,\alpha}=\beta_{i,\alpha}/X_{i}$
is known. The two kinds must be handled differently to avoid a bias
in the outcome of the fit. However, it is a common practice in many
experiments to publish the correlated systematic uncertainties as
the relative percentage errors $\sigma_{i,\alpha}$  
regardless of whether the systematic error is additive or
multiplicative, rather than the
absolute values $\beta_{i,\alpha}$.

To reconstruct the correlation matrix as 
\begin{equation}
\beta_{i,\alpha}=\sigma_{i,\alpha}X_{i},\label{eq:betaD-1}
\end{equation}
one selects the reference central value $X_{i}$ for each datum. Since
the additive errors are supposedly independent of the theory predictions,
it is natural, although not necessary, to use the experimental central
values $D_{i}$ as the references, $\beta_{i,\alpha}=\sigma_{i,\alpha}D_{i}$.

For a multiplicative error, the reference to $D_{i}$ is generally
unacceptable, as random fluctuations in $D_{i}$ tend to bias the
best-fit parameters. 
A well-known multiplicative bias described by D'Agostini arises
in the treatment of the normalization of the data \cite{D'Agostini:1999cm,D'Agostini:1993uj}.
A fit that references the normalization error to $D_{i}$ would underestimate
the true cross section. This downward bias is prevented if smoothly
behaving $X_{i}$ values are used as the references, such as the theoretical
values $T_{i}$ at each data point ($\beta_{i,\alpha}=\sigma_{i,\alpha}T_{i}$). 

Most of the time the experimental paper does not distinguish
between the additive and multiplicative errors. In this case,
the global fit has to choose between several
ways, and several trade-offs, for computing $\beta_{i,\alpha}$. When
preparing the CT10 NNLO PDFs, we explored various procedures for the
computation of $\beta_{i,\alpha}$ that have been identified in recent
literature. {[}See also a related discussion in the appendix of Ref.~\cite{Ball:2012wy}.{]}
\begin{enumerate}
\item Method $D,$ or the ``experimental normalization'' method, normalizes
\emph{all correlated errors} to the experimental central values: that
is, we compute $\beta_{i,\alpha}=\sigma_{i,\alpha}D_{i},$ for all
additive and multiplicative errors alike. In this method, the correlated
errors are independent of theory predictions. However, they are
affected by irregular fluctuations of the central data points, which may
result in a pronounced D'Agostini bias in the best-fit parameters when
the fluctuations are large.
\item Method $T$ normalizes  the \emph{multiplicative
errors }to the theoretical values that are updated in every
fitting iteration. ``The $T^{(0)}$ method'' \cite{Ball:2009qv}
is a variation on this approach in which the multiplicative errors
are updated once in many iterations. The $T^{(0)}$ method was proposed
to prevent nonlinear behavior of $\chi^{2}$ that may occur in
method $T$. \emph{Additive errors} in these methods 
remain normalized to $D_{i}.$ Both methods $T$ and $T^{(0)}$ 
are free of D'Agostini's bias, as the multiplicative errors are
estimated by using a smooth function. Method $T^{(0)}$
converges to the true solution after several updates of the $T_{i}^{(0)}$
values, provided the partial derivatives $\partial\chi^{2}/\partial T_{i}^{(0)}$
are negligible, and may fail to do so otherwise. Our comparisons
follow the original implementation of the $T$ and $T^{(0)}$ 
methods~\cite{Ball:2009qv}, in which only the luminosity errors (but
not other multiplicative errors such as for the jet energy scale)
were referenced to $T_{i}$ ($T_{i}^{(0)}$).
\item An extended version of method $T$ normalizes \emph{both additive
  and multiplicative errors} 
to the current theoretical values, $\beta{}_{i,\alpha}=\sigma_{i,\alpha}T_{i}.$ Similarly, the extended method $T^{(0)}$ normalizes all correlated
errors to fixed theoretical values, $\beta{}_{i,\alpha}=\sigma_{i,\alpha}T_{i}^{(0)}.$
The advantage of these methods is that $\beta_{i,\alpha}$ shows the
smoothest behavior among all considered. 
\end{enumerate}

The extended method $T$ is used by default in our NLO fits, including 
all CTEQ6.X series and CT10(W) NLO. At NNLO 
the implementation of systematic effects became even more important, 
given the reduction in other uncertainties. In Sec.~\ref{sec:corrjet} 
we compare five procedures for implementation of
systematic errors in inclusive jet
production, where the systematic effects are among the most
pronounced. We use the extended $T$ method for other scattering
processes.

\section{Experimental data sets and theoretical updates
\label{sec:DATA}}

\subsection{Data selection: vector boson and jet production}
The experimental data sets included in the CT10 NNLO fit are listed
in Table~\ref{tab:EXP_bin_ID}. With a small number of exceptions,
they were chosen to be the same as in the CT10 NLO fit. At NLO, we
presented two PDF sets, designated as CT10 and CT10W. The distinction
between them concerns the inclusion of 
the D\O\ Run-2 data for the rapidity asymmetry ($A_{\ell}$ )  
of the charged lepton from $W$ boson decay. 
This data was included in the CT10W analysis, with an extra weight, but not 
included in the CT10 analysis.  More specifically, 
the CT10W analysis  includes the $A_{\ell}$ data points in three 
ranges of the lepton transverse momentum ($p_{T\ell}$)  in the electron
decay channel, and one $p_{T\ell}$ bin in the muon decay channel.
After the publication of the CT10 and CT10W analyses, the  D\O\ collaboration 
has recommended not to include those two less inclusive data sets in $p_{T\ell}$,
{\it i.e.}, with $25\leq p_{T\ell}\leq35$ GeV and $p_{T\ell}\geq35$ GeV.
In the CT10NNLO analysis,  only the most inclusive data 
with $p_{T\ell}\geq25$ GeV, in both the electron and muon
decay channels, are included in the analysis. 
Since CT10NNLO includes only a part of the
D\O\ $A_{\ell}$ data that distinguishes between CT10 and CT10W
PDFs, it can be treated as an NNLO counterpart to either the CT10 or
CT10W NLO PDF sets.

The Tevatron Run-2 inclusive jet data have a wider rapidity coverage 
(in the case of CDF) and smaller statistical and systematic errors, 
as compared to the Tevatron Run-1 data. 
Their effects on the global analysis have been extensively discussed in 
Ref.~\cite{Pumplin:2009nk}.
We concluded that since the Run-1 jet data are not perfectly consistent
with the Run-2 data sets, one may ask if the Run-1
jet contributions should be retained. In the 
CT10NNLO analysis, we do not include the Tevatron Run-1 
inclusive jet data.  This choice has
implications for the large-$x$ gluon PDF, as will be shown later. 

\begin{table}[tb]
\begin{tabular}{|l|l|l|l|}
\hline 
 \textbf{Experimental data set} & $N_{pt}$ & CT10NNLO & CT10W\tabularnewline
\hline
\hline 
 Combined HERA1 NC and CC DIS \cite{Aaron:2009aa} & 579  & 1.07 & 1.17\tabularnewline
\hline 
 BCDMS $F_{2}^{p}$ \cite{Benvenuti:1989rh} & 339  & 1.16 & 1.14 \tabularnewline
\hline 
 BCDMS $F_{2}^{d}$ \cite{Benvenuti:1989fm} & 251  & 1.16 & 1.12 \tabularnewline
\hline 
 NMC $F_{2}^{p}$ \cite{Arneodo:1996qe} & 201  & 1.66 & 1.71 \tabularnewline
\hline 
 NMC $F_{2}^{d}/F_{2}^{p}$ \cite{Arneodo:1996qe} & 123  & 1.23 & 1.28\tabularnewline
\hline 
 CDHSW $F_{2}^{p}$ \cite{Berge:1989hr} & 85  & 0.83 & 0.66 \tabularnewline
\hline 
 CDHSW $F_{3}^{p}$ \cite{Berge:1989hr} & 96  & 0.81 & 0.75 \tabularnewline
\hline 
 CCFR $F_{2}^{p}$ \cite{Yang:2000ju} & 69  & 0.98 & 1.02 \tabularnewline
\hline 
 CCFR $xF_{3}^{p}$ \cite{Seligman:1997mc}  & 86  & 0.40 & 0.59 \tabularnewline
\hline 
 NuTeV neutrino dimuon SIDIS \cite{Mason:2006qa} & 38  & 0.78 & 0.94 \tabularnewline
\hline 
 NuTeV antineutrino dimuon SIDIS \cite{Mason:2006qa} & 33  & 0.86 & 0.91 \tabularnewline
\hline 
 CCFR neutrino dimuon SIDIS \cite{Goncharov:2001qe} & 40  & 1.20 & 1.25 \tabularnewline
\hline 
 CCFR antineutrino dimuon SIDIS \cite{Goncharov:2001qe} & 38  & 0.70 & 0.78 \tabularnewline
\hline 
 H1 $F_{2}^{c}$ \cite{Adloff:2001zj}  & 8  & 1.17 & 1.26 \tabularnewline
\hline 
 H1 $\sigma_{r}^{c}$ for $c\bar{c}$ \cite{Aktas:2004az,Aktas:2005iw} & 10  & 1.63 & 1.54 \tabularnewline
\hline 
 ZEUS $F_{2}^{c}$ \cite{Breitweg:1999ad} & 18  & 0.74 & 0.90 \tabularnewline
\hline 
 ZEUS $F_{2}^{c}$ \cite{Chekanov:2003rb}  & 27  & 0.62 & 0.76 \tabularnewline
\hline 
 E605 Drell-Yan process, $\sigma(pA)$ \cite{Moreno:1990sf} & 119  & 0.80 & 0.81 \tabularnewline
\hline 
 E866 Drell Yan process, $\sigma(pd)/(2\sigma(pp))$ \cite{Towell:2001nh} & 15  & 0.65 & 0.64 \tabularnewline
\hline 
 E866 Drell-Yan process, $\sigma(pp)$ \cite{Webb:2003ps} & 184  & 1.27 & 1.21 \tabularnewline
\hline 
 CDF Run-1 $W$ charge asymmetry \cite{Abe:1996us} & 11  & 1.22 & 1.24 \tabularnewline
\hline 
 CDF Run-2 $W$ charge asymmetry \cite{Acosta:2005ud} & 11  & 1.04 & 1.02 \tabularnewline
\hline 
 D\O\ Run-2 $W\rightarrow e\nu_{e}$ charge asymmetry \cite{Abazov:2008qv}  & 12  & 2.17 & 2.11 \tabularnewline
\hline 
 D\O\ Run-2 $W_{.}\rightarrow\mu\nu_{\mu}$ charge asymmetry \cite{Abazov:2007pm}  & 9  & 1.65 & 1.49 \tabularnewline
\hline 
 D\O\ Run-2 Z rapidity distribution \cite{Abazov:2006gs} & 28  & 0.56 & 0.54 \tabularnewline
\hline 
 CDF Run-2 Z rapidity distribution \cite{Aaltonen:2010zza}  & 29  & 1.60 & 1.44 \tabularnewline
\hline 
 CDF Run-2 inclusive jet production \cite{Aaltonen:2008eq} & 72  & 1.42 & 1.55 \tabularnewline
\hline 
 D\O\ Run-2 inclusive jet production \cite{Abazov:2008ae} & 110  & 1.04 & 1.13 \tabularnewline
\hline
\hline 
{\bf Total:} & \textbf{2641} & \textbf{1.11} & {\bf 1.13} \tabularnewline
\hline
\end{tabular}\caption{Experimental data sets examined in the CT10NNLO and
CT10W NLO analyses, together with their $\chi^2$ values. 
\label{tab:EXP_bin_ID} }
\end{table}

The data selection choices in regard to the Tevatron $W$ asymmetry 
and inclusive jet production affect mainly certain combinations of the
PDFs at $x > 0.1$: specifically the ratio $d/u$  (down- versus up-quark PDFs)  
and the gluon PDF. These combinations change by amounts that are mostly 
comparable to the PDF uncertainties obtained in the NLO analysis. 

\subsection{Data selection: fixed-target DIS experiments}
As in CT10(W) NLO, the inclusive proton data from BCDMS, NMC, CDHSW,
CCFR experiments are included in the form of structure functions 
$F_2(x,Q)$. Those were derived by the experimental groups 
from the measured cross sections using
the ratio $R=\sigma_L/\sigma_T$ of the longitudinal to transverse
cross sections for virtual photon DIS that is known better now than at the
time of the experimental publications.
In the past candidate fits, we have reconstructed $F_2(x,Q)$ 
using an alternative $R$ parametrization or replaced it entirely
by a reduced DIS cross section when available. The
resulting modifications in the PDFs in these trials have not exceeded 
the published PDF uncertainty.

In the CT10 NNLO analysis, the replacement of the BCDMS $F_{2}(x,Q)$ by the
respective reduced cross sections\footnote{The reduced cross sections
are reconstructed from the BCDMS data on $F_{2}^{p,d}(x,Q)$ determined
assuming $R=0$.} (without refitting)
results in about the same fit quality: $\chi^2/N_{pt}=$1.12 and 1.24
for the BCDMS proton and deuteron reduced cross
sections, to be compared against 1.16 and 1.16 for the data on
$F_{2}^{p,d}(x,Q)$.   

Recently, the ABM group found that the replacement of the NMC 
reduced cross section for inclusive proton DIS 
by the corresponding $F_2^p(x,Q)$ data modifies the
preferred gluon distribution and $\alpha_s(M_Z)$ in their NNLO fit, 
as well as the Higgs cross section that depends on them \cite{Alekhin:2011ey}.
We performed a similar comparison in the context of CT10 NNLO study,
by replacing $F_2^p(x,Q)$ for NMC (our default choice) by the NMC
reduced cross section. It was known for a long time that the NMC
$F_2^{p}$ data are not fitted well in CTEQ analyses
\cite{Pumplin:2002vw}, for the reasons that are not completely understood.
The CT10 NNLO fit results in $\chi^2/N_{pt}=1.67$ for the NMC
$F_{2}^p(x,Q)$ data, while for the NMC reduced cross section
 we get $\chi^2/N_{pt}=1.88$ (1.80) 
if the PDFs are fitted to the NMC $F_{2}^p(x,Q)$
(NMC reduced cross section). It is not possible to get a good fit to
the NMC proton data set even if its statistical weight in the $\chi^2$
is increased or other fixed-target DIS experiments (BCDMS) are dropped
from the fit. 

For $F_2^p(x,Q)$ data, the NNLO fit to the NMC data behaves
similarly to the CTEQ6 NLO fit discussed in Appendix B.2 of
\cite{Pumplin:2002vw}. The distribution of data-theory residuals is 
consistent with larger-than-normal fluctuations of the
data. All 12 nuisance parameters for experimental systematic shifts
are below 1.5 and contribute $\chi^2_\lambda\approx 
6$ to $\chi^2$. In contrast, a fit to the NMC reduced cross section
results in theory typically overshooting the data and requires a
$6\sigma $ and $2.8\sigma$ shifts in the experimental radiative correction 
and luminosity for the NMC set at the global minimum of $\chi^2$.

For the world-average $\alpha_s(M_Z)=0.118$ adopted by
CT10 NNLO, the above replacement induces essentially 
no change in the best-fit PDFs. 
When $\alpha_s(M_Z)$ is varied, the NMC reduced cross section 
is still poorly fitted and prefers 
$\alpha_s(M_Z)=0.112-0.113$, {\it i.e.} 
much lower than the world-average value $\alpha_s(M_Z)=0.1184\pm 0.0007$
\cite{Beringer:1900zz}.  We therefore choose to fit the NNLO PDFs to the 
NMC $F_2^p(x,Q)$ data for consistency with CT10 NLO, 
given that the replacement of $F_2^p(x,Q)$ by the reduced cross
section neither improves the fit nor modifies the PDFs. 
Besides the abovementioned observations about the NMC data made
in the CTEQ6 NLO study, the MSTW \cite{Thorne:2011kq}
and NNPDF \cite{NNPDF:2011aa} groups have reached  
a similar conclusion that some NMC data sets are poorly
fitted, and the replacement of the structure functions by the
reduced cross sections for them is inconsequential for the NNLO
PDFs.

\subsection{Theory developments}
Besides implementing the NNLO QCD contributions in neutral-current DIS
and vector boson production, we updated the theoretical treatment of several
experiments. In the coefficient functions for deep inelastic scattering,
we have updated the definitions of the electroweak couplings describing $Z$ and
$\gamma^*-Z$ interference contributions. This update mildly modifies
the $d$ quark PDFs at large $x$.

In the previous global analyses, the NLO jet cross sections utilized 
tables of point-by-point ratios of NLO/LO cross
sections ($K$ factors) computed with the \textsc{EKS} code
\cite{Ellis:1992en}.  The 
original \textsc{EKS} calculation was published in the 1990's and 
had limited accuracy when comparing it to the latest precision jet data.
A deeply modified version of the \textsc{EKS} code (\textsc{MEKS}) was
prepared to provide advanced predictions for jet cross sections
\cite{Gao:2012he}.  The \textsc{MEKS} calculation is entirely independent 
from \textsc{NLOJET++} \cite{Nagy:2003tz}, another frequently used
program for computation of NLO jet cross sections. 
A comparison to \textsc{MEKS} validated predictions from
\textsc{FastNLO} \cite{Kluge:2006xs,ftnlo:2010xy,Wobisch:2011ij} 
and \textsc{APPLGRID} \cite{Carli:2010rw}, 
the interfaces for fast interpolation of \textsc{NLOJET++} cross 
sections. The specific input settings for which the \textsc{MEKS},
\textsc{FastNLO}, and 
\textsc{APPLGRID} programs agree to a few percent 
were documented \cite{Gao:2012he,Ball:2012wy}. The magnitude of
theoretical uncertainties in inclusive jet cross sections was also
estimated. It was observed, for example, that the current NLO QCD scale
dependence is too large at the highest $P_{Tj}$ and $y_j$ 
of the LHC inclusive jet data set to provide meaningful constraints on
the relevant PDFs. This situation is expected to be improved when the
NNLO inclusive jet cross section calculation becomes available.  

In the CT10NNLO fit, we compute the NLO inclusive jet
cross sections with \textsc{FastNLO} version 1.0 
and cross-validate the results using \textsc{MEKS}. Some NNLO fits 
such as MSTW and NNPDF approximate the two-loop
contribution to Tevatron inclusive jet production by its logarithmic expansion
obtained in threshold resummation \cite{Kidonakis:2000gi}. Our jet
cross sections are evaluated at NLO and 
do not include the threshold resummation contribution, as 
collider jet production is not dominated by the
threshold kinematic region, and there are known examples when 
the exact NNLO correction is very
different from its  threshold approximation because of the substantial
power-like finite contributions arising in the exact NNLO result (see,
{\it e.g.}, \cite{Cacciari:2011hy,Czakon:2013goa}. 
Furthermore, at the LHC, the NLO cross section combined with a
two-loop threshold contribution provided by FastNLO disagrees 
with the inclusive jet data \cite{Ball:2012wy}. 

In the jet cross sections, 
both the factorization and renormalization scales are chosen to be equal to
the transverse momentum $P_{Tj}$ of each individual jet. 
This choice is different from the one adopted in the CTEQ6.X and CT10
NLO analyses, where both scales were set to $P_{Tj}/2$.

\section{The CT10NNLO parton distribution functions
\label{sec:pdfplots}}

With a similar setup as in the CT10/CT10W NLO analysis, 
the CT10NNLO fit results in about the same quality of the fit 
expressed in terms of the total chi-square. We obtain $\chi^2$ of order 2950
for 2641 data points, with minor variations dependent on the setup of
the fit. In the rest of the paper, we discuss representative 
results from the new fit, assuming NNLO PDFs, unless stated otherwise.

\figALLBANDS

The central fit and PDF eigenvector sets are shown in
Fig.~\ref{fig:ALLBANDS}.
In addition to obtaining the {\em central fit}, 
there is also a need to understand the uncertainty ranges of the
PDFs, resulting from the experimental uncertainties of the data 
included in the fit. For this purpose, we
generate 50 alternate fits
using the Hessian method. Figure \ref{fig:ALLBANDS} indicates the
PDFs for the alternate fits, also called the ``error PDFs'', in
addition to the central fit. In each graph, PDFs for 4 parton
combinations are shown, by plotting $x f(x,Q)$ versus $x$ for a
fixed value of $Q$. The parton combinations are 
$f=u_{\textrm{valence}},~d_{\textrm{valence}},~g,~q_{\textrm{sea}}$. The values of $Q$ are
$2, 3.16, 8, 85$\ GeV. The dashed curves are the corresponding
functions for the CT10 NLO PDFs. Those can differ  by significant
amounts from the NNLO 
ones in some regions of the $(x,Q)$ space, especially where $x$ 
and $Q$ are small. 

\figbestnnlovsnloA
\figbestnnlovsnloB

To better assess these differences, Fig.~\ref{fig:bestnnlovsnloA}
shows the ratios 
of various parton distributions from the CT10NNLO central fit to
those from the CT10W NLO central fit, at $Q=2$\ GeV.
A few changes are noticeable:
(1) At $x<10^{-2}$, the ${\cal O}(\alpha_{s}^{2})$ evolution
in CT10NNLO suppresses $g(x,Q)$ and
increases $q(x,Q)$, as compared to CT10W PDFs.
(2) The heavy charm $c(x,Q)$ and bottom $b(x,Q)$ partons change as a result of
adopting the ${\cal O}(\alpha_{s}^{2})$ GM VFN scheme in the CT10NNLO analysis.
(3) In the large-$x$ region, $g(x,Q)$ and $d(x,Q)$ are reduced
by the removal of the
 Tevatron Run-1 inclusive jet data, the revised electroweak couplings,
the alternative treatment of correlated systematic errors, and 
the scale choices.

A similar comparison, but for $Q=85$\ GeV, is shown in
Fig.~\ref{fig:bestnnlovsnloB}. 
It also indicates that at $x>0.1$, $d(x,Q)$, $\bar{u}(x,Q)$ and $\bar{d}(x,Q)$
are all reduced in CT10NNLO, compared to CT10W.
Likewise, $g(x,Q)$ is also reduced in the  large $x$ region.

\figNNLOvsNLO
The uncertainties of CT10NNLO PDFs for various parton flavors are
shown in Figure \ref{fig:NNLOvsNLO},  
compared to the CT10W (NLO) uncertainties, at $Q = 2$ GeV.

\fignnlovsmstw

The CT10NNLO central fit is compared to that of
MSTW2008NNLO~\cite{Martin:2009iq} 
in Figure~\ref{fig:nnlovsmstw}.  It shows that 
the CT10NNLO gluon and quarks are larger as $x \rightarrow 0$. 
 At the smallest $x$ values and at the 
initial scale $Q_0$, the CT10 NNLO gluon remains positive, although
consistent with zero, as a result of the chosen form of
parametrization. We also note that CT10NNLO strangeness is larger at
$x \sim 10^{-3}$. 

\section{Comparisons to individual experiments
\label{sec:comparisons}}

As was already observed, the goodness of the
NNLO fit to most experiments is about the same as for CT10(W) NLO, and the 
changes in going to the NNLO analysis are comparable to 
the experimental uncertainties.
The differences in $\chi^2_E$ rarely exceed the expected statistical
fluctuations of order $\sqrt{2 N_{pt}}$. 
Some improvement is observed in the fit to 
the HERA-1 combined data on DIS, the Tevatron Run-2 jet production data,
and the CCFR and NuTeV dimuon SIDIS. In this section, we present examples
of the description of the data by the CT10 theoretical
predictions, focusing on select precise measurements in DIS, $W$ boson
production, and jet production. 

\figHERAdata

\subsection{Deep-inelastic scattering at HERA}

The HERA combined data set for $e^\pm p$ neutral current (NC) and
charged current (CC) deep-inelastic scattering~\cite{Aaron:2009aa}
has 579 measurements of the reduced cross section
$\sigma_{\rm r}(x,Q)$,
after applying our restrictions $Q \geq 2$\ GeV 
and $W \geq 3.5$\ GeV, where the invariant mass of the
hadronic final state is $W=Q\sqrt{1/x -1}$.

In Fig.~\ref{fig:HERAdata} we show  the $e^+p$
NC DIS reduced cross sections (366 data points) 
along with their {\em uncorrelated} error bars.
Data are divided into two blocks:
small $x$ is shown on the left, {\it i.e.}, $0 < x < 0.002$, and 
large $x$ is shown on the right, {\it i.e.}, $0.002 \leq x \leq 0.65$.
Fig.~\ref{fig:HERAdata} also shows theory curves
based on the CT10NNLO parton distribution functions,
superimposed on the the HERA combined data.
Note that there are systematic differences
between the central data and the theory,
as there should be due to the systematic uncertainties of the data.
Our detailed comparison of data and theory must indicate 
whether the systematic differences are consistent with
the published errors~\cite{HERAPDF}.

\figHERADoverT

Data divided by theory for $e^+p$ NC DIS 
is shown in Figure \ref{fig:HERADoverT}, 
in which we have a clearer illustration of the differences.
The horizontal lines are marked by the corresponding values of $x$.
The black dots are the central experimental values, shown along 
with uncorrelated error bars.
The red dots are the optimally shifted data values,
determined by the Hessian analysis.
We observe that the systematic shifts
bring data and theory into better agreement, as expected.

A histogram of the residuals of the shifted data, defined by
\begin{equation}\label{eq:residuals}
{\rm Residual}_{i} = \left( D_{i}
- \sum_{\alpha=1}^{N_\lambda}\beta_{i\alpha}\overline{\lambda}_{\alpha}
- T_{i}\right)\ /\,s_{i} \equiv \left( D_{sh,i} - T_{i}\right)\ /\,s_{i}, 
\end{equation}
is shown in Figure \ref{fig:HERAresiduals}.
The $i$-th residual is defined by the difference between 
the optimally shifted data and the
theory, normalized to the uncorrelated error. 
The total number of systematic errors is $N_{\lambda}=114$.
The curve shown in Fig.\,\ref{fig:HERAresiduals} is the standard normal
distribution, with no adjustable parameters:
it has a mean = 0, standard deviation = 1, and integral = 1.
The close similarity between the histogram of residuals
and the Gaussian curve implies that the data and theory agree
within the experimental errors, and that the experimental errors 
have been correctly estimated.

\figHERAresiduals
\figHERAshifts

The best-fit nuisance parameters $\{\overline{\lambda}_{\alpha}\}$ 
for the combined 
HERA data are shown in the histogram of Figure \ref{fig:HERAshifts}.
113 sources of experimental systematic errors 
contribute $\chi^2_\lambda=60$ for this experiment, so that 
many correlated systematic uncertainties have
no effect on the agreement with theory and lead to null shifts in
their respective nuisance parameters. This is reflected in the histogram of
$\{\overline{\lambda}_{\alpha}\}$, which is clearly narrower than the
standard normal distribution.

\subsection{W charge asymmetry and vector boson
  production \label{sec:TevatronWASY}} 
\figTevWasyResiduals
The $W$ charge asymmetry measured at the Tevatron is primarily sensitive to 
the shape of the ratio $d(x,M_W)/u(x,M_W)$ at $x$ values above 0.1. The NNLO
fit produces essentially the same quality of the agreement with the included
Tevatron $W$ asymmetry sets as the CT10W NLO PDF set. The
$\chi^2/N_{pt}$ values for these data sets  are listed in
Table~\ref{tab:EXP_bin_ID}. As an illustration, 
Fig.~\ref{fig:TevWasyResiduals} shows the CT10W NLO and CT10 NNLO 
data residuals for the 12 points
of the D0 Run-2 lepton asymmetry data in the electron 
\cite{Abazov:2008qv} and muon \cite{Abazov:2007pm} decay channels. 
The residuals are very close for the two PDF
sets. The quality of the NLO and NNLO fits to other data sets on
vector boson production is also comparable.

\subsection{Inclusive jet production from Run-2 at the Tevatron}

The CT10NNLO global analysis only includes the Tevatron Run-2 inclusive jet 
data for the reasons explained in Sec.~\ref{sec:DATA}.
In what follows we examine the goodness of CT10NNLO fits 
to the Run-2 inclusive jet data. 

\figDzeroJthy

In Fig.~\ref{fig:DzeroJthy} we show the differential cross section
$d^{2}{\sigma}/(dy_{j}\,dP_{Tj})$ for inclusive jet production at 
D\O\ Tevatron Run-2 \cite{Abazov:2008ae}. The various curves correspond 
to the six rapidity intervals $0-0.4$, $0.4-0.8$, $0.8-1.2$, $1.2-1.6$, 
$1.6-2.0$, and $2.0-2.4$. 
The points are the central data values, while the error bars
(too small to be seen in most cases) represent uncorrelated errors.
In the same figure, we also show the theory calculations 
based on the CT10NNLO parton distribution functions.
In this figure we do not show systematic errors.
With the systematic shifts, we obtain 
$\chi^2_E/N_{pt} = 115/110$ for D\O\ Run-2, 
where 15 units are contributed by the penalty term $\chi^2_\lambda$ 
for the 23 nuisance parameters ($\lambda_\alpha$) controlling the systematic
shifts, cf. the definition in Eq.~(\ref{Chi2sys}).
 The overall effect of the correlated systematic shifts
 is significant, compared to the small uncorrelated errors, and is 
of the same order as in the CT10 NLO fits. 
More detailed comparisons at NLO, 
illustrating the impact of the systematic shifts 
on the jet data, can be found in \cite{Pumplin:2009nk}.

To check the consistency of the comparison of the shifted data with 
CT10NNLO predictions,  Fig.~\ref{fig:DzeroJresiduals} shows 
histograms of the residuals, as defined in 
Eq. (\ref{eq:residuals}), for six bins of the jet rapidity ($y$).  
The distributions observed in the histograms 
appear to be reasonably consistent with the superimposed 
standard normal distribution, within the limitations 
of the small number of data points in each $y_j$ bin.

\figDzeroJresiduals
We also check the size of the optimal shift parameters 
$\left\{\overline{\lambda}_{\alpha} ; \alpha = 1, 2, 3,
...,23\right\}$ 
and find that the histogram of the parameters shown in
Fig.~\ref{fig:DzeroJshifts}  
is also consistent with the standard normal distribution.
According to the figure, no unreasonable shifts are required 
to create agreement between the theory and the data.
Some of the systematic errors have a larger impact on the jet cross section 
than others, such as those related to the jet energy scale uncertainties.

\figDzeroJshifts

A comparable measurement of inclusive jet production 
was performed  by the CDF collaboration 
in  Run-2 \cite{Aaltonen:2008eq}. This jet 
measurement has different systematic uncertainties than D\O , and hence provides
an independent constraint on the PDF parameters.
A similar investigation indicates that the CT10 PDFs agree well with the
CDF measurement. Fig.~\ref{fig:CDFJthy} compares the CDF
data set for inclusive jet production and the
CT10 prediction, and 
the corresponding histogram 
of the residuals is shown in  Fig.~\ref{fig:CDFJresiduals}.
The CDF value of $\chi^2_E/N_{pt}=102/72 = 1.42$ 
is higher than for D\O , but the histograms of the residuals 
do not reveal pronounced systematic disagreements.  
The contribution of 25 systematic shifts to the $\chi^2_E$ function of the
CDF measurement is $\chi^2_\lambda = 18$. The corresponding histogram 
of the optimal systematic shifts is included in Fig.~\ref{fig:CDFJshifts}.

\figCDFJthy

\figCDFJresiduals

\figCDFJshifts

\subsection{Correlated systematic errors in Tevatron jet production\label{sec:corrjet}}

The results of the previous subsection show the importance of
experimental systematic uncertainties in describing the Tevatron jet data.
These uncertainties can be generally of two kinds, additive and multiplicative.
For example, the underlying event and pileup
uncertainties are additive errors. They are determined independently
of the jet cross section. The luminosity uncertainty and jet energy
scale (JES) uncertainty are multiplicative errors. From the start,
they are estimated as the percentage of the jet cross section in each
bin. In jet production, there is no reason 
to handle the luminosity uncertainty differently
from the JES and other multiplicative errors, as most of them
have a comparable impact on the 
fit.\footnote{In the CTEQ6.X series and some other previous fits, the
  normalization 
of each experiment was fitted as a separate free parameter with a
$\chi^{2}$ penalty imposed on its deviation from the nominal value.
The procedure used was similar to the ``normalization penalty trick''
discussed in Ref.~\cite{Ball:2009qv}. 
In the CT10 fits, the normalization uncertainty is instead included
in the correlation matrix $\beta_{i,\alpha}$ and treated similarly
to other multiplicative errors.} 
Often, it is unknown if the error is additive or multiplicative. 
All of them are reconstructed from the published
percentage values according to one of the methods that were reviewed in 
Sec.~\ref{sec:CorSysErrors}.

\begin{figure}
\begin{centering}
\includegraphics[width=4in]{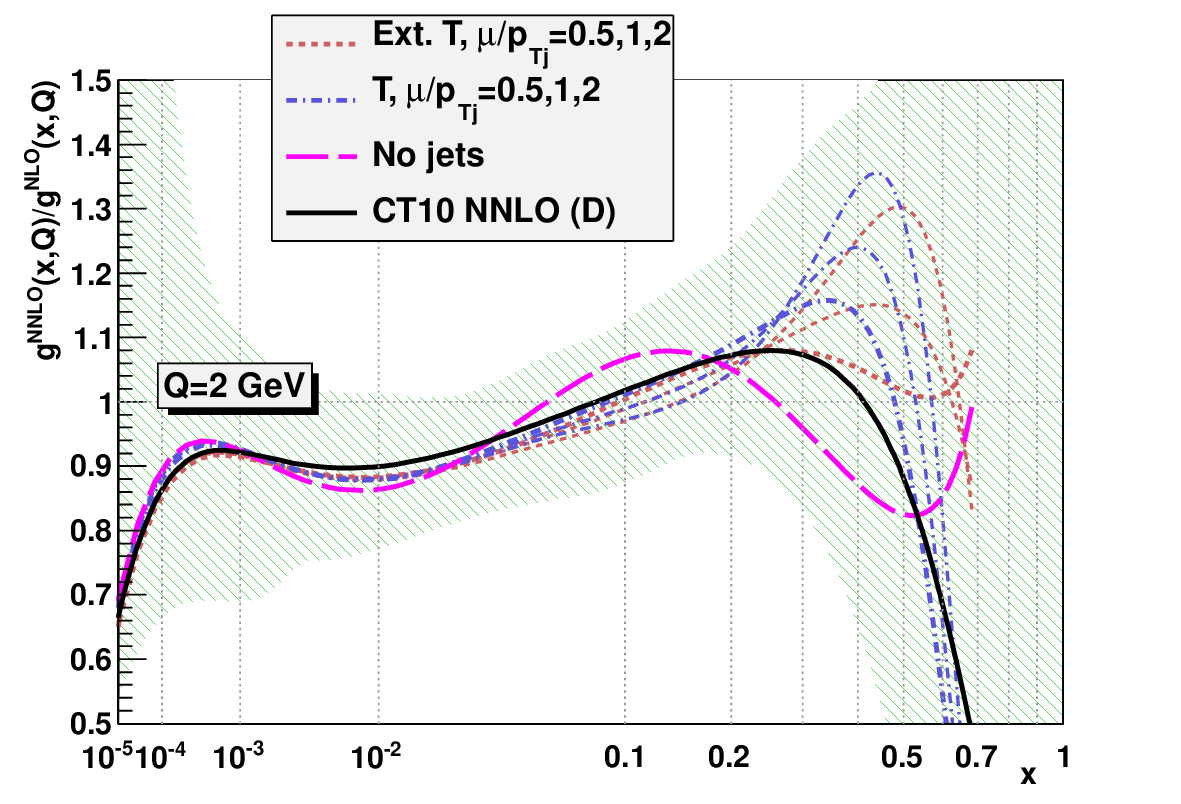}
\par\end{centering}

\caption{\label{fig:glue_variations} Comparison of NNLO gluon PDF
  parametrizations 
obtained using different procedures for including jet production data
sets in the global fit. }
\end{figure}

Both the QCD scale dependence of NLO jet cross sections and the detailed 
implementation of the systematic errors were explored when
preparing our NNLO set of PDFs.
An assortment of NNLO gluon PDF shapes, obtained by various 
treatments of jet production data in the global analysis and 
normalized to the shape of the CT10W NLO gluon PDF, is
illustrated in Fig.~\ref{fig:glue_variations}. We compare 
five methods for the computation of $\beta_{i,\alpha}$ in jet
production (cf.~Sec.~\ref{sec:CorSysErrors}) and several choices of
the renormalization and factorization scale. The CT10 central
set and its PDF error band, both obtained by assuming 
method $D$ for computing $\beta_{i,\alpha}$ and the QCD scale $\mu=P_{Tj}$,  are
shown by the solid black line and shaded green region, respectively,
with $P_{Tj}$ being the transverse momentum of an individual jet
in each $P_{T}$ bin.
For data sets other than the jet data in the global fit, we consistently use 
the extended $T$ definition of $\beta_{i,\alpha}$.

If the Run-2 jet cross sections are not included in 
the calculation of the global 
$\chi^{2}$ (or
included with a small weight, such as 0.01), the fit generally prefers
a softer gluon PDF at $x>0.1,$ as indicated by the magenta long-dashed
line in the figure. If $\alpha_{s}(M_{Z})$ is fitted as well, its
NNLO value in a fit without the jet data sets is also low, of order
0.114.

With the jet data included, there is a general preference for an enhanced
$g(x,Q)$ at large $x,$ with some 
possible outcomes shown in Fig.~\ref{fig:glue_variations}.
The curves were obtained using the extended $T$, $T,$ and
$D$ methods, for the factorization scales $\mu=P_{Tj}/2,P_{Tj},2P_{Tj}$. 
Both the original method $T$
(applied only to the luminosity error) and extended method $T$
(applied to all correlated errors) produce a
robust increase in $g(x,Q)$ at $x>0.2$, but the shape and magnitude of
the enhancement vary with $\mu$. The $D$ method (black solid curve) 
results in a softer shape of the best-fit $g(x,Q)$ which lies between
the fits with and without the jet data. All these curves lie well
within the PDF error band of the public CT10 NNLO set.

\begin{figure}[h!]
\begin{centering}
\includegraphics[width=4in]{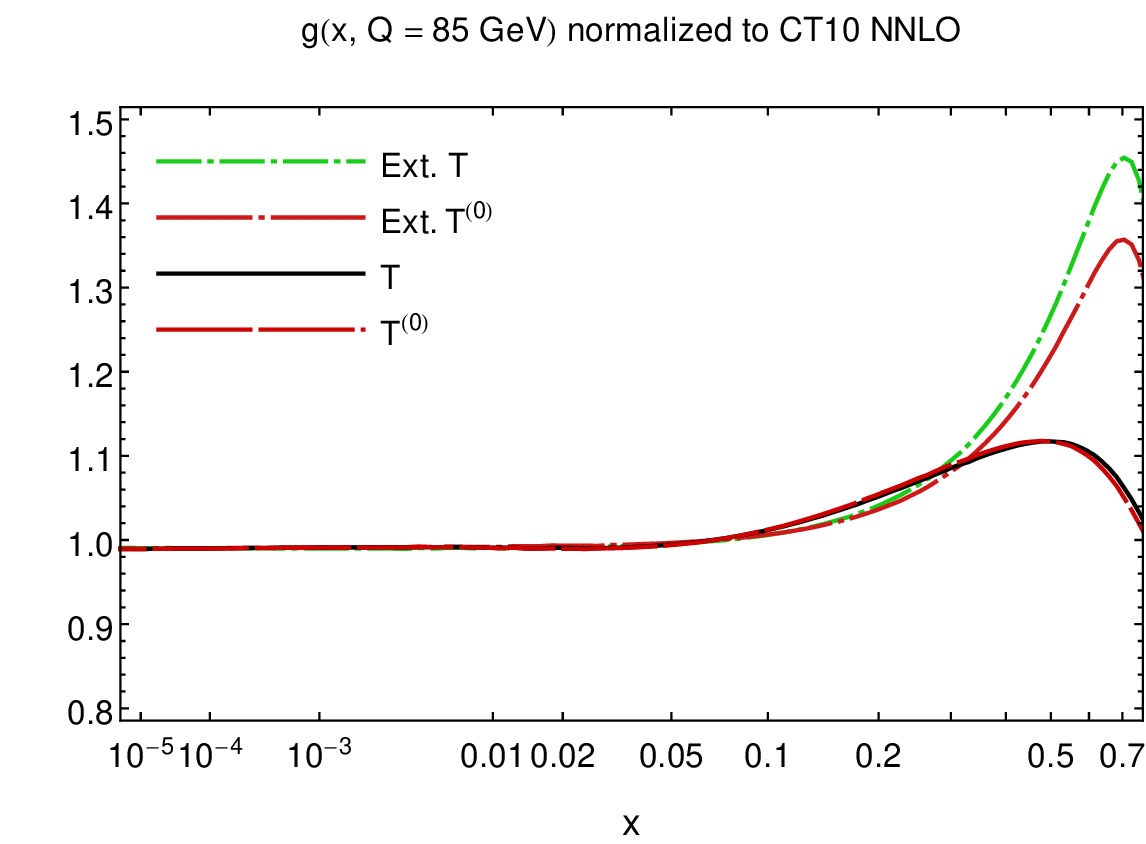} 
\end{centering}
\caption{\label{fig:gluonbeta2} 
Dependence of the gluon PDF on various definitions
of the correlation matrices $\beta_{i\alpha}$ for the Tevatron Run
2 jet data in the global fit.}
\end{figure}

A complementary perspective is provided by Fig.~\ref{fig:gluonbeta2},
comparing the gluon PDF $g(x,Q)$ at $Q=85$ GeV obtained with the 
five methods for computation of $\beta_{i\alpha}$ and the same factorization
scale $\mu=P_{Tj}$ in \textsc{FastNLO} jet cross sections.  
All the curves are normalized to the CT10 NNLO prediction (based on
the $D$ method).
Here, we observe some differences between the (extended) $T$ and $T^{(0)}$
prescriptions. The extended $T$ method results 
in a drastic enhancement of $g(x,Q)$ at $x>0.3,$ which 
is tempered when using the extended $T^{(0)}$ method. 
The $T$ and $T^{(0)}$ methods produce a stronger enhancement at
$0.05<x<0.3$, but are flatter for $x>0.3$. 

\begin{table}[h!]
\begin{centering}
\begin{tabular}{|c|c|ccccc|}
\hline 
 & Type of  & \multicolumn{5}{c|}{Method for computing $\beta_{i,\alpha}$}\tabularnewline
Parameter  & the fit  & $D$  & $T$  & $T^{(0)}$  & Ext. $T$  & Ext. $T^{(0)}$ \tabularnewline
\hline 
$\overline{\chi}^{2}_E/N_{pt}$ for CDF ($N_{pt}=72$)  & PDF only & 1.42, 1.04  & 1.23, 1.05  & 1.21, 1.04  & 1.48, 1.03  & 1.49, 1.03\tabularnewline
and D$\O$ ($N_{pt}=110)$  & PDF+$\alpha_{s}$  & 1.33, 0.95  & 1.24, 1.07  & 1.23, 1.06  & 1.48, 1.06  & 1.51,1.08 \tabularnewline
\hline 
  & PDF only & 1.2,1.3  & -1.4, -1.2  & -1.5, -1.3  & 0.53, 0.90  & 0.61, 0.96 \tabularnewline
\
$\overline{\lambda}_{\rm lumi}$ & PDF+$\alpha_{s}$  & 2.7,2.8  & -0.71,
-0.51  & -0.75, -0.55  & 1.2,1.5  & 1.4,1.8\tabularnewline
\hline 
 & PDF only & 0.118  & 0.118  & 0.118  & 0.118  & 0.118 \tabularnewline
$\overline{\alpha}_{s}(M_{Z})$  & PDF+$\alpha_{s}$  & 0.1137  & 0.1159  & 0.1159  & 0.1162  & 0.1159\tabularnewline
\hline 
\end{tabular}
\par\end{centering}

\caption{\label{jetfits}Best-fit values of
  $\chi^2_E/N_{pt}$
 and the luminosity nuisance parameter $\lambda_{\rm lumi}$ for the Tevatron Run-2 
 inclusive jet production, and for $\alpha_{s}(m_{Z})$ in a fit with
 floating $\alpha_s(M_Z)$. The columns are obtained with the five methods
 for including correlated systematic errors in jet production. The
 first (second) numbers in each column correspond to the CDF (D\O)
 data sets, respectively.}
\end{table}

In all of the listed fits to the jet data, the quality of the fits are generally
comparable. Table~\ref{jetfits} shows representative output
quantities for five methods: $\chi^{2}/N_{pt}$ for the CDF Run-2
and D$\O$ Run-2 jet data sets, the best-fit luminosity parameter $\lambda_{\rm lumi}$, and the 
best-fit $\alpha_{s}(M_{Z})$ (if fitted). In all
 fits, the values of $\chi^{2}/N_{pt}$ are similar, with the
$T/T^{(0)}$ methods showing a marginally better agreement with the
CDF jet data ($\chi^{2}/N_{pt}\approx1.23$ vs. 1.4-1.5). If the
QCD coupling strength is fixed in the fit to be 
$\alpha_{s}(M_Z)=0.118,$ all luminosity shifts do not exceed 
1.5 standard deviations, see the lines referring to ``PDF only'' fits.

However, if $\alpha_{s}(M_{Z})$ is also fitted (the ``PDF+$\alpha_{s}$''
fits), in the $D$ method we obtain a {\it low} $\alpha_{s}(M_{Z})$
of 0.1137, accompanied by a {\it large} luminosity shift
$\overline{\lambda}_{\rm lumi}\approx 2.7$. The $D$ method also results in
a low $\chi^{2}/N_{pt}$ combination of 1.33 and 0.95. 

With four other methods, we obtain higher best-fit values 
of $\alpha_{s}(M_{Z})\approx0.116$, higher $\chi^{2}$ values, 
and the luminosity shifts below $2\sigma.$ This peculiarity 
of the ``PDF+$\alpha_{s}$'' fit employing the $D$ method 
is suggestive of a marked D'Agostini bias. 
It does not occur at the same level in the ``PDF only''
fits that use a fixed $\alpha_{s}(M_{Z})$, including CT10 NNLO. 

\begin{figure}
\begin{centering}
\includegraphics[width=4in]{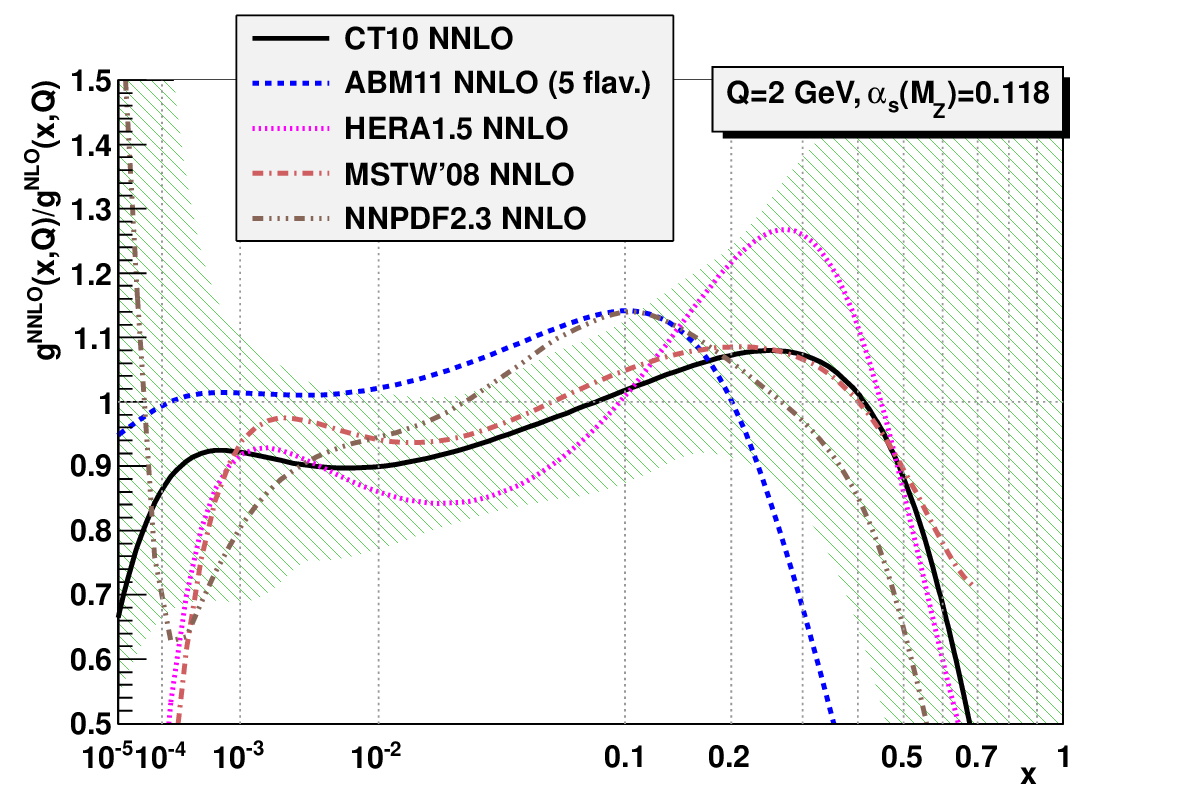}
\par\end{centering}

\caption{\label{fig:glue_variations2} Comparison of NNLO gluon PDF
  parametrizations obtained by different groups at the QCD scale
  $Q=2$ GeV and for $\alpha_s(M_Z)=0.118$. }
\end{figure}

Figure~\ref{fig:glue_variations2} compares the CT10 NNLO uncertainty
band for the gluon PDF (normalized to the CT10W NLO $g(x,Q)$) to the
central NNLO gluon PDF parametrizations from ABM11, HERAPDF1.5, MSTW, and
NNPDF2.3, evaluated with the same $\alpha_s(M_Z)=0.118$ as in CT10
NNLO. With the exception of ABM11, the central PDFs by other groups
agree with CT10 NNLO within the 90\% c.l. uncertainty. The ABM11 gluon
is distinctly smaller at $x>0.2$ and larger at $0.005<x<0.1$, most
likely because of a different heavy-quark scheme (FFN) 
that the ABM11 NNLO set uses and its preference for a low
$\alpha_s(M_Z)=0.1135$ that cause modifications in
the preferred gluon shape \cite{Thorne:2012az,Ball:2012wy,Ball:2013gsa}.

As can be seen from Fig.~\ref{fig:glue_variations}, by using method
$D$ to determine the published CT10 NNLO set and the QCD 
scale $\mu=P_{Tj}$, we obtain a PDF
uncertainty band that also covers the gluon PDF parametrizations found
with the alternative methods for the correlation matrix and other choices of QCD
scales. The softer gluon behavior than in CT10W NLO 
is consistent with the preference
of the non-jet experiments and results in lower $\chi^2$ for the
combined HERA DIS set. It may also simulate the anticipated effect
of missing NNLO corrections on the jet cross sections. 
While the method $T$ is generally preferable for the normalization of
multiplicative normalization errors, and before additional
recommendations on the separation of additive and multiplicative
errors in inclusive jet measurements  are provided by experimentalists, 
the uncertainty band for $g(x,Q)$ in Fig.~\ref{fig:glue_variations} 
represents the current
level of uncertainty in predicting the inclusive jet cross sections, due to 
variations in the fit procedure. 
These variations are typically smaller than the PDF error determined by the 
Hessian method. 

\clearpage
\section{Collider phenomenological predictions\label{sec:predictions}}

A number of standard model cross section measurements from the LHC
have been published, at center-of-mass energies of 7 and 8 TeV.
Differential distributions are available from the 7 TeV data (taken
in 2010 and 2011), while the current 8 TeV results (2012) are mostly
in the form of inclusive cross section measurements. 
These data sets have not been used in the determination
of the CT10 PDFs, but will be used in future fits.

An extensive 
comparison of theoretical predictions based on CT10NNLO and other PDF
sets to a variety of LHC cross sections (for production of electroweak
gauge bosons, inclusive jets, and top pairs) has been recently presented
in \cite{Ball:2012wy}. We refer the reader to that paper for detailed
comparisons of PDFs and parton luminosities by ABM, CT, HERAPDF, MSTW,
and  NNPDF groups; comparisons of theoretical predictions based on
these PDFs and the LHC data;  and juxtaposition of the current 
experimental and theoretical uncertainties with the goal to quantify 
projections for constraining the PDFs in near-future LHC measurements.

In this section, we collect several comparisons of theoretical
predictions for the LHC based on CT10 NNLO that support the
discussion in the other sections and complement the
benchmark comparisons in Ref.~\cite{Ball:2012wy}. We focus on
the predictions for LHC total cross sections, vector boson production,
and inclusive jet production that will likely be included 
in the upcoming PDF fits. As a variety of comparisons with different
PDF sets are published by the experimental groups, 
we include only select comparisons to illustrate general trends.
In some figures, the 
CT10 predictions are compared
to those from NNPDF2.3~\cite{Ball:2012cx} and MSTW2008~\cite{Martin:2009iq}
that follow a similar methodology of the global QCD
fit.\footnote{Comparisons to the ABM11 and HERAPDF1.5 predictions 
can be found in Ref.~\cite{Ball:2012wy}.}.
All calculations use the native value
of $\alpha_s(m_Z)$ for the PDF set. The PDF uncertainties are
shown at the 68\% c.l., and the $\alpha_s$ uncertainties are not considered.

The LHC standard model cross sections are calculated with the following
settings.
\begin{itemize}
\item For electroweak gauge boson production, the total cross sections
  are computed at NNLO using 
the program \textsc{FEWZ2.1} \cite{Gavin:2010az,Gavin:2012sy} 
and the factorization scale  $Q=M_V$. The differential cross
sections and asymmetries with cuts on transverse  momenta of the
leptons are computed with the code \textsc{ResBos}
\cite{Balazs:1997xd,Landry:2002ix} that realizes the NNLL resummation 
of logarithms at small transverse momentum. The settings of the \textsc{ResBos}
computation and nonperturbative function are the same 
as in \cite{Guzzi:2012jc}, including the hard QCD scale that is equal to
$M_V/2$. This calculation provides a close approximation 
to the NNLO resummed cross section. The needed two-loop coefficients 
are obtained from the ${\cal O}(\alpha_s^2)$ cross sections 
at large transverse momentum \cite{Arnold:1988dp,Arnold:1989ub} 
and by requiring 
that the \textsc{ResBos} total rate coincides with 
the inclusive NNLO rate \cite{Hamberg:1990np} 
computed by the \textsc{Candia} code \cite{Cafarella:2007tj,Cafarella:2008du}.

\item The cross section for top quark pair production is computed
by the \textsc{top++} code (version 1.4) \cite{Czakon:2011xx}.
The factorization and renormalization scale is chosen as $Q = m_t$,
the top quark mass is $m_t=172.5$ GeV \cite{Beringer:1900zz},
and the other settings are as in Ref.~\cite{Cacciari:2011hy}. 
The calculation includes the exact NNLO corrections to the $q\bar{q}
\rightarrow \bar{t} t$ production \cite{Baernreuther:2012ws} and an 
approximate NNLO combined with NNLL threshold logarithms in the 
other channels. Hence, the calculation provides a close to exact NNLO prediction
at the Tevatron, where the $q\bar q$ channel dominates, and an 
approximate NNLO prediction at the LHC, where top production 
takes place primarily from $gg$ initial states. 

\item Higgs boson production cross sections in the gluon-gluon fusion
channel are computed at NNLO  using the \textsc{HNNLO}
code~\cite{Catani:2007vq,Grazzini:2008tf}. The factorization and renormalization scale 
was taken to be $Q=m_H=125\mbox{ GeV}$, consistent with the recommendation of 
the LHC Higgs cross section working group~\cite{Dittmaier:2011ti}. 
Higgs production in the
bottom quark fusion channel is computed at NNLO using the 
\textsc{bbh@nnlo v1.3} code~\cite{Harlander:2003ai}. 

\item LHC inclusive jet cross sections are computed at NLO with the
\textsc{FastNLO} code~\cite{Kluge:2006xs,ftnlo:2010xy,Wobisch:2011ij}
and cross checked against the \textsc{MEKS} code in
~\cite{Gao:2012he,Ball:2012wy}. 
The central scale in these predictions is chosen to be $\mu=P_{Tj}$,
the transverse momentum of each jet. 
\end{itemize}

\subsection{W and Z total cross sections}

\figZLxsecB
\figZLxsecC

In Fig.~\ref{ZL-xsecB},  we show predictions at NNLO for $W^{\pm}$
production at 7, 8 and 14 TeV, compared to ATLAS data 
at 7 TeV \cite{Aad:2011dm} and
CMS data at 8 TeV \cite{CMS:2012zya}. 
The CT10 NNLO predictions are in good agreement
both with the data and predictions from the other NNLO PDFs
listed. The central values of the data are below all NNLO predictions,
but are within the uncertainties. For simplicity, the error
bars for the theoretical predictions indicate the 
symmetric PDF uncertainties at 68\% c.l. Generally, the PDF errors 
of the $W^+$ and $W^-$ are strongly correlated \cite{Nadolsky:2008zw}, 
but the 68\% c.l. error ellipses overlap well for all PDF sets.

In Figure~\ref{ZL-xsecC}, we show the NNLO predictions for the
$W^\pm$ and $Z$ total cross sections at 7, 8 and 14 TeV, again
compared to ATLAS data at 7 TeV \cite{Aad:2011dm} 
and CMS data at 8 TeV \cite{CMS:2012zya}.
The conclusions are broadly similar to those for the $W^+$ and $W^-$
comparisons.

\subsection{Top quark pair production, total cross sections}
\figTxsecA

In Fig.~\ref{T-xsecA}, we show the combined Tevatron total cross
section for $t\bar t$ production \cite{TevttbarCombined2012} 
and ATLAS \cite{ATLAS:2012fja,:2012xh,ATLAS:2012gpa,:2012ia,ATLAS8ttbar} 
and CMS \cite{:2012cj} total cross sections, compared to
approximate NNLO predictions. The
experimental central values and uncertainty ranges are indicated by 
vertical lines and colored rectangles. For each collider energy, 
all cross sections are normalized to the respective CT10NNLO
prediction. Again, we observe good agreement between CT10NNLO and 
other theoretical predictions as well as the data. 

\subsection{W and Z rapidity distributions}
\figZLxsecE

More detailed PDF information can be gained by comparisons to $W/Z$
rapidity distributions and $W$ charge lepton asymmetry 
from ATLAS \cite{Aad:2011dm}, CMS \cite{Chatrchyan:2012xt}, 
and LHC-B \cite{Aaij:2012vn} at 7 TeV. 
For differential observables, we use \textsc{ResBos} to account 
for multiple soft gluon radiation that may be non-negligible
in the presence of constraints on the charged lepton's 
transverse momentum $p_{T\ell}$.
Typical $W/Z$ production measurements 
require $p_{T\ell}$ to be above 20-25 GeV in order to suppress
charged leptons from background processes, and because of 
trigger requirements. 

 In some measurements, 
the lower $p_{T\ell}$ cut is as high as 35 GeV, approaching the position of
the Jacobian peak in $d\sigma/dp_{T\ell}$ at $p_{T\ell}\approx M_W/2 \approx
40$ GeV (for $W$ boson production), 
where large logarithms $\ln(Q_T/Q)$ dominate. 
($Q$ and $Q_T$ denote the invariant mass and transverse momentum of the 
weak boson, respectively.)
With such high $p_{T\ell}$ cuts, the  \textsc{ResBos}  resummation
calculation,  which resums large  logarithmic contributions to all orders in 
strong coupling, is expected to provide a better description of the data 
than the fixed-order calculations
\cite{Melnikov:2006kv,Gavin:2010az,Catani:2009sm}. 
 In practical comparisons, differences between the
resummed and fixed-order NNLO calculations are most pronounced with an
{\it upper} cut close to $p_{T\ell} \approx Q/2$ \cite{Guzzi:2011sv}. 
The resummed and NNLO asymmetries are 
closer in general when only a {\it lower} cut is imposed. For
instance, the \textsc{ResBos} and FEWZ 3.0 for the latest CMS $W$
asymmetry agree both in the $p_T^\ell > 25$ and 35 GeV bins with the
FEWZ's theoretical uncertainty \cite{CMSWasy2012}.

In Fig.~\ref{ZL-xsecE} we show a comparison between predictions
for the $Z$ and $W^\pm$ lepton rapidity distributions and
the data measured by the ATLAS collaboration using
$35 \mbox{ pb}^{-1}$ of integrated luminosity at 7 TeV \cite{Aad:2011dm}.
In all three figures the green band represents the PDF
uncertainty, while the red solid line shows the central 
prediction. For the $Z$ rapidity distribution, the
central \textsc{ResBos} prediction overshoots the data by a few
percent, but is within the PDF uncertainties from the data.
These LHC $Z$/$W^\pm$ data are yet to be included in the fits and may
modify the quark PDFs at small $x$. It is
interesting to note that ATLAS found that the tight constraints on
the relevant normalization between the $W$ and $Z$ data sets
resulted in a larger-than-expected strange quark density in the $x$
range represented by the  $Z$ and $W^\pm$ lepton rapidity
distributions \cite{Aad:2012sb}. 
We shall include these (and upcoming) data sets,  
with correlated systematic error analysis, 
in our next run of global analysis.

\figWasyA
\figWasyB
\figWasyCorr

In Fig.~\ref{WasyA},  a \textsc{ResBos} prediction is compared to
the lepton charge asymmetry from ATLAS and LHC-B \cite{Aaij:2012vn} at
7 TeV as a function of the lepton pseudorapidity $\eta_{lep}$.  
The agreement of CT10 NNLO with both asymmetry sets 
is good, {\it e.g.} $\chi^2_E/N_{pt}= 0.45$ for ATLAS $W$ asymmetry.
The LHC-B experiment probes a different kinematic range (more
forward coverage) than either ATLAS or CMS, and thus can provide
unique information on large-$x$ quark distributions.
In this figure, the central prediction and PDF error are indicated by the
black solid and blue solid lines, respectively.

The LHC $W$ charge asymmetry measurements have the potential to test 
some combinations of the PDFs that are still poorly constrained. 
For example, the
measurement of the charge asymmetry in the rapidity bins of CMS in 
Fig.~\ref{WasyB} probes {\it valence} quark PDFs,
$u_v(x,Q)\equiv u(x,Q)-\bar u(x,Q)$ and $d_v(x,Q)\equiv d(x,Q)-\bar
d(x,Q)$ in the small-$x$ region that is not constrained by the previous
data. 

Fig.~\ref{WasyCorr} shows the PDF-induced correlation cosine
$\cos\varphi$ \cite{Nadolsky:2008zw,Nadolsky:2001yg} between the LHC
$W$ charge asymmetry in several bins of lepton rapidity $\eta_{lep}$ and
valence $u$ and $d$ quark PDFs and the ratio $d(x,Q)/u(x,Q)$ of the
CT10NNLO PDF set. In the intervals of $x$ where $\cos\varphi$ 
is close to $\pm 1$, the $W$
asymmetry data at the shown $\eta_{lep}$ values is sensitive to the
PDF uncertainty on the indicated parton distributions. A strong
correlation can be observed with $u_v$ at $0.01<x<0.15$ and $d_v$ at
$0.005<x<0.05$. Both valence PDFs show a strong anticorrelation at
$x<0.001$ as a consequence of the valence sum rules. 
The $d/u$ ratio shows a strong correlation at $0.0005<x<0.005$ and
anticorrelation at $0.05 < x < 0.2$. The other PDF flavors are not strongly
correlated with the $W$ asymmetry in these rapidity ranges. 

The current CMS $W$ asymmetry data shown in the left panel of Fig.~\ref{WasyB} 
agrees reasonably with the CT10 NNLO PDFs, given the present PDF
uncertainties and experimental uncertainties. The CMS data is slightly
below the central CT10 prediction at $|\eta_{lep}| < 1$, but within the
PDF uncertainty band. The disagreement is increased at  $|\eta_{lep}| <
2$, where we also observe less regular behavior of the data.

We note that the reconstruction of the present CMS data has a systematic
uncertainty associated with the discrimination from background
processes. In the Tevatron $W$ asymmetry measurements, it was essential
to require a sufficiently high missing $E_T$ (MET), of order 20 GeV or more,
to discriminate the $W$ boson decay events from the significant
background.  The MET requirement has not been imposed in the
current CMS measurement, which applied a different technique to suppress
the background.

In the ResBos theoretical calculation, we cannot exactly implement the
experimental background subtraction technique when comparing to the
CMS data. At the lowest order in perturbative QCD 
(when $Q_T =0$), the additional
MET cut is not needed, as the condition $p_{Tl} > 35\mbox{ GeV}$
automatically implies $\mbox{MET} > 35\mbox{ GeV}$. At higher orders,
the $p_{Tl}$ and MET cuts are no longer equivalent. The predictions
for $W$ charge asymmetry depend on the assumed MET cut, cf. the right panel 
in Fig.~\ref{WasyB}. This behavior 
suggests that subtle effects in the separation of $W$ boson events
from the background may be comparable to the observed
differences between CT10 NNLO theory and CMS data in Fig.~\ref{WasyB}.

In the absence of applicable
experimental constraints in the CT10 fit, the small-$x$ behavior of
the $u_{v}$ and $d_{v}$ PDFs is assumed to be governed by a shared $x$ power,
$u_v(x,Q_0), d_v(x,Q_0) \sim x^{A_{1v}}$ in the $x\rightarrow 0$ 
with the same $A_{1v} \approx -1/2$. 
This Regge-inspired assumption partly explains the correlations of PDFs at small $x$. It will be tested by including the LHC 
$W$ asymmetry in the upcoming fits.

\subsection{ATLAS inclusive jet distribution}
\figjetatlasA
\figjetatlasB

In Figs.~\ref{fig:jetatlasA} and \ref{fig:jetatlasB}, we compare the
inclusive cross section predictions 
to the ATLAS measurements using the anti-$k_T$ jet
algorithm with $R=0.6$.  In Fig.~\ref{fig:jetatlasA}, the comparisons
are made to the ATLAS raw, unshifted data; in  Figs.~\ref{fig:jetatlasB}, the
optimal systematic shifts have been applied to the ATLAS central values.
The scale uncertainty in the theory prediction is calculated by
varying the QCD scale in the range $P_{Tj}/2 \leq \mu
\leq 2 P_{Tj}$. The uncertainties of the data points shown in
Figures~\ref{fig:jetatlasA}  and ~\ref{fig:jetatlasB} are evaluated
by adding the statistical and uncorrelated systematic errors in
quadrature. 

The overall agreement of the shifted data 
with the (NLO) predictions utilizing CT10NNLO PDFs is good even without 
theory uncertainty: $\chi^2_E/N_{pt}=0.74\ (0.78)$ for
$R=0.6\ (0.4)$.  Some differences in the shape and
normalization observed for the unshifted data disappear 
upon application of the systematic error shifts.
The $\chi^2_E$ function is defined according to the
extended $T$ convention for the ATLAS correlation matrix. 
Slightly higher values of $\chi^2_E$ would be obtained with
alternative conventions for $\beta_{i,\alpha}$ \cite{Ball:2012wy}.

\subsection{SM Higgs boson total cross sections}

\figHxsecA

Now we turn to the NNLO total cross sections for Higgs boson production 
at both the Tevatron and LHC via gluon-gluon fusion
\cite{Catani:2007vq,Grazzini:2008tf} and $b\bar b$
annihilation~\cite{Harlander:2003ai}, shown as ratios 
to the CT10NNLO prediction in Fig.~\ref{H-xsecA}. 
At the Tevatron, PDF uncertainties
of the Higgs boson production cross section via gluon-gluon fusion
are large and cover all the central predictions. At the LHC,
the CT10NNLO gluon-gluon fusion predictions are in a good agreement
with those from the other PDF groups, 
except for NNPDF2.3 at 7 and 8 TeV, which
predicts larger cross sections. For Higgs boson production 
via $b\bar b$ annihilation, the NNLO
predictions obtained with all PDFs are consistent within the quoted PDF
uncertainties. 
\section{Discussion and Conclusion
\label{sec:conclusion}}

We have presented next-to-next-to-leading order (NNLO) parton distribution 
functions (PDFs) from the CTEQ-TEA group. These CT10NNLO PDFs have
been determined  
based on essentially the same global data sets used in the previous
CT10 and CT10W NLO PDF  
analyses. In this new analysis, the effects of finite quark masses 
have been implemented in the S-ACOT-$\chi$ scheme
at NNLO accuracy. We obtain a similar quality of agreement with
the fitted experimental data sets in the NNLO fit as at NLO. 

We find that at low $x$ (below $10^{-2}$), the NNLO gluon distribution
is suppressed, while the quark distributions increase, compared to the
same distributions at NLO. The ${\cal O}(\alpha_{s}^{2})$  GM VFN
scheme used in the NNLO fit results in changes to the heavy quark
distributions, both charm and bottom.  The large-$x$ gluon and $d$-quark
distributions are reduced due to (1) the removal of the Tevatron Run-1
inclusive jet data, (2) the
alternate treatment of correlated systematic errors and 
choices of renormalization and factorization scales in jet cross
sections, and (3) revised electroweak couplings in DIS cross sections.

Compared to the MSTW2008 NNLO PDFs, the gluon and quark distributions
are larger as $x$ approaches zero (the CT10 parameterization requires
a positive gluon), while the strangeness distribution is larger over
most of the $x$ range. The differences tend to decrease as $Q^2$
increases.  

We have compared NNLO predictions using CT10 to available LHC data and
have found good agreement. Numerical comparisons presented here
complement a more detailed benchmark study of the dependence of 
LHC predictions on PDFs and $\alpha_s(M_Z)$ 
that has been recently released~\cite{Ball:2012wy}. The available LHC
data is already providing important information on PDFs, and future
data will provide even stronger constraints. This will be developed
further in a future publication.  

\begin{acknowledgments}

This work was supported by the U.S. DOE Early Career Research Award
DE-SC0003870 and by Lightner-Sams Foundation; by the U.S. Department
of Energy under Grant No. DE-FG02-96ER40969; by the U.S. National
Science Foundation under Grant No. PHY-0855561; by the National
Science Council of Taiwan under Grant Nos. NSC-98-2112-M-133-002-MY3
and NSC-101-2112-M-133-001-MY3. 

\end{acknowledgments}


\end{document}